\documentclass[aps, prd, reprint, amsfonts,
  amssymb, amsmath, showpacs, preprintnumbers, letterpaper,
  nofootinbib,floatfix,superscriptaddress]{revtex4-1}

\usepackage{braket,bm}
\usepackage[pass]{geometry} 
\usepackage{color}
\usepackage{mathrsfs,mathtools}
  \DeclareMathAlphabet{\mathpzc}{OT1}{pzc}{m}{it}

\usepackage{multirow}
  
\usepackage{colortbl}

\usepackage{soul}

\newcommand{\fref}[1]{Fig.~\ref{#1}}

\newcommand{\veryshortarrow}[1][3pt]{\mathrel{%
   \hbox{\rule[\dimexpr\fontdimen22\textfont2-.2pt\relax]{#1}{.4pt}}%
   \mkern-4mu\hbox{\usefont{U}{lasy}{m}{n}\symbol{41}}}}

\makeatletter

\setbox0\hbox{$\xdef\scriptratio{\strip@pt\dimexpr
    \numexpr(\sf@size*65536)/\f@size sp}$}

\newcommand{\scriptveryshortarrow}[1][3pt]{{%
    \hbox{\rule[\scriptratio\dimexpr\fontdimen22\textfont2-.2pt\relax]
               {\scriptratio\dimexpr#1\relax}{\scriptratio\dimexpr.4pt\relax}}%
   \mkern-4mu\hbox{\let\f@size\sf@size\usefont{U}{lasy}{m}{n}\symbol{41}}}}

\makeatother

\begin{document}


\title{Fluctuations about the Fubini-Lipatov instanton for \\
false vacuum decay in classically scale invariant models}

\author{Bj\"{o}rn Garbrecht}
\email{garbrecht@tum.de}
\affiliation{Physik-Department T70, Technische Universit\"{a}t M\"{u}nchen,\\ James-Franck-Stra\ss e, 85748 Garching, Germany}

\author{Peter Millington}
\email{p.millington@nottingham.ac.uk}
\affiliation{School of Physics and Astronomy, University of Nottingham,\\
Nottingham NG7 2RD, United Kingdom}


\preprint{TUM-HEP-1136-18}


\begin{abstract}

For a scalar theory whose classical scale invariance is broken by quantum effects,
we compute self-consistent bounce solutions and Green's functions. Deriving analytic expressions,
we find that the latter
are similar to the Green's functions in the archetypal thin-wall model for tunneling
between quasi-degenerate vacua. The eigenmodes and eigenspectra are, however, very
different. Large infrared effects from the modes of low angular momentum $j=0$ and $j=1$,
which include the approximate dilatational modes for $j=0$,
are dealt with by a resummation of one-loop effects.
For a parametric example, this resummation is carried out numerically.

\end{abstract}

\maketitle


\section{Introduction}

A sufficiently large lifetime of metastable vacuum states~\cite{Coleman:1977py, Callan:1977pt} is an important criterion
for the viability of models of electroweak symmetry breaking~\cite{Isidori:2001bm,Branchina:2013jra,Chigusa:2017dux,Andreassen:2017rzq}. For other sectors that are more or less closely tied to the electroweak one, false vacuum decay
can play an essential role in cosmology~\cite{Kusenko:1996xt}. Since tunneling events do not correspond
to extrema of the Minkowskian action, the calculation of the decay rates relies on Euclidean
solitons, which are saddle-point configurations often referred to as \emph{bounces}~\cite{Coleman:1977py}.

Since the bounce action enters the decay rate exponentially, leading-order calculations are
often sufficient to check the metastability. Higher precision can be achieved when
including the first quantum corrections, which also establish the correct dimensionful prefactors
for a decay rate per unit volume by trading the zero modes associated with the spontaneous breakdown
of spacetime symmetries for integrals over collective coordinates.

The fluctuation modes around the solitons differ from those in calculations of effective
potentials because they include the gradient corrections from the varying background, whereas, for the
effective potential, one assumes a field configuration that is constant throughout spacetime.
Nevertheless, effective potentials can be useful in order to include radiative effects, provided
it can be justified that gradient corrections are of higher order and pathologies, such as
imaginary parts appearing in concave regions of the tree potential, can be ignored.

One-loop corrections to the action that account for the gradients can be expressed through the
functional determinant of the quadratic fluctuations around the bounce. A powerful method
to deal with this apparently complicated task is provided by the Gel'fand-Yaglom theorem, which reduces the problem to one of solving field equations for certain boundary conditions~\cite{Gelfand:1959nq,Coleman:1985:book}.

When dealing with bounce configurations
that are not perturbatively close to a tree-level solution (see Ref.~\cite{Garbrecht:2015cla}), we encounter a significant limitation of this method. Such a situation occurs for models of radiative
symmetry breaking, where the true vacuum only appears through loop corrections~\cite{Garbrecht:2015yza} or, as in the
case of interest for the present work, for approximately scale-invariant models, where the
scale of the radius of the nucleating bubbles is not known before consistently accounting for quantum effects. In particular, for a scalar theory with a negative quartic self-coupling, the classical solution, known as the Fubini-Lipatov instanton~\cite{Fubini:1976jm,Lipatov:1976ny},
contains a scale parameter that determines the radius of the nucleating bubbles, which is
not fixed at tree level.
In this case, one needs to find the bounce solutions by computing radiative corrections
to the equations of motion self-consistently within the bounce field configuration, which is the main
objective of the present paper.
We must emphasize that there are many ways in which the classical
scale invariance can be broken, and we make here a specific choice that we find most suitable
for our present methodical developments. Our particular setup is therefore presented in
Sec.~\ref{sec:setup}, and additional technical details are provided in App.~\ref{app:CW}.

Methodically, we compute the loop corrections from Green's functions in the bounce background. This approach has already been applied to examples in the thin-wall limit in Refs.~\cite{Garbrecht:2015oea,Garbrecht:2015yza,Garbrecht:2017idb}.
Formally, these Green's functions are the inverse of the quadratic fluctuation operator, where
the aforementioned zero modes must be projected out in order to leave the inversion well defined.
We find that analytic solutions are available and that these seem to suggest a decomposition into zero and positive- or negative-definite modes in a form that is not consistent with what we obtain when computing these contributions explicitly. This may be
a concern, and a way of checking
this and of gaining further insight is to obtain the Green's function from its explicit representation
in terms of a spectral sum. This requires the knowledge of the eigenmodes, the complete set of which, however, does not appear
to be available in terms of analytic expressions for
the Fubini-Lipatov case. In Sec.~\ref{sec:spectrum}, we therefore return
to the archetypal case of tunneling between quasi-degenerate vacua, originally considered
by Coleman and Callan~\cite{Coleman:1977py, Callan:1977pt}, where the spectral sum representation can be worked out explicitly.
It turns out there that the contributions from the discrete modes cannot simply be ``read off''
from the analytic solution for the Green's function, such that we may not
expect such a straightforward decomposition for the Fubini-Lipatov case either, which should address the above concern.

In Sec.~\ref{sec:zeroes}, we describe a procedure for treating the zero modes
pertaining to the spontaneous breakdown of translational invariance in the bounce background.
Particular attention is paid to combine this with the resummation of one-loop corrections
in order to regulate the infrared divergences that occur in this sector.

These developments are subsequently applied to the case of a Fubini-Lipatov instanton with corrections
from scalar loops in Sec.~\ref{sec:selfcon}. We also explain in that section how we
deal with infrared effects due to dilatational zero modes associated with the classical scale invariance. These modes are not normalizable in the proper sense because they appear at
the end point of a continuum spectrum. Our calculation makes use of the physical regularization
of the infrared divergences, i.e.~due to the radiative breakdown of scale invariance, by
resumming the one-loop corrections within the Green's function. This method therefore
presents an alternative to the subtraction of the dilational sector in favour
of a collective coordinate for the scale transformations.

Finally, in Sec.~\ref{sec:results}, we apply the methods presented in this
work to a parametric example of scalar theory with quartic interactions and extra
couplings to scalar fields. This serves to illustrate and to test the methods
presented here such that further developments and parametric studies can follow in the future.
Our conclusions and possible future directions are presented in Sec.~\ref{sec:conclusions}.


\section{Setup}
\label{sec:setup}

We work with the following Euclidean Lagrangian, comprising a real scalar field $\Phi_x\equiv\Phi(x)$:
\begin{equation}
  \label{eq:lag}
  \mathcal{L}\ =\ \frac{1}{2}\,(\partial_\mu \Phi)^2\:+\:U(\Phi)\:+\:\delta U(\Phi)\;,
\end{equation}
where the classical potential is
\begin{equation}
  \label{eq:U:classical}
  U(\Phi)\ =\ \frac{1}{2}\,m^2\Phi^2 \:+\:\frac{1}{3!}\,g\,\Phi^3\:+\: \frac{1}{4!}\,\lambda\,\Phi^4\;,
\end{equation}
and
\begin{equation}
  \label{eq:pot}
  \delta U(\Phi)\ =\ \frac{1}{2}\,\delta m^2\,\Phi^2\:+\:\frac{1}{3!}\,\delta g\,\Phi^3\:+\:\frac{1}{4!}\,\delta \lambda\,\Phi^4
\end{equation}
contains the mass and coupling-constant counterterms. In Eq.~\eqref{eq:lag}, $\partial_{\mu}\equiv\partial/\partial x_{\mu}$ is the derivative with respect to the Euclidean coordinate $x_{\mu}\equiv(\mathbf{x},x_4)$.

When $\lambda<0$, the potential $U(\varphi)$ exhibits a false vacuum at $\varphi\equiv\braket{\Phi}=0$ and is unbounded from below for $\varphi\to\pm\,\infty$. At the level of the classical potential, and when setting $m=0$ and $g=0$, transitions from the false vacuum at $\varphi=0$ proceed via quantum-mechanical tunneling, and the bounce solution
to the equation of motion
\begin{equation}
\label{eq:bounce}
-\:\frac{{\rm d}^2}{{\rm d}r^2}\,\varphi\:-\:\frac{3}{r}\,\frac{{\rm d}}{{\rm d}r}\,\varphi\:+\:U'(\varphi)\ =\ 0
\end{equation}
is the Fubini-Lipatov instanton~\cite{Fubini:1976jm,Lipatov:1976ny}:
\begin{equation}
\label{eq:FL}
\varphi(r)\ =\ \frac{\varphi(0)}{1+r^2/R^2}\;,
\end{equation}
where $\varphi(0)=\sqrt{-\,48/\lambda/R^2}$, $r=\sqrt{x\cdot x}$ is the four-dimensional radial coordinate and $R$ is a constant that characterizes the radius of the nucleating
bubble. The bounce action is given by~\cite{Linde:1981zj}
\begin{equation}
  \label{eq:B}
  B\ =\ 2\pi^2\int_0^{\infty}\!{\rm d}r\;r^3\bigg[\frac{1}{2}\,\big(\partial_r\varphi\big)^2+\frac{\lambda}{4!}\,\varphi^4\bigg]\ =\ -\,\frac{16\pi^2}{\lambda}
\end{equation}
and is independent of $R$, as a consequence of classical scale invariance.
These results should give a first approximation to the tunneling rate per unit volume
\begin{equation}
  \varGamma/V\ \sim\ e^{-B/\hbar}\;,
\end{equation}
whenever $m\ll R^{-1}$, which is what we will also find explicitly when including radiative corrections, cf.~the results for the bounce solutions shown in Sec.~\ref{sec:results}.

Now, radiative corrections will, in general, break classical scale invariance, and the field experiences these already in the false vacuum, where $\varphi=0$.  We choose a nonvanishing mass at the symmetric point as a renormalization condition and thereby as the unique renormalization scale, i.e.~we take $m\neq 0$ in Eq.~\eqref{eq:U:classical} and the renormalization condition $U^{{\rm ren}\prime\prime}_{\rm eff}(\varphi)|_{\varphi\,=\,0}=m^2$.
More details on this procedure are presented in App.~\ref{app:CW}.
The existence of a bounce solution then requires the presence of radiative corrections~\cite{Linde:1981zj} and, when these are small, there is the hierarchy $m\ll R^{-1}$, such that $m R$ is perturbatively small.
Since the inverse radius is much larger than the scale $m$, we can consider the present setup as a classically scale-invariant model, albeit with a small breaking provided by the parameter $m$. We will confirm the perturbative deviation from scale invariance
explicitly in the numerical studies of Sec.~\ref{sec:results}.

From this perspective, and without a symmetry to protect the vanishing of the mass of the scalar field when $\varphi=0$, $m$ appears as a dimensionful scale that is necessarily introduced by radiative corrections, similar to the scale $\mu$ at which Coleman and Weinberg fix the quartic coupling in their original work on radiative symmetry breaking~\cite{Coleman:1973jx}. We note that
the present setup would not change substantially when choosing the tree-level parameter $m^2=0$
, while still maintaining $U^{{\rm ren}\prime\prime}_{\rm eff}(\varphi)|_{\varphi\,=\,0}=m^2$, but we
prefer the above choice for calculational simplicity.

The present model is reminiscent of approximately scale-invariant theories that may be trapped in a false vacuum at some point during the cosmological evolution. Depending on the exact masses of the Higgs boson and the top quark, the Standard Model may be the most important example of such a theory~\cite{Buttazzo:2013uya}.
Nevertheless, there are important qualitative differences in the running of the quartic coupling
and in the way in which the potential barrier between false and true vacuum is generated. We will
comment on these matters in the Conclusions.

We also include here the effect from $N_\chi$ extra fields $\chi_i$ of mass $m_\chi$:
\begin{align}
\label{eq:speclag}
{\cal L}\ \rightarrow\ {\cal L}\:+\:\sum\limits_{i\,=\,1}^{N_\chi}\left\{\frac{1}{2}\,(\partial_\mu\chi_i)^2\:+\:\frac{1}{2}\, m_\chi^2\chi_i^2\:+\:\frac{1}{4}\,\alpha\,\Phi^2\chi_i^2\right\}\;.
\end{align}
These additional fields are useful for controlling the amount of radiative breaking of the scale invariance.
The effective potential resulting from our choice of renormalization conditions is given by (see App.~\ref{app:CW})
\begin{align}
  \label{eq:U:eff:ren}
  U^{\rm ren}_{\rm eff}(\varphi)\ &=\ \frac{1}{2}\,m^2\varphi^2\:+\:\frac{\lambda}{4!}\,\varphi^4\nonumber\\
  &\quad +\:\frac{1}{256\pi^2}\Big\{\big(2m^2\:+\:\lambda\varphi^2\big)^2\ln\frac{2m^2\:+\:\lambda\varphi^2}{2m^2}\nonumber\\
  &\qquad-\:2\lambda\, m^2\varphi^2\:-\:\frac{3}{2}\,\lambda^2\varphi^4\Big\}\nonumber\\
  &\quad+\:\frac{N_\chi}{256\pi^2}\Big\{\big(2m_\chi^2\:+\:\alpha\varphi^2\big)^2\ln\frac{2m_\chi^2\:+\:\alpha\varphi^2}{2m_{\chi}^2}\nonumber\\
  &\qquad -\:2\,\alpha\, m_\chi^2\varphi^2\:-\:\frac{3}{2}\,\alpha^2\varphi^4\Big\}\;.
\end{align}
We note that since $\lambda<0$, this radiative contribution to the effective potential develops an imaginary part for $\varphi>\sqrt{-\,2/\lambda}$. This is due to the tachyonic instability of the field $\Phi$ in that region. Nevertheless, the effective action remains real when evaluated at the bounce solution. Namely, the imaginary part present in the effective potential --- which assumes a constant, homogeneous field configuration --- is removed when we account for the gradients of the field.


\section{Spectrum of fluctuations}
\label{sec:spectrum}

We aim to compute radiative corrections to tunneling transitions using Green's functions.
While these can be obtained as direct solutions to their defining
equations, additional insights can be gained through consideration of the particular contributions
from the fluctuation spectrum.
Unfortunately, in the Fubini-Lipatov case, where  $m^2=0$ and $\lambda<0$, the problem of finding the complete eigenspectrum and all eigenmodes of the fluctuation operator
is not fully analytically tractable.

It is therefore instructive to  compare with the archetypal example of tunneling in quantum field theory: the case $m^2=-\,\mu^2<0$ and $\lambda>0$, as studied in the seminal works by Coleman and Callan~\cite{Coleman:1977py, Callan:1977pt}.
When the $\mathbb{Z}_2$-breaking term ($g\,\Phi^3/3!$) is small, the minima are quasi-degenerate and the radius of the critical bubble is extremely large compared to the width of the bubble wall: $\mu R=2\sqrt{3}\mu \lambda/g\gg 1$. This is the \emph{thin-wall} regime, which leads to simplifications such that
the full spectral decomposition can be carried out in an analytic calculation.
Moreover, it turns out that the Green's functions for the Fubini-Lipatov and the
thin-wall cases agree up to prefactors and the
dependence of the radial part on the total angular momentum quantum number $j$. We will therefore present a transformation
from the basis of the fluctuation operator for the thin-wall problem to the Fubini-Lipatov
case. We shall refer to the respective bases as the \emph{thin-wall} and \emph{Fubini-Lipatov
bases}.

In order to deal with the fluctuations in the four-dimensional
spherically symmetric situation,
we separate the angular dependence as
\begin{align}
\phi_{\lambda^X j \{\ell\}}(x)\ =\ \phi_{\lambda^Xj}(r)Y_{j\{\ell\}}(\mathbf{e}_x)\;.
\end{align}
The radial eigenfunctions $\phi_{\lambda^Xj}(r)$ carry labels for the radial eigenvalue $\lambda^X$
(which may be discrete or part of a continuum), as well as the total angular momentum $j$. The $Y_{j\{\ell\}}$ are hyperspherical harmonics, where $\ell\equiv \{\ell_1,\ell_2\}$, with $\ell_1=0,1,\ldots, j$ and
$\ell_2=-\ell_1,-\ell_1+1,\ldots,\ell_1$, and $\mathbf{e}_x\equiv\mathbf{x}/|\mathbf{x}|$ is a four-dimensional unit vector. The radial eigenvalue equation is
\begin{align}
\label{eq:EV}
&\bigg[
-\:\frac{{\rm d}^2}{{\rm d}r^2}\:-\:\frac{3}{r}\,\frac{\rm d}{{\rm d} r}\:+\:\frac{j(j+2)}{r^2}\nonumber\\&\qquad \qquad +\:U''(\varphi)
\bigg]\phi_{\lambda^X j}\ =\ \lambda^X \phi_{\lambda^X j}\;.
\end{align}
This applies to both the Fubini-Lipatov and the thin-wall cases, which differ in the particular
form of $U(\varphi)$. In the Fubini-Lipatov case, the operator on the left-hand side of
Eq.~\eqref{eq:EV} has a set of eigenfunctions that can be expressed in either the thin-wall basis $X={\rm TW}$ or the Fubini-Lipatov basis $X={\rm FL}$, see Sec.~\ref{sec:specsum}.

In both cases $m^2=-\,\mu^2<0$, $\lambda>0$ (thin wall) and $m=0$, $\lambda<0$ (Fubini-Lipatov), we can exploit the $O(4)$ symmetry of the bounce. Working in hyperspherical coordinates, this allows us to expand the Green's functions
as follows:
\begin{align}
\label{eq:Green:jsum}
G(x,x')\ =\ \frac{1}{2\pi^2}\sum\limits_{j\,=\,0}^\infty (j+1)U_j(\cos\theta) G_j(r,r^\prime)\;,
\end{align}
where $\cos\theta=x\cdot x'/(|x|\,|x'|)$ and the $U_j$ are Chebyshev polynomials of the second kind. The hyperradial Green's function $G_j(r,r')$ satisfies the equation
\begin{align}
\label{eq:hyperradeq}
&\bigg[
-\:\frac{{\rm d}^2}{{\rm d}r^2}\:-\:\frac{3}{r}\,\frac{\rm d}{{\rm d} r}\:+\:\frac{j(j+2)}{r^2}\:+\:U''(\varphi)
\bigg]
G_j(r,r')\nonumber\\&\qquad =\ \frac{\delta(r-r')}{r'^3}\;.
\end{align}

In the presence of zero modes, the Green's function can only be defined in the subspace perpendicular to these,
implying that such modes need to be subtracted. Therefore, and in order to gain further insight into
the nature of the radiative effects, it is useful to represent the Green's function as a
spectral sum, which can be written in the following form:
\begin{align}
\label{eq:specsum}
&G(x,x')\ =\ \frac{1}{2\pi^2}\sum_{j\,=\,0}^{\infty}(j+1)U_j(\cos\theta)\nonumber\\&\qquad\times\left[\sum_{\lambda^X\,\in\, L^X_{{\rm d}j}}\:+\!\!\int\limits_{\lambda^X\,\in\, L^X_{{\rm c}j}}\!\!\!\!\!\!\frac{{\rm d}\lambda^{X}}{2\pi}\right]\frac{\phi_{\lambda^X j}^{-}(r')\phi^{+}_{\lambda^X j}(r)}{\lambda^X}\;,
\end{align}
where the sum runs over the set of discrete eigenvalues $L^X_{{\rm d}j}$ and the integral over the continuum $L^X_{{\rm c}j}$. The $+$ and $-$ indicate the basis and reciprocal basis. Note that the discrete and continuum eigenmodes have different dimensionality, since
\begin{subequations}
\begin{align}
\int{\rm d}r\;r^3\;\phi^{-}_{\lambda^{X\prime}j}(r)\phi^{+}_{\lambda^X j}(r)\ &=\ \delta_{\lambda^X \lambda^{X\prime}}\nonumber\\&\qquad \textnormal{for}\;\lambda^X,\lambda^{X\prime}\: \in\: L^X_{{\rm d}j}\;,\\
\int{\rm d}r\;r^3\;\phi^{-}_{\lambda^{X\prime} j}(r)\phi^{+}_{\lambda^X j}(r)\ &=\ {2\pi}\delta(\lambda^X-\lambda^{X\prime})\nonumber\\&\qquad \textnormal{for}\;\lambda^X,\lambda^{X\prime}\: \in\: L^X_{{\rm c}j}\;,
\end{align}
\end{subequations}
i.e.~the former are normalizable and the latter normalizable in the improper sense. For the real scalar field considered here, the basis and reciprocal basis are the same in the case of the discrete modes, and we will therefore drop the distinction and the superscripts $+$ and $-$ in what follows. Note that the basis and its reciprocal are distinct for the continuum modes.


\subsection{Zero and negative modes}

We first consider the eigenmodes with zero and negative eigenvalues. The zero modes are
associated with the spontaneous breakdown of symmetries: translations in both the Fubini-Lipatov
and thin-wall cases, and, in addition, dilatations around the tree-level Fubini-Lipatov instantons.
A negative eigenmode is a hallmark of metastable states, reflecting the fact that
the bounce is a saddle-point solution, such that it is present in both cases. We will show
that the negative mode can approximately be associated with a dilatation of the bounce in the thin-wall case,
whereas, about the Fubini-Lipatov instanton, it does not correspond to dilatations, which yield, in contrast, a nonnormalizable
zero mode.

\subsubsection{Thin wall}

 In the \emph{thin-wall} regime, the gradients of the bounce are negligible everywhere except in the vicinity of the bubble wall. We can therefore make the following series of approximations for the damping term in the equation of motion~\eqref{eq:bounce}:
\begin{equation}
  -\:\frac{3}{r}\,\frac{{\rm d}}{{\rm d}r}\,\varphi(r)\ \approx\ -\:\frac{3}{R}\,\frac{{\rm d}}{{\rm d}r}\,\varphi(r)\ \approx\ 0\;,
\end{equation}
and the bounce is given by the well-known kink solution
\begin{equation}
  \varphi(r)\ =\ v\,\mathrm{tanh}[\gamma(r-R)]\;,\qquad \gamma\ \equiv\ \mu/\sqrt{2}\;.
\end{equation}

The four-dimensional translational invariance of the action leads to four eigenmodes of zero eigenvalue. Since the multiplicity of the $j$-th angular momentum mode is $(j+1)^2$ (in four dimensions), these translational zero modes must have the angular quantum number $j=1$. This may readily be verified by acting on the equation of motion for the bounce with the infinitesimal generator of translations $P_{\mu}=-\,i\partial_{\mu}$:
\begin{align}
\label{eq:Pmu}
  P_{\mu}\bigg[-\:\frac{{\rm d}^2}{{\rm d}r^2}\,\varphi\:-\:\frac{3}{r}\,\frac{{\rm d}}{{\rm d}r}\,\varphi\:+\:U'(\varphi)\bigg]\ &=\ 0\nonumber\\
  \Leftrightarrow\ \bigg[-\:\frac{{\rm d}^2}{{\rm d}r^2}\:-\:\frac{3}{r}\,\frac{{\rm d}}{{\rm d}r}\:+\:\frac{3}{r^2}\:+\:U''(\varphi)\bigg]P_{\mu}\varphi(r)\ &=\ 0\;.
\end{align}
It follows that the four zero modes have the explicit forms
\begin{equation}
  \phi_{\mu}(r)\ \propto\ P_{\mu}\varphi(r)\ \propto\ \frac{x_{\mu}}{r}\,\mathrm{sech}^2[\gamma(r-R)]\;,
\end{equation}
where $\mu\in\{1,2,3,4\}$.

The negative eigenmode owes its existence to the instability.
More specifically, in the thin-wall case, it
arises because the bounce action is a \emph{maximum} with respect to the radius of the critical bubble $R$. In fact, the negative eigenvalue is given by~\cite{Garbrecht:2015oea}
\begin{equation}
\label{eq:negev}
  \lambda^{\rm TW}_{20}\ =\ \frac{1}{B}\,\frac{\delta^2 B}{\delta R^2}\ = \ -\:\frac{3}{R^2}\;,
\end{equation}
where TW stands for ``thin wall'', and the subscripts will be clarified in the remainder of this section.
We may therefore anticipate that the negative eigenmode is to be associated with dilatations of the critical bubble in the present case, that is with $\partial_r\varphi=  -\,\partial_R\varphi$. Acting on the equation of motion for the bounce with the operator $\partial_r$, we find
\begin{align}
  \partial_r\bigg[-\:\frac{{\rm d}^2}{{\rm d}r^2}\,\varphi\:-\:\frac{3}{r}\,\frac{{\rm d}}{{\rm d}r}\,\varphi\:+\:U'(\varphi)\bigg]\ &=\ 0\nonumber\\
  \Leftrightarrow\ \bigg[-\:\frac{{\rm d}^2}{{\rm d}r^2}\:-\:\frac{3}{r}\,\frac{{\rm d}}{{\rm d}r}\:+\:\frac{3}{r^2}\:+\:U''(\varphi)\bigg]\partial_r\varphi(r)\ &=\ 0\;,\label{eq:diln}
\end{align}
which we can immediately relate to the eigenvalue equation of the translational zero modes ($j=1$), cf.~Eq.~\eqref{eq:Pmu}. When we
apply the thin-wall approximation for the centripetal term, replacing in Eq.~\eqref{eq:diln}
\begin{equation}
  \frac{3}{r^2}\,\partial_r\varphi(r)\ \approx\ \frac{3}{R^2}\,\partial_r\varphi\;,
\end{equation}
(as is appropriate because the field gradients are peaked at $r\sim R$)
and comparing with Eq.~\eqref{eq:EV},
we indeed recover the negative eigenvalue~\eqref{eq:negev}. We therefore conclude that the negative eigenmode can be associated with dilatations in the thin-wall regime.

To understand further whether dilatations generally pertain to the negative eigenvalue, we
act on the bounce with the infinitesimal generator of dilatations.
A geometric dilatation is generated by $\bar D = x_{\mu}\partial_{\mu}$, and
taking account of the scaling dimension {\it one} for the scalar field, a scale
transformation is generated by $D=1+\bar D= 1+ x_{\mu}\partial_{\mu}$, which we refer to
as a dilatation (without the adjective geometric), in the same way this term is used
in recent literature~\cite{Andreassen:2017rzq,Chigusa:2018uuj}. Spherical symmetry then
implies that
$\bar D\varphi(r)=r\partial_r\varphi(r)\approx -R\partial_R\varphi(r)$, where the latter approximation holds only in the thin-wall regime, since $\mathrm{sech}^2[\gamma(r-R)]$ is strongly peaked at $r\sim R$, which confirms that the
shape of the negative mode is close to a dilatation. Note further that in the thin-wall regime, where
$R\gg\gamma^{-1}$, $D\varphi(r)\approx \bar D\varphi(r)$.
However, acting instead on the equation of motion for the bounce, we find
\begin{align}
  \label{eq:dilatations}
  \bar D\bigg[-\:\frac{{\rm d}^2}{{\rm d}r^2}\,\varphi\:-\:\frac{3}{r}\,\frac{{\rm d}}{{\rm d}r}\,\varphi\:+\:U'(\varphi)\bigg]\ &=\ 0\nonumber\\
  \Leftrightarrow\ \bigg[-\:\frac{{\rm d}^2}{{\rm d}r^2}\:-\:\frac{1}{r}\,\frac{{\rm d}}{{\rm d}r}\:+\:\frac{4}{r^2}\:+\:U''(\varphi)\bigg]\bar D\varphi\ &=\ 0\;,
\end{align}
illustrating that one cannot conclude that geometric dilatations correspond to eigenmodes of any eigenvalue.

Finally, note that in the thin-wall limit, the negative and zero eigenmodes
do not contribute to the Green's function because of a vanishing integration measure. Nevertheless,
for each $j$, there are two discrete eigenvalues and a continuum starting for positive energies
in the corresponding quantum mechanical problem~\cite{Garbrecht:2015oea}.

\subsubsection{Fubini-Lipatov instanton}
\label{subsubsec:FL}

We now turn our attention to the spectrum of fluctuations over the Fubini-Lipatov instanton in Eq.~\eqref{eq:FL}. For this given background, the eigenvalue equation~\eqref{eq:EV} has the form
\begin{align}
\label{eq:EV:FL}
&\bigg[-\:\frac{{\rm d}^2}{{\rm d} r^2}\:-\:\frac{3}{r}\,\frac{{\rm d}}{{\rm d}r}\:+\:\frac{j(j+2)}{r^2}\nonumber\\&\qquad-\:\frac{24R^2}{(r^2+R^2)^2}\:-\:\lambda^{\rm FL}\bigg]\phi_{\lambda^{\rm FL} j}(r)\ =\ 0\;.
\end{align}
The classical action is invariant under both four-dimen\-sional translations and dilatations, and the
dilatational and translational zero modes are given in terms of the associated Legendre polynomials $P_n^{\omega}(z)$ of degree $n=2$
and order $\omega=j+1=1,2$.

The translational symmetry leads to four zero eigenmodes, as before, and we can readily verify that these are given by the action of the infinitesimal generator $P_{\mu}=-i\partial_{\mu}$ on the Fubini-Lipatov instanton itself. Specifically, we find
\begin{align}
\label{eq:transmode}
  &\phi_{\mu}(r)\: \propto\: P_{\mu}\,\varphi(r)\: \propto\: \frac{x_{\mu}}{r^2}\,P_2^2\bigg(\frac{1-r^2/R^2}{1+r^2/R^2}\bigg)\nonumber\\&\qquad\qquad\qquad \propto\: \frac{x_{\mu}}{(1+r^2/R^2)^2}\;,
\end{align}
where we use the spacetime index $\mu$ in favour of the pair $\{\lambda^{\rm FL}, j\}$, as in Eq.~\eqref{eq:EV}, in order to
label the translational modes.
These modes have no nodes in the radial direction, such that there are no lower-lying modes for $j=1$,
and, as per the original argument by Coleman and Callan~\cite{Callan:1977pt}, there must therefore exist a lower eigenmode, whose eigenvalue is negative and whose angular momentum is $j=0$.

The action is also invariant under scale transformations, and there exists an additional, but nonnorm\-alizable, zero mode with $j=0$:
\begin{equation}
  \label{eq:dilzero}
  D \varphi
  \: \propto\:
  \partial_R\,\varphi(r)\: \propto\: \frac{1}{r}\,P_2^1\bigg(\frac{1-r^2/R^2}{1+r^2/R^2}\bigg)\: \propto\: \frac{1-r^2/R^2}{(1+r^2/R^2)^2}\;,
\end{equation}
associated with dilatations of the critical bubble, but having nothing to do with geometric dilatation transformations, as per Eq.~\eqref{eq:dilatations}.
We note also that applying the dilatation to the equation of motion and comparing with Eq.~\eqref{eq:EV}
for $j=0$ and $\lambda^{\rm FL}=0$, we find
\begin{align}
&D\left[\left(-\:\frac{\rm d^2}{{\rm d}r^2}\:-\:\frac{3}{r}\,\frac{\rm d}{{\rm d}r}\right)\varphi\:+\:U^\prime(\varphi)\right]
\notag\\&
\quad -\:\left(-\:\frac{\rm d^2}{{\rm d}r^2}\:-\:\frac{3}{r}\frac{\rm d}{{\rm d}r}\:+\:U^{\prime\prime}(\varphi)\right)D\varphi
\notag\\
&\quad =\ 2\,\frac{\rm d^2}{{\rm d}r^2}\,\varphi\:+\:\frac{6}{r}\,\frac{\rm d}{{\rm d}r}\,\varphi\:+\:U^\prime(\varphi)\:-\:\varphi\, U^{\prime\prime}(\varphi)\ =\ 0\;.
\end{align}
Using the equation of motion~\eqref{eq:bounce}, we then obtain
\begin{align}
3\,U^\prime(\varphi)\ =\ \varphi \,U^{\prime\prime}(\varphi)\;,
\end{align}
as we expect because the quartic potential is the unique scale-invariant interaction
term in four dimensions.

\begin{figure}[!t]
\centering
\includegraphics[scale=0.6]{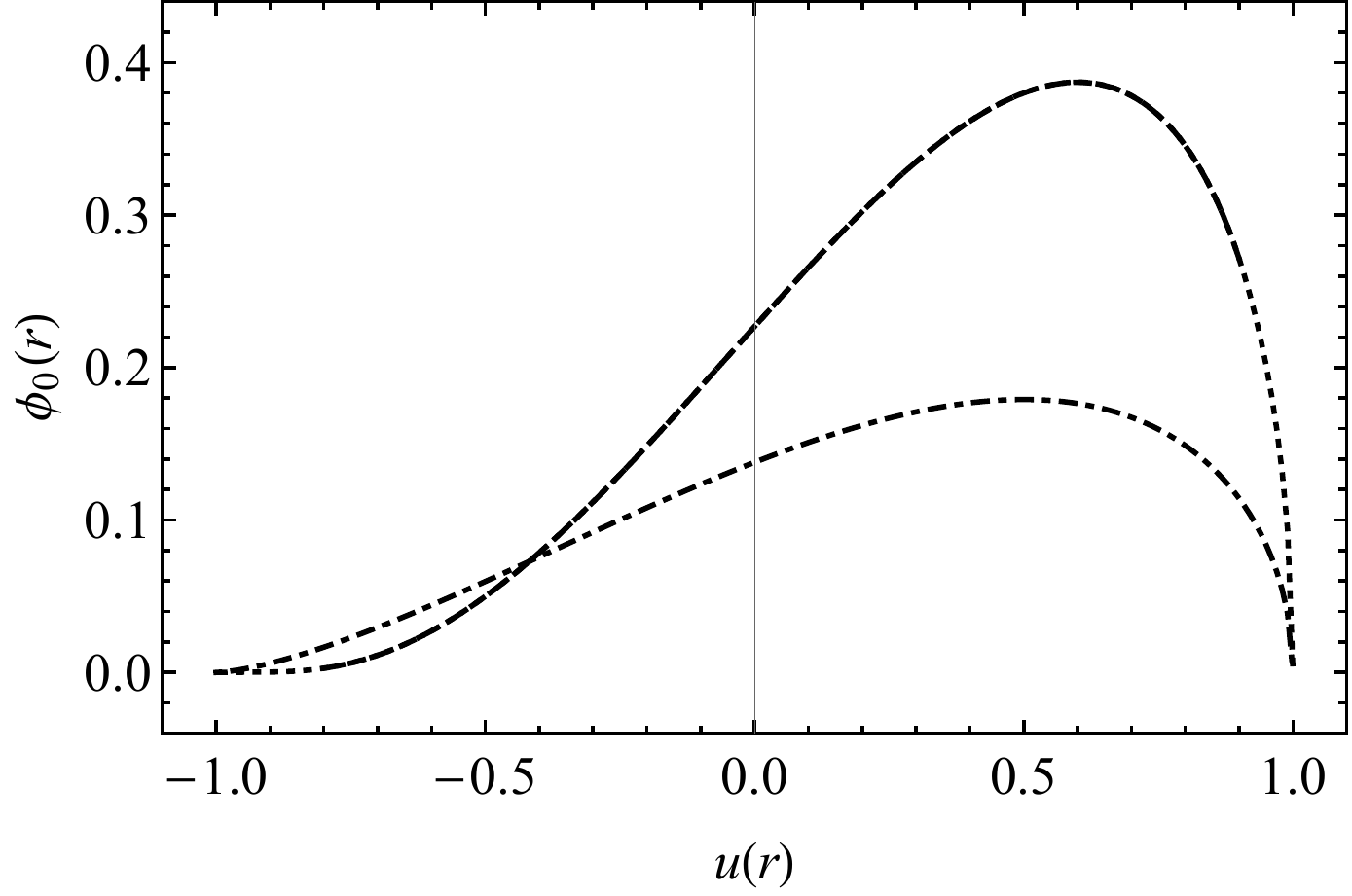}
\caption{\label{fig:negative} Plot of the $[10/10]$ Pad\'{e} approximant of the normalized negative eigenmode (dotted) versus the numerical estimate (dashed) and the ``apparent'' negative mode, as appears in the Green's function (dot-dashed). The value of the negative eigenvalue for the approximate analytic solution was taken from the numerical estimate.}
\end{figure}

We note that, because the mode~\eqref{eq:dilzero} has a single node and zero eigenvalue, there also
is, as anticipated, a negative mode for $j=0$, which is associated with the metastability
of the false vacuum state~\cite{Battarra:2013rba}.
This negative eigenmode satisfies the eigenvalue equation
\begin{equation}
  \bigg[-\:\frac{{\rm d}^2}{{\rm d} r^2}\:-\:\frac{3}{r}\,\frac{{\rm d}}{{\rm d}r}\:-\:\frac{24R^2}{(r^2+R^2)^2}\:+\:|\lambda^{\rm FL}_{20}|\bigg]\phi_{0}(r)\ =\ 0\;.
\end{equation}
For the sake of notational congruence with its thin-wall counterpart~\eqref{eq:negev},
we have attached the subscript $20$ to the negative eigenvalue $\lambda^{\rm FL}_{20}$ and, for simplicity, we label the pertaining lowest-lying mode with the index $0$ rather than the
pair $\{\lambda^{\rm FL},j\}$.
Introducing the dimensionless variable $z=r/R$ and defining $f(z)\equiv(1+z^2)^2\phi_0(r)$, we are looking for the solution to
\begin{align}
  &-\,z\,(1+z^2)f''(z)\:+\:(5z^2-3)f'(z)\nonumber\\&\qquad\qquad +\:z(|\lambda^{\rm FL}_{20}|-8+|\lambda^{\rm FL}_{20}|\,z^2)f(z)\ =\ 0\;.
\end{align}
By the Frobenius method, we obtain the following series solution:
\begin{equation}
  f(z)\ =\ N\sum_{n\,=\,0}^{\infty}a_n\,z^n\;,
\end{equation}
where $N$ is a constant and
with non-vanishing even coefficients satisfying the fourth-order homogeneous recurrence relation
\begin{subequations}
\begin{gather}
 a_0\ = \ 1\;,\\
 a_2\ =\ \frac{|\lambda^{\rm FL}_{20}|-8}{8}\,a_0\;,\\
 a_{n+4}\ =\ \frac{[|\lambda^{\rm FL}_{20}|-n(n-2)]\,a_{n+2}\:+\:|\lambda^{\rm FL}_{20}|\,a_n}{(n+4)(n+6)}\;.
\end{gather}
\end{subequations}
For $\lambda^{\rm FL}_{20}=0$, this series truncates at first order, and we recover the dilatational zero mode in Eq.~\eqref{eq:dilzero}. For a range of $\lambda^{\rm FL}_{20}<0$, the coefficients of this series alternate in sign. While we have been unable to find the solution to this recurrence relation in closed form, we can extend the radius of convergence of a finite truncation of this series by forming the Pad\'{e} approximant. Figure~\ref{fig:negative} shows the approximate analytic (dotted) and numerical (dashed) estimates of the normalized negative eigenmode, compared to the form one might naively extract from the Green's function (dot-dashed) (see Sec.~\ref{sec:Green}).

We note that in the Fubini-Lipatov case, for each $j$, there is a continuum of positive eigenvalues
starting at zero because these are the positive-energy solutions in the corresponding Schr\"odinger problem. It follows that the translational and dilatational zero modes are not discrete. Moreover,
we notice that for $j>1$, the effective potential in the Schr\"odinger problem is positive semi-definite,
such that there cannot be any negative modes. For $j=0$, we observe that the
dilatational mode has one node and the negative mode has zero, such that these
cover all modes with eigenvalues $\leq 0$. For $j=1$, the translational modes have
eigenvalue zero and no nodes, such that there can be now lower-lying modes. In summary,
the spectrum that appears in Eq.~\eqref{eq:specsum} takes the form:
\begin{subequations}
\begin{align}
L^{\rm FL}_{{\rm d}0}\ &=\ \left\{\lambda_{20}^{\rm FL}\right\}\;,\\
L^{\rm FL}_{{\rm d}j}\ &=\ \emptyset\;\;\textnormal{for}\ j\: >\: 0\;,\\
L^{\rm FL}_{{\rm c}j}\ &=\ \left\{\lambda^{\rm FL}\;|\;\lambda^{\rm FL}\: >\: 0\right\}\;,
\end{align}
\end{subequations}
where the superscript ${\rm FL}$ indicates the basis of the operator in the eigenvalue equation~\eqref{eq:EV:FL}
for fluctuations around the Fubini-Lipatov instanton. The above spectrum is also summarized
and compared with the thin-wall case in Table~\ref{tab:eigen}.

For
a self-consistent solution with radiative corrections and $m^2>0$, the continua begin
in contrast at the point $m^2$. While in that case, the translational eigenvalues that are
protected by Goldstone's theorem remain zero, the first nonnegative eigenvalue for $j=0$,
i.e.~the one associated with an approximate dilatation, is lifted to $m^2$. The negative
mode that is present for $j=0$ due to the metastability remains negative and discrete also after taking account of the radiative corrections. In that situation, the above spectrum is modified
to
\begin{subequations}
\begin{align}
L^{\rm FL}_{{\rm d}0}\ &=\ \left\{\lambda_{20}^{\rm FL}\right\}\;,\\
L^{\rm FL}_{{\rm d}1}\ &=\ \left\{0\right\}\;,\\
L^{\rm FL}_{{\rm d}j}\ &=\ \emptyset\;\;\textnormal{for}\ j\:>\:1\;,\\
L^{\rm FL}_{{\rm c}j}\ &=\ \left\{\lambda^{\rm FL}\;|\;\lambda^{\rm FL}\:>\:m^2\right\}\;.
\end{align}
\end{subequations}
%


\subsection{Green's function}
\label{sec:Green}

\subsubsection{Thin wall}
\label{sec:Green:TW}
In the thin-wall regime, we can make the approximations (see~Ref.~\cite{Garbrecht:2015oea})
\begin{subequations}
\begin{gather}
\frac{j(j+2)}{r^2}\ \longrightarrow\ \frac{j(j+2)}{R^2}\;,\\
-\:\frac{3}{r}\,\frac{\rm d}{{\rm d}r}\ \longrightarrow\ -\:\frac{3}{R}\,\frac{\rm d}{{\rm d}r}\ \longrightarrow\ 0\;,
\end{gather}
\end{subequations}
and consistently replace
\begin{equation}
\frac{\delta(r-r')}{r^{\prime3}}\ \longrightarrow\ \frac{\delta(r-r')}{R^3}\;.
\end{equation}
Subsequently, we move to a coordinate aligned with the kink solution itself, via the change of variables
\begin{equation}
\label{eq:thinwallcoords}
u^{(\prime)}\ \equiv\ u^{(\prime)}(r^{(\prime)})\ =\ \tanh[\gamma(r^{(\prime)}-R)\big]\ \in\ (-1,1)\;,
\end{equation}
which yields
\begin{align}
\label{eq:thinwalludiff}
&\Bigg[\frac{\rm d}{{\rm d}u}\,(1-u^2)\,\frac{\rm d}{{\rm d}u}\:-\:\frac{\omega^2}{1-u^2}\:+\:6\Bigg]F_j(u,u')\nonumber\\&\qquad\qquad =\ -\,\delta(u-u')\;,
\end{align}
where $F_j(u,u')\equiv \gamma R^3 G_j(r,r')$ and
\begin{equation}
\label{eq:m1}
\omega\ =\ \bigg(4+\frac{j(j+2)}{\gamma^2R^2}\bigg)^{\!1/2}\;.
\end{equation}
In this way, one finds the Green's function to be~\cite{Garbrecht:2015oea}
\begin{align}
\label{eq:archsol}
  & G_j(r,r') \ = \ \frac{1}{2 \gamma R^3 \omega}\bigg[ \vartheta(u - u') 
  \bigg(\frac{1 - u}{1 + u}\bigg)^{\! \! \tfrac{\omega}{2}} \! 
  \bigg(\frac{1 + u'}{1 - u'}\bigg)^{\! \! \tfrac{\omega}{2}}
  \nonumber \\ 
  & \quad \times \bigg( \! 1 - 3 \,
  \frac{(1 - u)(1 + \omega + u)}{(1 + \omega)(2 + \omega)} \bigg)
  \nonumber \\
  & \quad \times \bigg( \! 1 - 3 \,
  \frac{(1 - u')(1 - \omega + u')}{(1 - \omega)(2 - \omega)} \bigg)
  + (u \leftrightarrow u') \bigg] \; ,
\end{align}
where $\vartheta(z)$ is the generalized unit-step function.

It is helpful, however, to proceed slightly differently, by substituting for the function $\tilde{G}_j(r,r')\equiv r^3G_j(r,r')$ \emph{before} making the thin-wall approximation. Doing the former yields the hyperradial equation
\begin{align}
\label{eq:hyperradeq2}
&\bigg[
-\:\frac{{\rm d}^2}{{\rm d}r^2}\:+\:\frac{3}{r}\,\frac{\rm d}{{\rm d} r}\:+\:\frac{j(j+2)-3}{r^2}\:+\:U''(\varphi)
\bigg]
\tilde{G}_j(r,r')\nonumber\\&\qquad =\ \frac{r^3}{r'^{3}}\,\delta(r-r')\;,
\end{align}
where we emphasize the change in sign of the first-order derivative, the shift in the centrifugal potential and the appearance of the dimensionless ratio of the two hyperradial coordinates on the right-hand side. Making the thin-wall approximations
\begin{subequations}
\label{eq:thinwallapprox}
\begin{gather}
\frac{j(j+2)-3}{r^2}\ \longrightarrow\ \frac{j(j+2)-3}{R^2}\;,\\
+\:\frac{3}{r}\,\frac{\rm d}{{\rm d}r}\ \longrightarrow\ +\:\frac{3}{R}\,\frac{\rm d}{{\rm d}r}\ \longrightarrow \ 0\;,
\end{gather}
\end{subequations}
and consistently replacing
\begin{equation}
\frac{r^3}{r'^3}\,\delta(r-r')\ \to\ \delta(r-r')\;,
\end{equation}
we recover the differential equation~\eqref{eq:thinwalludiff} and the solution in Eq.~\eqref{eq:archsol} but with
\begin{equation}
\label{eq:m2}
\omega\ =\ \bigg(4+\frac{j(j+2)-3}{\gamma^2R^2}\bigg)^{\!1/2}\;.
\end{equation}
While the difference between Eqs.~\eqref{eq:m1} and \eqref{eq:m2} is immaterial in the thin-wall regime (because there we take $R\to\infty$ while keeping the tangential momentum $k\approx j/R$ finite, implying that $j\to\infty$ for fixed $k$), the latter approach allows us to make contact with the true eigenspectrum. Specifically, the coincident limit of the Green's function is
\begin{align}
  \label{eq:Gu}
  G(r, r) \ & = \ \frac{\gamma}{4 \pi^2 R^3}
  \sum_{j\, =\, 0}^{\infty}  \frac{(j + 1)^2}{\omega}\bigg[ \frac{1}{\gamma^2} \nonumber\\&\qquad + 3
  \big( 1 - u^2 \big) \sum_{n\, =\, 1}^2
  \frac{(-1)^n (n - 1 - u^2) }{\gamma^2(\omega^2 - n^2)} \bigg] \;,
\end{align}
where the denominators
\begin{equation}
\lambda^{\rm TW}_{nj}\ = \ \gamma^2(\omega^2\:-\:n^2)\ =\ \gamma^2(4\:-\:n^2)\:+\:\frac{j(j+2)-3}{R^2}
\end{equation}
are the discrete eigenvalues [cf.~Eq.~\eqref{eq:thinwall:discrete:ev} below], as one would naively expect from the contribution of the associated modes to the spectral representation~\eqref{eq:specsum} of the Green's function. At this point, it is tempting to ``read off'' the functional forms of the discrete eigenfunctions. We will return to this point shortly. We notice that the contribution from $j=1$, $n=2$ is singular. This contribution corresponds to the four translational zero eigenmodes. As described in Sec.~\ref{sec:zeroes}, these eigenmodes should be excluded from the Green's function, and we will discuss this further for the Fubini-Lipatov instanton below. Note that, in the analysis of Ref.~\cite{Garbrecht:2015oea}, the sum over discrete angular momenta was traded for a continuous momentum integral in the planar-wall approximation. In this case, the contributions from the zero modes are measure zero, and it is therefore not necessary to exclude them explicitly, cf.~Ref.~\cite{Garbrecht:2015oea}.

\subsubsection{Fubini-Lipatov instanton}

For the classically scale-invariant model, we factor out the $1/r$ dependence of the Green's function, defining
\begin{align}
\label{eq:Gtilde}
\widetilde{G}_j(r,r')\ =\ rr'G_j(r,r')\;.
\end{align}
We then have
\begin{align}
&\Bigg[-\:\frac{{\rm d}^2}{{\rm d} r^2}\:-\:\frac{1}{r}\,\frac{{\rm d}}{{\rm d}r}\:+\:\frac{(1+j)^2}{r^2}\:-\:\frac{24R^2}{\big(r^2+R^2\big)^2}\Bigg]\widetilde{G}_j(r,r')\nonumber\\&\qquad =\ \frac{r}{r^{\prime 2}}\,\delta(r-r')\;.
\end{align}
We again move to coordinates aligned with the background field configuration --- this time the Fubini-Lipatov instanton --- making the change of variables
\begin{equation}
\label{eq:cvars}
u^{(\prime)}\ \equiv\ u^{(\prime)}(r^{(\prime)})\ =\ \frac{1-r^2/R^2}{1+r^2/R^2}\ =\ 2\,\frac{\varphi(r)}{\varphi(0)}\:-\:1\ \in\ (-1,1]\;,
\end{equation}
with $u\to -1$ corresponding to $r\to\infty$. Rearranging Eq.~\eqref{eq:cvars}, we have
\begin{equation}
r^{(\prime)}\ =\ R\bigg(\frac{1-u^{(\prime)}}{1+u^{(\prime)}}\bigg)^{1/2}\;.
\end{equation}
This change of variables leads to
\begin{align}
\label{eq:thickwalltransdiff}
&\Bigg[\frac{\rm d}{{\rm d}u}\,(1-u^2)\,\frac{\rm d}{{\rm d}u}\:-\:\frac{\omega^2}{1-u^2}\:+\:6\Bigg]F_j(u,u')\nonumber\\&\qquad =\ -\,\bigg(\frac{1-u}{1+u}\bigg)\bigg(\frac{1+u'}{1-u'}\bigg)\delta(u-u')\;,
\end{align}
where $F_j(u,u')\equiv R^2\,\widetilde{G}_j(r,r')$ and
\begin{equation}
\label{eq:mthickwall}
\omega\ = \ j\:+\:1\;.
\end{equation}
We note that the homogeneous part of the differential equation~\eqref{eq:thickwalltransdiff} is of precisely the same form as for the thin-wall case in Eq.~\eqref{eq:thinwalludiff}. Proceeding as described in App.~\ref{app:Greens}, we find the hyperradial Green's function
\begin{align}
\label{eq:Gj}
&G_j(r,r') = \frac{1}{2R^2\omega}\Bigg[\vartheta(u-u')\bigg(\frac{1-u}{1+u}\bigg)^{\!\!\frac{j}{2}}\bigg(\frac{1+u'}{1-u'}\bigg)^{\!\!\frac{j+2}{2}}\nonumber\\&\ \times\bigg(1-3\,\frac{(1-u)(1+\omega+u)}{(1+\omega)(2+\omega)}\bigg)\nonumber\\&\ \times\bigg(1-3\,\frac{(1-u')(1-\omega+u')}{(1-\omega)(2-\omega)}\bigg)\nonumber\\&\ +\:(u\:\leftrightarrow u')\Bigg]\;.
\end{align}
It follows then that the coincident limit of the Green's function is
\begin{align}
  \label{eq:Gu2}
  G(r,r) \ & = \ \frac{1}{4 \pi^2 R^2}\bigg(\frac{1+u}{1-u}\bigg)
  \sum_{j\, =\, 0}^{\infty}  \frac{(j + 1)^2}{\omega}\bigg[ 1 \nonumber\\&\qquad + 3
  \big( 1 - u^2 \big) \sum_{n\, =\, 1}^2
  \frac{(-1)^n (n - 1 - u^2) }{\omega^2 - n^2} \bigg] \;,
\end{align}
differing from Eq.~\eqref{eq:Gu} in the form of $\omega$, the dimensionful prefactors and an overall factor of $1/r^2$.
From this expression, one might suspect that there are discrete eigenvalues proportional to
\begin{equation}
\omega^2\:-\:n^2\ =\ (j+1)^2\:-\:n^2\;.
\end{equation}
However, this is misleading because it is in disagreement with the spectrum discussed in Sec.~\ref{subsubsec:FL}.
We may recognize zero modes for $j=0$, $n=1$ and $j=1$, $n=2$, which correspond to the five Goldstone modes that arise from the spontaneous
breakdown of symmetries in presence of the bounce solutions: one ($j=0$) arises from broken dilatational and four ($j=1$) from broken translational invariance. In addition,
there is one negative mode for $j=0$, $n=2$, but we cannot directly
infer its value from Eq.~\eqref{eq:Gu2}.
We emphasize, however, that the zero modes are not discrete, and there are no discrete modes at all for
$j>0$, whereas Eq.~\eqref{eq:Gu2} might suggest otherwise.


\subsection{Spectral sum}
\label{sec:specsum}

In order to understand the apparent mismatch between the eigenspectra and the form
of the coincident Green's function in the Fubini-Lipatov background~\eqref{eq:Gu2}, it would be interesting to compare with an explicit construction via a spectral sum in the form of Eq.~\eqref{eq:specsum}. However, this cannot be carried out analytically for the Fubini-Lipatov
case in terms of the eigenmodes of the operator in Eq.~\eqref{eq:EV}, i.e.~in the \emph{Fubini-Lipatov basis}.

However, we notice that the left-hand side of Eq.~\eqref{eq:thickwalltransdiff}
for the Green's function of the Fubini-Lipatov case
is of precisely the same form as Eq.~\eqref{eq:thinwalludiff}
for the thin-wall case. As a result, we can transform the problem of finding the spectral sum in the Fubini-Lipatov case to a \emph{thin-wall basis}, such that it becomes analytically tractable. Moreover, this relation between the two problems clarifies
why the \emph{apparent} modes in the Fubini-Lipatov Green's function~\eqref{eq:Gu2}
agree with those of the thin-wall case~\eqref{eq:Gu}.
It is important though to emphasize that the eigenfunctions of the thin-wall basis are not directly related to the true eigenfunctions over the Fubini-Lipatov instanton, and we give an overview of the basis transformation in App.~\ref{app:spectral}. This is the case for the true negative eigenfunction, and the coefficients of the transformation cannot be found in analytic form.

In the thin-wall case, we will see that there are nontrivial cancellations between contributions from the true discrete
and continuum spectra, which obscures the origin of the particular contributions to the
Green's function. This strongly hints at the possibility that similar cancellations occur for the
Fubini-Lipatov case also without applying the transformation to the thin-wall basis,
such that this may explain why the true negative eigenmode is not directly apparent in 
Eq.~\eqref{eq:Gu2}.

We can cast the problem of finding the spectral sum representation for the thin-wall case in a convenient form by making the following (judicious) rescaling of the eigenfunctions:
\begin{equation}
\tilde{\phi}^{\pm}_{\lambda^{\rm TW} j}(u)\ \equiv\  (r/R)^3\gamma^{1/2}R^{3/2}\phi^{\pm}_{\lambda^{\rm TW} j}(r)\;.
\end{equation}
The eigenvalue equation~\eqref{eq:EV} then takes the form
\begin{align}
&\bigg[
-\:\frac{{\rm d}^2}{{\rm d}r^2}\:+\:\frac{3}{r}\,\frac{\rm d}{{\rm d} r}\:+\:\frac{j(j+2)-3}{r^2}\nonumber\\&\qquad \qquad\qquad +\:U''(\varphi)
\:-\:\lambda^{\rm TW}\bigg]
\tilde \phi^{\pm}_{\lambda^{\rm TW} j}(u)\ =\ 0\;,
\end{align}
where we again emphasize the change of sign in the damping term and the shift in the numerator of the centrifugal potential. After making the thin-wall approximation\footnote{Note that $(r/R)^3\phi^{\pm}_{\lambda^{\rm TW} j}\to \phi^{\pm}_{\lambda^{\rm TW} j}$ in the thin-wall approximation.} via Eq.~\eqref{eq:thinwallapprox} and changing coordinates via Eq.~\eqref{eq:thinwallcoords}, the eigenproblem becomes
\begin{equation}
\label{eq:Legendrediff}
\bigg[\frac{\rm d}{{\rm d}u}\,(1-u^2)\,\frac{\rm d}{{\rm d}u}\:-\:\frac{\varpi^2}{1-u^2}\:+\:6\bigg]\tilde{\phi}_{\lambda^{\rm TW} j}^{\pm}(u)\ =\ 0\;,
\end{equation}
where
\begin{equation}
\label{eq:varpidef}
\varpi^2\ \equiv\ \omega^2\:-\:\bar{\lambda}^{\rm TW}\;,
\end{equation}
$\bar{\lambda}^{\rm TW}\equiv \lambda^{\rm TW}/\gamma^2$ are the dimensionless eigenvalues and $\omega$ is given by Eq.~\eqref{eq:m2}. Equation~\eqref{eq:Legendrediff} is the associated Legendre differential equation, and the solutions are the associated Legendre functions of degree 2 and order $\varpi$. Setting $\varpi=n\in\{1,2\}$, we immediately recover the discrete eigenmodes discussed earlier. The order of these functions becomes imaginary when $\bar{\lambda}^{\rm TW} >\omega^2$, $\varpi^2<0$, implying positive energies in the corresponding Schr\"odinger problem, and this marks the beginning of the positive continuum.\footnote{In Ref.~\cite{Garbrecht:2015oea}, it is incorrectly stated that the continuum begins at $\lambda_{10}=2\gamma^2$. In the thin-wall regime, the continuum instead begins at $\lambda^{\rm TW}_{00}\approx \lambda^{\rm TW}_{01}=4\gamma^2$, cf.~Ref.~\cite{Konoplich:1987yd}.} However, rather than dealing with the associated Legendre functions of imaginary order, we can define
\begin{equation}
f^{\pm}_{\lambda^{\rm TW} j}(u)\ =\ \bigg(\frac{1-u}{1+u}\bigg)^{\!\pm\varpi/2}\tilde{\phi}^{\pm}_{\lambda^{\rm TW} j}(u)
\end{equation}
and recast the eigenvalue problem in terms of the Jacobi differential equation
\begin{equation}
\label{eq:Jacobidiff}
\bigg[(1-u^2)\,\frac{{\rm d}^2}{{\rm d}u^2}\:-\:2\big(u\mp\varpi\big)\,\frac{{\rm d}}{{\rm d}u}\:+\:6\bigg]f^{\pm}_{\lambda^{\rm TW} j}(u)\ =\ 0\;.
\end{equation}
With these manipulations, the spectral sum representation of the Green's function is given by
\begin{align}
\label{eq:specsumTW}
&G^{\rm TW}(x,x')\ =\ \frac{1}{2\pi^2\gamma R^3}\sum_{j\,=\,0}^{\infty}(j+1)U_j(\cos\theta)\nonumber\\& \times\left[\sum_{\bar{\lambda}^{\rm TW}\,\in\, \bar{L}^{\rm TW}_{{\rm d}j}}\:+\!\!\int\limits_{\bar{\lambda}^{\rm TW}\,\in\, \bar{L}^{\rm TW}_{{\rm c}j}}\!\!\!\!\!\!\!\!\frac{{\rm d}\bar{\lambda}^{\rm TW}}{2\pi}\right]\!\!\bigg(\frac{1+u}{1-u}\bigg)^{\!+\frac{\varpi}{2}}\bigg(\frac{1+u'}{1-u}\bigg)^{\!-\frac{\varpi}{2}}\nonumber\\& \times\frac{f_{\lambda^{\rm TW} j}^{-}(u')f^{+}_{\lambda^{\rm TW} j}(u)}{\bar{\lambda}^{\rm TW}}\;.
\end{align}

In the Fubini-Lipatov case, the (judiciously) rescaled eigenfunctions are
\begin{equation}
\tilde{\phi}^{\pm}_{\lambda^{\rm FL} j}(u)\ \equiv\ r\phi^{\pm}_{\lambda^{\rm FL} j}(r)\;.
\end{equation}
After transforming to the thin-wall basis (see App.~\ref{app:spectral}), we can write the spectral sum representation of the Green's function as
\begin{align}
\label{eq:specsumFL}
&G^{\rm FL}(x,x')\ =\ \frac{1}{2\pi^2 R^2}\sum_{j\,=\,0}^{\infty}(j+1)U_j(\cos\theta)\nonumber\\& \times\left[\sum_{\bar{\lambda}^{\rm TW}\,\in\, \bar{L}^{\rm TW}_{{\rm d}j}}\:+\!\!\int\limits_{\bar{\lambda}^{\rm TW}\,\in\, \bar{L}^{\rm TW}_{{\rm c}j}}\!\!\!\!\!\!\!\!\frac{{\rm d}\bar{\lambda}^{\rm TW}}{2\pi}\right]\bigg(\frac{1+u}{1-u}\bigg)^{\!+\frac{\varpi+1}{2}}\nonumber\\& \times\bigg(\frac{1+u'}{1-u}\bigg)^{\!-\frac{\varpi-1}{2}}\frac{f_{\lambda^{\rm TW} j}^{-}(u')f^{+}_{\lambda^{\rm TW} j}(u)}{\bar{\lambda}^{\rm TW}}\;.
\end{align}
Here, the dimensionless eigenvalues are $\bar{\lambda}^{\rm TW}\equiv \lambda^{\rm TW} R^2$, and the superscript TW indicates that we are working in the thin-wall basis. These $\bar{\lambda}^{\rm TW}=\omega^2-\varpi^2$ of the thin-wall \emph{basis} should not be confused with those of the thin-wall \emph{case}, which differ in the value of $\omega$. We emphasize that the Fubini-Lipatov value of $\omega=j+1$ [cf.~Eq.~\eqref{eq:m2} for the thin-wall case] does not change under the basis transformation.

In order to see that the eigenfunctions of the thin-wall basis are not directly related to the true eigenfunctions over the Fubini-Lipatov instanton,
we note that the eigenvalue equation in Eq.~\eqref{eq:EV} becomes
\begin{align}
&\bigg[\frac{\rm d}{{\rm d}u}(1-u^2)\frac{\rm d}{{\rm d}u}\:-\:\frac{\omega^2}{1-u^2}\notag\\ &\qquad\qquad +\:\frac{\bar{\lambda}^{\rm FL}}{(1+u)^2}\:+\:6\bigg]\tilde{\phi}_{\lambda^{\rm FL} j}^{\pm}(u)\ =\ 0\;.
\end{align}
This differs from Eq.~\eqref{eq:Legendrediff} for the thin-wall problem, due to the extra dependence on $u$ in the term pertaining to the eigenvalue $\lambda^{\rm FL}$. Hence, while the Green's functions for the Fubini-Lipatov and the thin-wall cases are identical
up to  algebraic prefactors and the different form of $\omega$ in Eqs.~\eqref{eq:m2} and~\eqref{eq:mthickwall}, their true eigenspectra are very different, with the exception of the zero modes. When $\bar{\lambda}^{\rm FL}=0$, we see from Eq.~\eqref{eq:Legendrediff} and the definition of $\varpi$ in Eq.~\eqref{eq:varpidef} that $\bar{\lambda}^{\rm TW}$ also vanishes and the eigenvalue equations coincide in the thin-wall and Fubini-Lipatov bases. Thus, for the Fubini-Lipatov case, the zero modes (translational and dilatational) coincide in the thin-wall and Fubini-Lipatov bases.

The apparent disparity between the eigenspectra of the two bases originates from their differing normalizations; namely, for the discrete modes,
\begin{subequations}
\label{eq:innerprod}
\begin{gather}
R^2\int\!\frac{\rm{d}u}{1-u^2}\,\tilde{\phi}^-_{\lambda^{\rm TW}j}(u)\tilde{\phi}^+_{\lambda^{\rm TW}j}(u)\ =\ 1\;,\\
R^2\int\!\frac{\rm{d}u}{(1+u)^2}\,\tilde{\phi}^-_{\lambda^{\rm FL}j}(u)\tilde{\phi}^+_{\lambda^{\rm FL}j}(u)\ =\ 1\;,
\end{gather}
\end{subequations}
with analogous expressions holding for the continuum modes, which are normalized in the improper sense. This difference in normalization is the reason why the dilatational mode, which is nonnormalizable in the Fubini-Lipatov basis, becomes an apparent normalizable mode in the thin-wall basis. Moreover, it indicates that the transformation, described here and in App.~\ref{app:spectral} for illustrative purposes, is not a basis transformation in the proper sense, since the two bases span different Hilbert spaces.

Returning to the problem of finding the spectral sum representations in the thin-wall basis, the solutions to Eq.~\eqref{eq:Jacobidiff} are the Jacobi polynomials of degree 2: \smash{$P_2^{(\mp\varpi,\pm\varpi)}(u)$}. For $\varpi=n\in\{1,2\}$, we can show that these are normalizable (see App.~\ref{app:Jacfuncim})
\begin{align}
\label{eq:Jacorthogdisc}
\int_{-1}^{+1}\frac{{\rm d}u}{1-u^2}\;\bigg(\frac{u+1}{u-1}&\bigg)^{\!+\frac{n}{2}}\bigg(\frac{u+1}{u-1}\bigg)^{\!-\frac{n'}{2}}P_2^{(-n,+n)}(u)\nonumber\\\times \:P_2^{(+n',-n')}(u)\ &=\ \frac{(-1)^n\pi}{4\sin(n\pi)}\,(4-n^2)(1-n^2)\,\delta_{nn'}\nonumber\\& =\ \begin{cases}-\,\frac{3}{2}\,\delta_{nn'}\;,\quad &n\ = \ 1\\3\,\delta_{nn'}\;,\quad &n \ = \ 2
\end{cases}\;.
\end{align}
For $\bar{\lambda}^{\rm TW}>\omega^2$, the continuum is specified by $\varpi=i\xi$ ($\xi\in \mathbb{R}$), and we can show that (see App.~\ref{app:Jacfuncim})
\begin{align}
\label{eq:Jacorthog}
&\int_{-1}^{+1}\frac{{\rm d}u}{1-u^2}\;\bigg(\frac{u+1}{u-1}\bigg)^{\!+\frac{i\xi}{2}}\bigg(\frac{u+1}{u-1}\bigg)^{\!-\frac{i\xi'}{2}}P_2^{(-i\xi,+i\xi)}(u)\nonumber\\&\times P_2^{(+i\xi',-i\xi')}(u)\ =\ \frac{\pi}{2}\,(4+\xi^2)(1+\xi^2)\,\delta(\xi-\xi')\;.
\end{align}
Hence, the sets of (dimensionless) discrete and continuous eigenvalues that appear in the spectral sum in the thin-wall basis can be specified as follows:
\begin{subequations}
\begin{align}
\bar{L}^{\rm TW}_{{\rm d}j}\ &=\ \left\{\bar{\lambda}^{\rm TW}_{2 j},\bar{\lambda}^{\rm TW}_{1 j}\right\}\;,\\
\bar{L}^{\rm TW}_{{\rm c}j}\  &=\ \left\{\bar{\lambda}^{\rm TW}\; |\; \bar{\lambda}^{\rm TW} \:>\: \omega^2 \right\}\;,
\end{align}
\end{subequations}
where
\begin{align}
\label{eq:thinwall:discrete:ev}
\bar{\lambda}^{\rm TW}_{nj}\ =\ \omega^2\:-\:n^2\;.
\end{align}
The \emph{apparent} discrete modes, which are the true eigenmodes in the thin-wall case, contribute to the Green's functions as
\begin{align}
\label{eq:discretecont}
&G_{\rm d}(r,r)\ = \frac{1}{4\pi^2R^2}\begin{Bmatrix} \frac{1}{\gamma R}\\ \Big(\frac{1+u}{1-u}\Big)\end{Bmatrix}\sum_{j\,=\,0}^{\infty}(j+1)^2\nonumber\\&\qquad\times\:3(1-u^2)\sum_{n\,=\,1}^2\frac{(-1)^n(n-1-u^2)}{n(\omega^2-n^2)}\;,
\end{align}
where the upper alternative applies to the thin-wall case and the lower to the Fubini-Lipatov case.
The contributions of the apparent continuum modes to the Green's functions are
\begin{align}
G_{\rm c}(r,r)\ &=\ \frac{1}{2\pi^2R^2}\begin{Bmatrix} \frac{1}{\gamma R}\\ \Big(\frac{1+u}{1-u}\Big)\end{Bmatrix}\nonumber\\&\qquad\times\sum_{j\,=\,0}^{\infty}(j+1)^2\int_{-\infty}^{+\infty}\frac{{\rm d}\xi}{2\pi}\;\frac{4}{(4+\xi^2)(1+\xi^2)}\nonumber\\&\qquad\times\:\frac{P_2^{(-i\xi,+i\xi)}(u)P_2^{(+i\xi,-i\xi)}(u)}{\omega^2+\xi^2}\;.
\end{align}
Integrating over $\xi$ (see App.~\ref{app:spectral}, where the case $r'\neq r$ is also treated), we obtain
\begin{align}
\label{eq:continuumcont}
G_{\rm c}(r,r)\ &=\ \frac{1}{4\pi^2R^2}\begin{Bmatrix} \frac{1}{\gamma R}\\ \Big(\frac{1+u}{1-u}\Big)\end{Bmatrix}\sum_{j\,=\,0}^{\infty}(j+1)^2\bigg[\frac{1}{\omega}\nonumber\\&\qquad+\:3(1-u^2)\sum_{n\,=\,1}^{2}\frac{(-1)^n(n-1-u^2)}{\omega(\omega^2-n^2)}\nonumber\\&\qquad-\:3(1-u^2)\sum_{n\,=\,1}^{2}\frac{(-1)^n(n-1-u^2)}{n(\omega^2-n^2)}\bigg]\;.
\end{align}
Despite having integrated over the continuum part of the spectrum, there appear
terms involving sums over $n$ and matching denominators to the discrete eigenvalues.
Furthermore, we see that the final line of Eq.~\eqref{eq:continuumcont} cancels against the contribution from the discrete eigenmodes in Eq.~\eqref{eq:discretecont}, leaving the results already quoted in Eqs.~\eqref{eq:Gu} and~\eqref{eq:Gu2}. Therefore, while we can still ``read off'' the functional form of the discrete eigenmodes from the final Green's function in the thin-wall case, we see that there is a nontrivial interplay between the discrete and continuum parts of the spectrum. 

In the Fubini-Lipatov case, while we can read off the functional forms of the zero modes, we cannot read off the true functional form of any other modes due to the transformation of the eigenvalue problem. Notice, for instance, that the functional form of the negative mode that one would read off from the Green's function does \emph{not} coincide with the true negative eigenfunction, as can be seen in Fig.~\ref{fig:negative}. Moreover, by considering the transformed eigenproblem, one might be led to conclude that there are two infinite towers of discrete modes also for the Fubini-Lipatov case. This is not correct, and we reiterate that, in the Fubini-Lipatov case, the $\bar{\lambda}^{\rm TW}_{nj}$ in Eq.~\eqref{eq:thinwall:discrete:ev} do not compose the true discrete eigenspectrum. The eigenspectra for both cases are summarized in Table~\ref{tab:eigen}.

We remark that, while one might be tempted to remove the apparent discrete zero modes in the thin-wall basis, this will not eliminate the infrared divergences, since these reside also in the continuum, as is clear from the interplay of the apparent discrete and continuum modes described above. Since the one-loop fluctuation determinant can be related to the Green's function, the same subtraction in the transformed problem may also be problematic there, cf.~Ref.~\cite{Andreassen:2017rzq}, and we leave further study of this point for future work.

\begin{table}[t]

\begin{tabular}{| >{\centering\arraybackslash} p{2em} || >{\centering\arraybackslash} p{12em} | >{\centering\arraybackslash} p{12em} |}
\hline
\multirow{2}{*}{$j$}  & $U = -\tfrac{1}{2}|\mu^2|\Phi^2+\tfrac{1}{4!}|\lambda| \Phi^4$ & $U = -\tfrac{1}{4!}|\lambda| \Phi^4$\\ & $\lambda^{\rm TW}_{nj}=( 4-n^2)\gamma^2+\frac{j(j+2)-3}{R^2}$ & $\lambda^{{\rm FL}}_{nj}\propto (j+1)^2-n^2$\\
\hline\hline
\multirow{2}{*}{$0$} & $\lambda^{\rm TW}_{20}=-\frac{3}{R^2} < 0$ & $\lambda^{\rm FL}_{20}< 0$ \\
 & negative-definite mode & negative-definite mode \\\hline
\multirow{2}{*}{$0$} & $\lambda^{\rm TW}_{10}=3\gamma^2-\frac{3}{R^2}$ &  \cellcolor[gray]{0.9} $\lambda^{\rm FL}_{10}=0$\\ 
& positive-definite mode &  \cellcolor[gray]{0.9} dilatational mode\\ \hline
$0$ & \cellcolor[gray]{0.9}$\lambda^{{\rm TW}}\geq4\gamma^2-\frac{3}{R^2}>0$ & \cellcolor[gray]{0.9} $\lambda^{{\rm FL}}>0$\\ \hline
\multirow{2}{*}{$1$} & $\lambda^{\rm TW}_{21}=0$ & \cellcolor[gray]{0.9} $\lambda^{\rm FL}_{21}= 0$ \\ 
& translational zero mode & \cellcolor[gray]{0.9} translational zero mode\\\hline
\multirow{1}{*}{$1$} & $\lambda^{\rm TW}_{11}=3\gamma^2$ & \cellcolor[gray]{0.9} $\lambda^{{\rm FL}}> 0$ \\\hline
\multirow{1}{*}{$1$} & \cellcolor[gray]{0.9}$\lambda^{{\rm TW}}\geq4\gamma^2$ & \cellcolor[gray]{0.9} \\\hline
\multirow{1}{*}{$>1$} & $\lambda^{\rm TW}_{2j}=\frac{j(j+2)-3}{R^2}$ &  \cellcolor[gray]{0.9} $\lambda^{{\rm FL}}> 0$ \\\hline
\multirow{1}{*}{$>1$} & $\lambda^{\rm TW}_{1j}=3\gamma^2+\frac{j(j+2)-3}{R^2}$ & \cellcolor[gray]{0.9} \\\hline
\multirow{1}{*}{$>1$} & \cellcolor[gray]{0.9}$\lambda^{{\rm TW}}\geq4\gamma^2+\frac{j(j+2)-3}{R^2}$ & \cellcolor[gray]{0.9} \\\hline
\end{tabular}

\caption{\label{tab:eigen}Comparison of the eigenspectra for the archetypal (thin-wall case) and scale-invariant (Fubini-Lipatov case with $m=0$) theories over their respective tunneling configurations. Modes that are part of the discrete spectrum are in white cells; parts of the continuum spectrum are in grey-shaded cells. Note that when lifting the
mass parameter to $m^2>0$ and including radiative corrections, the translational zero modes become
discrete in the Fubini-Lipatov spectrum, whereas the approximate dilatational mode remains at the
end point of the continuum, with its eigenvalue lifted to $\lambda^{\rm FL}_{10}=m^2$.}
\end{table}


\section{Zero modes}
\label{sec:zeroes}

In order to deal with the translational zero modes, we first decompose the field as
\begin{equation}
\label{eq:decomp}
\Phi(x)\ =\ \varphi(x-y)\:+\:\sum_{\mu\,=\,1}^{4}a_{\mu}\phi_{\mu}(x-y)\:+\:\Phi^{\prime}(x-y)\;,
\end{equation}
where $\varphi(x-y)$ is the bounce, $y$ is its coordinate centre, \smash{$\Phi^{\prime}(x-y)$} contains the contributions of the negative- and positive-definite (discrete and continuum) eigenmodes, and %
\begin{equation}
\label{eq:zerodef}
\phi_{\mu}(x-y)\ =\ {\cal N}\partial_{\mu}^{(x)}\varphi(x-y)
\end{equation}
are the translational zero eigenmodes with the normalization factor (no sum over $\mu$)
\begin{align}
{\cal N}=\left[\int {\rm d}^4 x\; (\partial_\mu \varphi(x))^2\right]^{-1/2}\;.
\end{align}
Note that in the case of the tree-level bounce, ${\cal N}=B^{-1/2}$, where $B$ is the bounce action [see Eq.~\eqref{eq:B}]. The zero eigenmodes satisfy the orthonormality relations
\begin{subequations}
\label{eq:ortho}
\begin{align}
\int\!{\rm d}^4x\;\phi_{\mu}^{*}(x)\phi_{\nu}(x)\ &=\ \delta_{\mu\nu}\;,\\
\int\!{\rm d}^4x\;{\Phi^{\prime}}^*(x)\phi_{\mu}(x)\ &=\ 0\;.
\end{align}
\end{subequations}
We emphasize that, while all the eigenmodes depend on $y$,\footnote{This can be shown explicitly by considering the eigenvalue problem directly, which can be expressed entirely in terms of $x-y$.} the original field $\Phi\equiv\Phi(x)$ is independent of $y$. 

With this decomposition, the functional integral can be written in the form
\begin{equation}
\label{eq:funcint}
\int\mathcal{D}\Phi\ =\ \int\mathcal{D}\Phi^{\prime}\;\prod_{\mu\,=\,1}^4\Bigg[(2\pi\hbar)^{-1/2}\int\!{\rm d}a_{\mu}\Bigg]\;,
\end{equation}
where we have isolated the problematic integrals over the zero modes. These integrals can be performed by means of a Faddeev-Popov-type method~\cite{Gervais:1974dc,Aleinikov:1987wx,Olejnik:1989id,Vandoren:2008xg}. Specifically, we insert unity in the form
\begin{equation}
\label{eq:FP}
1\ =\ \prod_{\mu\,=\,1}^4\int\!{\rm d}y_{\mu}\;\big|\partial_{\mu}^{(y)}f_{\mu}(y)\big|\delta\big(f_{\mu}(y)\big)\;,
\end{equation}
in order to trade the integrals over the $a_{\mu}$ for integrals over the collective coordinates $y_{\mu}$. We take
\begin{equation}
f_{\mu}(y)\ =\ \int\!{\rm d}^4x\;\Phi(x)\partial_{\mu}^{(x)}\varphi(x-y)\;,
\end{equation}
from which it follows that
\begin{equation}
\partial_{\mu}^{(y)}f_{\mu}(y)\ =\ -\:\int\!{\rm d}^4x\;\Phi(x)\partial_{\mu}^{(x)}\partial_{\mu}^{(x)}\varphi(x-y)\;.
\end{equation}
By virtue of the orthogonality of the eigenmodes, we can quickly show that
\begin{equation}
f_{\mu}(y)\ =\ {\cal N}^{-1} a_{\mu}\;,
\end{equation}
such that the delta function in Eq.~\eqref{eq:FP} can be written as
\begin{equation}
\delta\big(f_{\mu}(y)\big)\ =\ {\cal N}\delta(a_{\mu})\;.
\end{equation}
As a result, the Jacobian becomes
\begin{align}
\label{eq:dfmu}
&\partial_{\mu}^{(y)}f_{\mu}(y)\big|_{a_{\mu}=0}\ =\ \int\!{\rm d}^4x\;\big[\partial_{\mu}^{(x)}\varphi(x-y)\big]^2\nonumber\\&-\:\int\!{\rm d}^4x\;\Phi^{\prime}(x-y)\,\partial_{\mu}^{(x)}\partial_{\mu}^{(x)}\varphi(x-y)\;.
\end{align}
The first term is just the normalization ${\cal N}^{-2}$, and we find
\begin{align}
&1\ =\ \prod_{\mu\,=\,1}^4\bigg[\int\!{\rm d}y_{\mu}\;{\cal N}^{-1}\delta(a_{\mu})\nonumber\\&\qquad\times\:\bigg(1-{\cal N}^2\int\!{\rm d}^4x\;\Phi^{\prime}(x-y)\partial^{(x)}_{\mu}\partial^{(x)}_{\mu}\varphi(x-y)\bigg)\bigg]\;.
\end{align}
Inserting this result into the original functional integral in Eq.~\eqref{eq:funcint} gives
\begin{align}
&\int\mathcal{D}\Phi\ =\ \bigg(\frac{1}{2\pi\hbar{\cal N}^2}\bigg)^{\!2}\int\mathcal{D}\Phi^{\prime}\int\!{\rm d}^4y\nonumber\\&\quad\times\:\prod_{\mu\,=\,1}^4\bigg(1-{\cal N}^2 \int\!{\rm d}^4x\;\Phi^{\prime}(x-y)\partial_{\mu}^{(x)}\partial_{\mu}^{(x)}\varphi(x-y)\bigg)\;.
\end{align}
The integral within the parentheses is independent of $y$, and we can write
\begin{align}
&\int\mathcal{D}\Phi\ =\ VT\bigg(\frac{1}{2\pi\hbar{\cal N}^2}\bigg)^{\!2}\int\mathcal{D}\Phi^{\prime}\nonumber\\&\qquad\times\:\prod_{\mu\,=\,1}^4\bigg(1-{\cal N}^2 \int\!{\rm d}^4x\;\Phi^{\prime}(x)\partial_{\mu}\partial_{\mu}\varphi(x)\bigg)\;,
\end{align}
where the four-volume factor $VT$ has arisen from the integral over the collective coordinates.

We now apply this decomposition in positive- or negative-definite and zero modes to the path integral in order to determine the radiative corrections to the bounce from the generating functional
\begin{align}
\mathcal{Z}[J]\ &=\ \int\!\mathcal{D}\Phi\;\exp\bigg[-\:\frac{1}{\hbar}\int\!{\rm d}^4x\;\bigg(\frac{1}{2}\,\partial_{\mu}\Phi(x)\,\partial_{\mu}\Phi(x)\nonumber\\&\qquad+\:\frac{\lambda}{4!}\,\Phi^4(x)\:-\:J(x)\Phi(x)\bigg)\bigg]\;.
\end{align}
For the evaluation, we first expand around the classical solution $\varphi^{(0)}$ as \smash{$\Phi=\varphi^{(0)}+\hbar^{1/2} \sum_{\mu\,=\,1}^4a_{\mu}\phi_{\mu}^{(0)}+\hbar^{1/2}\Phi'$}. A superscript $(0)$ has been attached here to the zero modes $\phi_\mu$ in order to indicate
that they are obtained from substituting the tree-level bounce $\varphi^{(0)}$ into
Eq.~\eqref{eq:zerodef}. We have also made explicit the bookkeeping factors of $\hbar^{1/2}$.

The Faddeev-Popov procedure described above then leads to
(We use the more compact index notation where, e.g., $f(x)\equiv f_x$ for functions
and $\int\!{\rm d}^4 x\equiv\int_x$ for integrals.)
\begin{align}
\mathcal{Z}[J]\ &=\ VT\,\frac{B^2}{(2\pi\hbar)^2}
\nonumber\\&\quad \times\Bigg[
\prod_{\mu\,=\,1}^4\bigg(1\:-\:\frac{\hbar}{B}\int_x\;\frac{\delta}{\delta J_x}\,\partial_{\mu}\partial_{\mu}\varphi^{(0)}_x\bigg)\Bigg]
\,\mathcal{Z}'[J]
\;,
\label{eq:Z}
\end{align}
where
\begin{align}
&\mathcal{Z}'[J]\ =\ \exp\Bigg[-\:\frac{1}{\hbar}\bigg(S[\varphi^{(0)}]\:-\:\int_x\;J_x\varphi_x^{(0)}\bigg)\Bigg]\nonumber\\&\qquad\times \int\mathcal{D}\Phi'\;\exp\Bigg[-\int_{xy}\;\frac12 \Phi^\prime_x {G}^{-1}_{xy}\Phi^\prime_y-\hbar^{-1/2} \int_x J_x \Phi^\prime_x \nonumber\\
&\qquad-\:\hbar^{1/2}\int_x\;\frac{\lambda}{3!}\,\varphi^{(0)}_x \Phi_x^{\prime 3}\:-\:\hbar\int_x\;\frac{\lambda}{4!}\,\Phi_x^{\prime 4}\Bigg]
\label{eq:z}
\end{align}
and
\begin{align}
\label{eq:modifiedKG}
G^{-1}_{xy}\ &=\ \delta^4_{xy}\big[-\:\partial_y^2\:+\:U''\big(\varphi^{(0)}_{y}\big)\big]
\end{align}
is the Klein-Gordon operator. In the full Hilbert space, this operator is not invertible because of the zero modes $\partial_\mu\varphi^{(0)}$. Nevertheless, the inversion is well defined in the subspace perpendicular to the
zero modes, where the subtracted two-point function $G^\perp$ is the solution to~\cite{Aleinikov:1987wx}
\begin{align}
\label{eq:Gperp:inv}
&\int_z\;G^{-1}_{xz}\;G^{\perp}_{zy}\ =\ \delta^4_{xy}-\:\,\sum_{\mu\,=\,1}^4\big(\phi^{(0)}_{\mu}\big)_x\,\big(\phi^{(0)}_{\mu}\big)_y\;,
\end{align}
with the additional requirement that
\begin{align}
\label{eq:Gperp:ortho}
\int_y G^{\perp}_{xy}\big(\phi^{(0)}_\mu\big)_y\ =\ 0\;.
\end{align}
We refer to $G^\perp$ and to its perturbatively improved variants as the \emph{subtracted Green's functions} because they may be thought of as emerging from their spectral sum representation with
the (divergent) contributions from the zero modes subtracted.

Making the shift \smash{$\Phi^\prime\to\Phi^\prime+\hbar^{-1/2}\int_y G^{\perp}_{xy} J_y$} (Note that the redefined $\Phi^\prime$ thus remains orthogonal
to the zero modes.),
we then recast the object~\eqref{eq:z} to
\begin{align}
&\mathcal{Z}'[J]\ =\ \exp\Bigg[-\:\frac{1}{\hbar}\bigg(S[\varphi^{(0)}]\:-\:\int_x\;J_x\varphi_x^{(0)}\bigg)\Bigg]\nonumber\\
&\ \times\int\mathcal{D}\Phi'\;\exp\Bigg[-\:\hbar^{2}\int_x\;\frac{\lambda}{3!}\,\varphi^{(0)}_x \frac{\delta^3}{\delta J_x^3}\:-\:\hbar^3\int_x\;\frac{\lambda}{4!}\,\frac{\delta^4}{\delta J_x^4}\Bigg]
\nonumber\\
&\ \times
\exp\Bigg[-\int_{xy}\;\frac{1}{2}\, \Phi^\prime_x {G}^{-1}_{xy}\Phi^\prime_y\:+\:\frac{1}{2\hbar}\int_{xy} J_x G^{\perp}_{xy} J_y \Bigg]
\;.
\label{eq:z2}
\end{align}

Next, we make use of the apparent decomposition of the generating functional~\eqref{eq:Z}
into
\begin{align}
\label{eq:Z:decomposition}
{\cal Z}[J]\ =\ {\cal Z}^\perp[J]\:+\:{\cal Z}^{\veryshortarrow}[J]\;,
\end{align}
where ${\cal Z}^\perp$ is the
contribution that arises from replacing the product term (over $\mu$) with
$1$ and ${\cal Z}^{\veryshortarrow}$ is the remainder involving tadpole corrections
$\propto \partial_\mu\partial_\mu\varphi^{(0)}$. We then define
\begin{subequations}
\begin{align}
\varphi^\perp_x\ &=\ \hbar\,\frac{1}{\mathcal{Z}^{\perp}[0]}\,\frac{\delta}{\delta J_x}\,\mathcal{Z}^\perp[J]\bigg|_{J\,=\,0}\,,\\
\label{eq:Gperpfunc}
\hbar\,{\cal G}^\perp_{xy}\ &=\ \hbar^2\,\frac{1}{\mathcal{Z}^{\perp}[0]}\,\frac{\delta^2}{\delta J_x\delta J_y}\,\mathcal{Z}^\perp[J]\bigg|_{J\,=\,0}\;,\\
\varphi^{\veryshortarrow}_{x}\ &= \ \hbar\,\frac{1}{\mathcal{Z}[0]}\,\frac{\delta}{\delta J_x}\,\mathcal{Z}[J]\bigg|_{J\,=\,0}\,,\\
\hbar\,{\cal G}^{\veryshortarrow}_{xy}\ &=\ \hbar^2\,\frac{1}{\mathcal{Z}[0]}\,\frac{\delta^2}{\delta J_x\delta J_y}\,\mathcal{Z}[J]\bigg|_{J\,=\,0}\;,
\end{align}
\end{subequations}
with the superscript $\veryshortarrow$ indicating the inclusion of translational modes.
In the following, we choose to drop the superscripts on
$\varphi^\perp$ and ${\cal G}^\perp$ (when no ambiguity results) because we find these to be the quantities most useful for the present calculations and want to keep notation compact.
Note that Goldstone's theorem implies that not only are $\partial_\mu \varphi^{(0)}$
zero modes of the inverse Green's function $G^{-1}$ at tree-level but that
the same holds true at each order in perturbation theory
up to the exact solutions $\partial_\mu \varphi^{\veryshortarrow}$ and
${\cal G}^{\veryshortarrow-1}$, as well as $\partial_\mu \varphi$ and
${\cal G}^{-1}$.

The large infrared contributions from certain fluctuation modes around the
approximately scale-invariant solitons require the resummation of loop corrections
to the Green's functions. In the subspace perpendicular to the zero modes, this
can be achieved by the coupled system of equations of motion for the one-point function and the Schwinger-Dyson equations for the correlations (Recall that we have dropped the superscripts
$\perp$.)
\begin{subequations}
\label{eqs:SD}
\begin{align}
\label{eqs:SD:onepoint}
&-\:\partial_x^2\varphi_x\:+\:U'(\varphi_x)\:+\:\hbar\,\Pi_{xx}\varphi_x\ =\ 0\,,\\
&\int_z\left[\delta_{xz}\left(-\:\partial_x^2\:+\:U''(\varphi_x)\:+\:\hbar\,\Pi_{xx}\right)\:+\:\hbar\,\Sigma_{xz}\right]{\cal G}_{zy}\notag\\
&\quad =\ \int_z{\cal G}^{-1}_{xz}{\cal G}_{zy}\ =\ \delta^4_{xy}\:-\:\sum\limits_{\mu\,=\,1}^4 \big(\phi_{\mu}\big)_{x}\big(\phi_{\mu}\big)_{y}\;,
\label{eqs:SD:twopoint}
\end{align}
\end{subequations}
where $\Pi_{xx}$ is the coincident and $\Sigma_{xy}$ the noncoincident contribution to the
proper self-energy. In place of relation~\eqref{eq:Gperp:ortho},
we now impose
\begin{align}
\int_y {\cal G}_{xy}\big(\phi_\mu\big)_y \ =\ 0\;.
\end{align}

Note that, in order to carry out the inversion of the Klein-Gordon operator appearing
in Eq.~\eqref{eqs:SD:twopoint}, it has been necessary to define the
Green's function ${\cal G}$ such that it acts in the subspace perpendicular
to the modes $\phi_\mu$, which include quantum corrections to the classical soliton, in contrast to the the tree-level Green's function that
acts in the space perpendicular to $\phi^{(0)}_\mu$.
The validity of this procedure can be confirmed order by order in the loop expansion of
the self energy.
To show this, we assume that, in Eq.~\eqref{eqs:SD},
$\Pi$ and $\Sigma$, as well as the solution $\varphi$ and
${\cal G}$, are given to a certain order in this expansion
and we consider an infinitesimal translation
$\varphi\to\varphi +\varepsilon \partial_\varrho \varphi$  of Eq.~\eqref{eqs:SD:onepoint} in the
$\varrho$ direction. In this way, we obtain
\begin{align}
\label{eq:trinv:varphirho}
&\int_z
\bigg[\delta_{xz}\left(-\:\partial_x^2\:+\:U''(\varphi_x)\:+\:\hbar\,\Pi_{xx}\right)\nonumber\\&\qquad+\:\varphi_x\,\frac{\delta}{\delta \varphi_z}\,\hbar\,\Pi_{xx}\bigg] \partial_{\varrho}\varphi_{z}\ =\ 0\;.
\end{align}
The last term can be calculated by applying the following variation to each propagator
appearing in $\Pi$:
\begin{align}
\label{eq:transl:Gperp}
\frac{\delta}{\delta\varphi_z}\,G^\perp_{xy}\ =\ 
-\:\int_{vw}G^{\perp}_{xv}\,\frac{\delta G^{-1}_{vw}}{\delta\varphi_z}\,G^{\perp}_{wy}
\ =\ -\:G^\perp_{xz}\lambda\varphi_z G^\perp_{zy}\;,
\end{align}
which is a standard identity for the derivative of an inverse operator that holds
in the present case because $G^{\perp}$ is the inverse of $G^{-1}$ in the
subspace perpendicular to the zero modes at a fixed location of the bounce.\footnote{
In contrast, when we perform the variation of Eq.~\eqref{eq:Gperp:inv} in the
form of
\begin{align}
\frac{\delta}{\delta\varphi_z} \left(
\int_{w} G^{-1}_{xw}G^\perp_{wy}\:+\:\sum\limits_{\mu\,=\,1}^4 \big(\phi_\mu\big)_{x}\big(\phi_\mu\big)_{y}
\right)
\ =\ \frac{\delta}{\delta\varphi_z}\,\delta_{xy}\ =\ 0\;,
\end{align}
we obtain
\begin{align}
\label{eq:var:Gperp}
&\frac{\delta G^\perp_{xy}}{\delta \varphi_z}\ =\ -\:G^\perp_{xz}\lambda\varphi_z G^\perp_{zy}
\:+\:\sum_{\nu\,=\,1}^4\int_w\bigg\{\frac{\delta}{\delta \varphi_z}\left((\phi_\nu\phi_\nu)_{xw} G^\perp_{wy}\right)\notag\\
&\quad-\:\left[\left(\frac{\delta}{\delta \varphi_z}(\phi_\nu\phi_\nu)_{xw}\right)G^\perp_{wy}\:
+\:G^\perp_{xw}\left(\frac{\delta}{\delta \varphi_z}(\phi_\nu\phi_\nu)_{wy}\right)\right]\bigg\}\;,
\end{align}
where we suppress the superscript $(0)$ on $\phi_\mu$ in this footnote.
Compared to Eq.~\eqref{eq:transl:Gperp}, there appear extra terms collected in curly brackets.
Within these, the first term vanishes
because of the condition~\eqref{eq:Gperp:ortho}. In order to interpret
the second term, we perform a spatial translation of
Eq.~\eqref{eq:Gperp:ortho}, which leads to
\begin{align}
\label{eq:Gperp:ortho:tran}
\int_y \left(G^{\perp}_{xy}\:+\:\int_z \frac{\delta G^\perp_{xy}}{\delta \varphi_z}\,\varepsilon\,\partial_\varrho \varphi_z \right)\big(\phi_\mu\:+\:\varepsilon\,\partial_\varrho \phi_\mu\big)_y\ =\ 0\;.
\end{align}
When substituting Eq.~\eqref{eq:var:Gperp} and evaluating to order $\varepsilon$,
the first term from Eq.~\eqref{eq:var:Gperp} leads here to a vanishing contribution
because of the condition~\eqref{eq:Gperp:ortho}
(just like the first term in curly brackets already commented on). The remaining
second term in curly brackets leads to the contribution
\begin{align}
&-\:\sum_{\nu\,=\,1}^4\int_{ywz}
\bigg[\left(
(\partial_\varrho \varphi)_{z}
\frac{\delta}{\delta \varphi_z}
(\phi_{\nu}\phi_{\nu})_{xw}\right)G^\perp_{wy}
\nonumber\\&\qquad\qquad
+\:G^\perp_{xw}\left((\partial_\varrho\varphi)_{z}\frac{\delta}{\delta \varphi_z}(\phi_\nu\phi_\nu)_{wy}\right)\bigg]\big(\phi_{\mu}\big)_{y}
\nonumber\\
& =\ -\:\int_w G^\perp_{xw}\big(\partial_\varrho \phi_{\mu}\big)_{w}\;, 
\label{eq:Gremainder:perp}
\end{align}
where we have repeatedly used Eq.~\eqref{eq:Gperp:ortho}, as well as the orthonormality \smash{$\int_x (\phi_{\mu})_{x}(\phi_{\nu})_{x}=\delta_{\mu\nu}$}. In Eq.~\eqref{eq:Gperp:ortho:tran},
this expression cancels the product of the first term in the first pair of brackets with
the second one in the second pair of brackets. The extra terms in Eq.~\eqref{eq:var:Gperp}
are therefore associated with a change of the Hilbert space over which the path integral
is performed, while, in a fixed space, Eq.~\eqref{eq:transl:Gperp} is the appropriate translation of
the Green's function.}

The variation of $\Pi$ appearing in Eq.~\eqref{eq:trinv:varphirho} therefore leads
to convolutions of the translational modes with all possible insertions in propagator 
lines of $\Pi$, i.e.
\begin{align}
\label{eq:vary:Pi}
\int_z\varphi_x\,\frac{\delta\Pi_{xx}}{\delta\varphi_z}\,\partial_{\varrho}\varphi_{z}
\ \equiv\ \int_z\Sigma_{xz}\,\partial_{\varrho}\varphi_{z}\;.
\end{align}
This implies that $\partial_{\mu}\varphi$ is a zero mode of the Klein-Gordon
operator in Eq.~\eqref{eqs:SD:twopoint} such that, indeed, we can invert it in the subspace perpendicular
to $\partial_{\mu}\varphi$. The system of equations~\eqref{eqs:SD} is then closed when
expressing $\Pi$ and $\Sigma$ as two-particle irreducible (2PI)
self-energies, i.e.~these are in a diagrammatic form derivable from
a 2PI effective action in terms of the resummed propagators ${\cal G}$ and expectation values $\varphi$.

We have so far focused on computing $\varphi$, ${\cal G}$ in the subspace
perpendicular to the zero modes.
The functions
$\varphi^{\veryshortarrow}$ or ${\cal G}^{\veryshortarrow}$ can be obtained to order $\hbar$ as
\begin{subequations}
\begin{align}
\label{eq:varphi:full}
\varphi^{\veryshortarrow}_x\ &=\ \varphi_x\:-\:\frac{\hbar}{B} \int_y{\cal G}_{xy}\,(\partial^2 \varphi_y^{(0)})\,,\\
{\cal G}^{\veryshortarrow}_{xy}\ &=\ {\cal G}_{xy}\:+\:\frac{\hbar}{B}\,\lambda\, \int_{vw} {\cal G}_{xv}\,{\cal G}_{vw}\,(\partial^2 \varphi_w^{(0)})\,\varphi_v\,{\cal G}_{vy}\;.
\label{eq:G:full}
\end{align}
\end{subequations}
We can check
explicitly that $\partial_{\mu}\varphi^{\veryshortarrow}$
is indeed a zero mode because
\begin{align}
{\cal G}^{\veryshortarrow-1}_{xy}\ &=\ G^{-1}_{xy}\:-\:\frac{\hbar}{B}\,\lambda \int_w \varphi_x\,\mathcal{G}_{xw}\big(\partial^2\varphi^{(0)}_w\big)\delta_{xy}
\nonumber\\
&\quad+\:\hbar\,\Pi_{xx}\delta_{xy}\:+\:\hbar\,\Sigma_{xy}
\:+\:{\cal O}(\lambda^2)\;,
\end{align}
which, to order $\lambda$, is orthogonal to
\begin{align}
\partial_\mu\varphi^{\veryshortarrow}_x\ =\ \partial_\mu \varphi_x
\:+\:\frac{\hbar}{B}\, \lambda \int_{yz}{\cal G}_{xz}\,\varphi_z \left(\partial_\mu\varphi_z\right)
{\cal G}_{zy}\left(\partial^2 \varphi_y^{(0)}\right)\;,
\end{align}
derived from Eq.~\eqref{eq:varphi:full}. It would be desirable to
derive a formally exact expression for ${\cal G}^{\veryshortarrow-1}$ that also
points to systematic approximations for this quantity to all orders.\footnote{
An expression for the proper self-energy that should appear in
${\cal G}^{\veryshortarrow-1}$ valid to all orders
would require some generalization
of the standard diagrammatic expansion because of the tadpole terms in the generating
functional~\eqref{eq:Z} that are not exponentiated.}
An apparent
direction to pursue is the construction of a 2PI effective action in the presence of the zero modes. We will address this interesting
formal step in future work.


\section{Self-consistent solutions in the classically scale-invariant model}
\label{sec:selfcon}

We now apply the more general considerations of the previous sections
in order to obtain solutions in a classically scale-invariant setup,
i.e.~self-consistent radiative corrections of the Fubini-Lipatov instanton.
For the contributions $j\geq2$, the Green's functions can be
computed straightforwardly in the background of either the classical or
loop-improved bounce, with a resulting difference of higher order in perturbation theory. Instead,
for $j=0$ and $j=1$, a resummation of the loop corrections is required
in order to arrive at a finite result.


\subsection{Modes $j\geq2$}
\label{sec:jgreq2}

For $j\geq 2$, we can proceed straightforwardly, i.e.~we can substitute the Fubini-Lipatov instanton,
which is the tree-level solution for $\varphi$ from Eq.~\eqref{eq:FL}, in Eq.~\eqref{eq:hyperradeq} for the Green's functions. The solutions $G_j$, to leading accuracy in the gradient corrections, are presented in Sec.~\ref{sec:Green} and App.~\ref{app:Greens}.


\subsection{Spectator fields}

In order to force the action to have an extremum at the bounce, we  add to the model in Eq.~\eqref{eq:lag} $N_\chi$ spectator fields $\chi$, as in Eq.~\eqref{eq:speclag}. At leading order, the Green's functions for each of the fields $\chi$ are determined as the solution to
\begin{align}
\label{eq:Green:spec}
&\bigg[
-\:\frac{{\rm d}^2}{{\rm d}r^2}\:-\:\frac{3}{r}\frac{\rm d}{{\rm d}r}\:+\:\frac{j(j+2)}{r^2}\:+\:\frac{\alpha}{2}\varphi^2(r)
\bigg]G_{\chi j}(r,r^\prime)\nonumber\\&\qquad =\ \frac{\delta(r-r^\prime)}{r^{\prime 3}}\;.
\end{align}
Summation over $j$ as in Eq.~\eqref{eq:Green:jsum} yields $G_\chi(x,x^\prime)$. Unlike the $\Phi$ Green's functions, there are no zero
or infrared-enhanced modes that require a particular procedure.


\subsection{Resummation of loop corrections, renormalization and the local approximation}
\label{sec:resum:renorm:local}

For the modes $j=0$ and $j=1$, it is necessary to account for infrared effects.
The required resummation is carried out through
the inclusion of self-energy terms in the equation of motion~\eqref{eq:bounce}
for the bounce and in Eq.~\eqref{eq:hyperradeq} for the Green's function,
which acquire loop corrections as
\begin{subequations}
\label{eq:SD:radial}
\begin{align}
-\:\frac{{\rm d}^2}{{\rm d}r^2}\,\varphi\:-\:\frac{3}{r}\,\frac{{\rm d}}{{\rm d}r}\,\varphi\:+\:U'(\varphi)\qquad\quad&
\nonumber\\
+\: \varphi\Big[\Pi^{\rm ren}\big(r(\varphi)\big)\:+\:\Pi^{\rm ren}_{\alpha}\big(r(\varphi)\big)\Big]\ &=\ 0\;,
\label{eq:bounce:loopcorrected}\\[1em]
\label{eq:GF:loopcorrected}
\tilde O_j\,{\cal G}_j(r,r^\prime)\:+\:\delta_{j1}\,\tilde \phi^{\rm tr}(r)\tilde \phi^{\rm tr}(r^\prime)\ &=\ \frac{\delta(r-r^\prime)}{r'}\;,
\end{align}
\end{subequations}
where we have introduced the one-loop-improved, re\-scaled
radial Klein-Gordon operator
\begin{align}
\label{eq:Oj}
\tilde O_{j}\ &=\ 
-\:\frac{{\rm d}^2}{{\rm d}r^2}\:-\:\frac{1}{r}\frac{\rm d}{{\rm d}r}\:+\:\frac{(j+1)^2}{r^2}\:+\:U''(\varphi)\nonumber\\&\qquad+\:\frac{\partial}{\partial\varphi}\,\varphi\Big[\Pi^{\rm ren}\big(r(\varphi)\big)\:+\:\Pi^{\rm ren}_{\alpha}\big(r(\varphi)\big)\Big]\;,
\end{align}
and where a separation of the angular part in analogy to Eq.~\eqref{eq:Green:jsum} is implied.
Compared to Eq.~\eqref{eqs:SD:twopoint}, the self-energy terms are expressed as a derivative
with respect to the background field $\varphi$. This amounts to a local approximation of the
self energy $\Sigma$, as we will discuss further at the end of the present subsection.
Moreover, we have introduced
the rescaled translational zero modes
\begin{align}
\tilde \phi^{\rm tr}(r)\ =\ {\cal N}r \partial_r\varphi(r)
\end{align}
that include the effects breaking the scale invariance and are to be subtracted for $j=1$.

\begin{figure}
\centering
\includegraphics[scale=0.97]{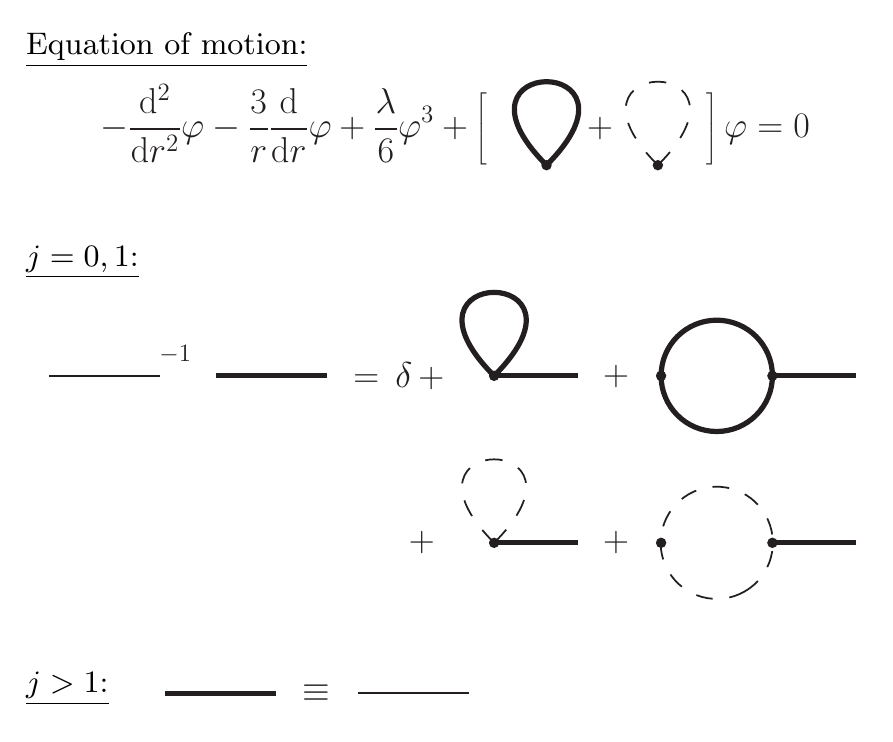}
\caption{
\label{fig:feyn:diags} Graphical representation of the approximation scheme applied in Secs.~\ref{sec:selfcon}
and~\ref{sec:results}. Solid lines represent the Green's function for the field $\Phi$; dashed
lines represent the fields $\chi$. Thin lines stand for tree-level Green's functions (in the
background of the bounce found by solving the equations of motion); thick lines stand
for the partly one-loop-resummed Green's functions.}
\end{figure}

The loop corrections in Eq.~\eqref{eq:bounce:loopcorrected} are contained within the coincident renormalized self-energies from
loops of $\Phi$ and $\chi$, i.e.~$\Pi^{\rm ren}$ and $\Pi^{\rm ren}_{\alpha}$.
The former is given by
\begin{align}
\label{eq:PiR}
\Pi^{\rm ren}(r)\ =\ \frac{\lambda}{2}\,{\cal G}^\perp(r,r)\:+\:\delta m^2\:+\:\frac{\delta\lambda}{6}\,\varphi^2(r)\;,
\end{align}
where we have introduced counterterms $\delta m^2$ and $\delta \lambda$. While we
have expressed this in terms of the exact Green's function ${\cal G}^\perp$,
a suitable approximation in order to capture the leading quantum corrections
for $j\geq2$
is to use the tree-level form from Sec.~\ref{sec:jgreq2}, while, for $j=0$ and $j=1$, we
need to solve the system~(\ref{eq:SD:radial}) self-consistently, thereby
resumming the one-loop effects. This approximation procedure is summarized
in Fig.~\ref{fig:feyn:diags}.

As renormalization conditions, we impose that the mass and the quartic coupling
of the one-loop effective potential in the false vacuum coincide with their tree-level values,
cf.~Eq.~\eqref{eq:counters}.
In the false vacuum, we evaluate the effective potential
at a homogeneous field configuration, for which there are closed expressions for the hyperspherical Green functions:
\begin{align}
G^{\rm hom}_j(m;r,r^\prime)\ & =\
\frac{1}{r r^\prime}\,\vartheta(r-r^\prime )
K_{j+1}
\left(
m r
\right)
I_{j+1}
\left(
m r^\prime
\right)\notag\\&\qquad +\:
(r\:\leftrightarrow\: r^\prime)\;,
\label{eq:Ghom}
\end{align}
where  $I_{\nu}$ and $K_{\nu}$ are the modified Bessel functions. When summing over $j$ according to Eq.~\eqref{eq:Green:jsum}, the full function $G^{\rm hom}(m;x,x^\prime)$ is obtained. The counterterms can be expressed conveniently in terms of the Green's functions in the homogeneous background by realizing that the coincident Green's functions are related to the derivatives of the Coleman-Weinberg potential~\eqref{eq:U:eff:ren} as $\varphi\,\Pi^{\rm ren}=\partial U^{\rm ren}_{\rm eff}/\partial\varphi$. This leads to
\begin{subequations}
\label{eqs:counterterms}
\begin{gather}
\delta m^2\ =\ -\:\frac{\lambda}{2}\,\frac{\partial}{\partial\varphi}\,\varphi\,G^{\rm hom}\big(\sqrt{m^2+\lambda\,\varphi^2/2};r,r\big)\Big|_{\varphi\,=\,0}\
\notag\\
=\ -\:\frac{\lambda}{2}\,G^{\rm hom}(m;r,r)\;,\\
\delta \lambda\ =\ -\:\frac{\lambda}{2}\,\frac{\partial^3}{\partial\varphi^3}\,\varphi\,G^{\rm hom}\big(\sqrt{m^2+\lambda\,\varphi^2/2};r,r\big)\Big|_{\varphi\,=\,0}\;.
\end{gather}
\end{subequations}
The derivatives can be evaluated by making use of the recursion relations of the modified Bessel functions, leading to closed-form
expressions for the counterterms.

Along the same lines,
the one-loop effects from the field $\chi$ enter Eqs.~\eqref{eq:bounce:loopcorrected} and \eqref{eq:GF:loopcorrected} through
\begin{align}
\Pi^{\rm ren}_{\alpha}(r)\ =\ N_\chi\frac{\alpha}{2}\,G_{\chi}(r,r)\:+\:\delta m_{\alpha}^2\:+\:\frac{\delta\lambda_{\alpha}}{6}\,\varphi^2(r)\;,
\end{align}
where we introduce the counterterms
\begin{subequations}
\label{eqs:counterterms:alpha}
\begin{align}
\delta m_{\alpha}^2\: &=\: -\:N_\chi\frac{\alpha}{2}\,G^{\rm hom}(m_{\chi};r,r)\;,\\
\delta \lambda_{\alpha}\: &=\: -\:N_\chi\frac{\alpha}{2}\,\frac{\partial^2}{\partial \varphi^3}\,\varphi\,G^{\rm hom}\Big(\!\sqrt{m_\chi^2+\alpha\,\varphi^2/2};r,r\Big)\Big|_{\varphi\,=\,0}\;.
\end{align}
\end{subequations}
Here, for all $j$ modes,
the propagator $G_\chi$ can be taken at tree level in the background of either the tree-level or the loop-improved bounce
in order to capture the leading quantum
corrections from the field $\chi$ in the system~\eqref{eq:SD:radial}, since no problematic
infrared effects occur. Again, the difference between using the tree-level and self-consistent bounces is of higher
perturbative order and will only matter when aiming for two-loop precision.

The counterterms specified above through Eqs.~\eqref{eqs:counterterms} and~\eqref{eqs:counterterms:alpha} are valid at one-loop order. However, the self-consistent solution to Eq.~\eqref{eq:GF:loopcorrected} inserts radiative corrections into the Green's function for $j=0$ and $j=1$.
This is reminiscent of 2PI effective actions, where it is well-known that
the one-loop counterterms are insufficient to render the one-loop resummed quantities finite. While there is a proof of principle that 2PI renormalization schemes are viable~\cite{Berges:2004hn,Berges:2005hc}, we currently do not find it feasible to apply these techniques to the present context. Instead of explicitly specifying local counterterms for $j=0$ and $j=1$, we therefore make the replacement
\begin{align}
{\cal G}_{j}(r,r)
\ &\longrightarrow\ 
{\cal G}_{j}(r,r)\:-\:{\rm Re}\big[G^{\rm hom}_{j}\big(M_\varphi(\varphi);r,r\big)\nonumber\\&\qquad\qquad -\:G^{\rm hom}_{j}\big(\sqrt{m^2+\lambda\,\varphi^2/{2}};r,r\big)\big]\;,
\end{align}
where we make use of the Green's function in the homogeneous background~\eqref{eq:Ghom} and
\begin{align}
&M^2_\varphi(\varphi)\ \equiv\ m^2\:+\:\frac{\lambda}{2}\,\varphi^2\nonumber\\&\qquad+\:\frac{\partial}{\partial\varphi}\,\varphi\Big[\Pi^{\rm ren}\big(r(\varphi)\big)\:+\:\Pi^{\rm ren}_{\alpha}\big(r(\varphi)\big)\Big]\;.
\end{align}
Thus, only the contributions due to the gradients in the field $\varphi$ are resummed, while those terms that are already present for $\varphi={\rm const.}$ are dropped. Since the gradient corrections are ultraviolet finite, we can therefore apply the one-loop counterterms while self-consistently regulating the infrared enhancement.

We now comment on the local approximation applied in Eq.~\eqref{eq:Oj}.
When compared with Eq.~\eqref{eqs:SD:twopoint}, it can be expressed as
\begin{align}
\int_{z} \Sigma_{xz}\,{\cal G}_{zy}\ &=\ \varphi_x\int_z\left(\frac{\delta}{\delta\varphi_{z}}\,\Pi_{xx}\right)
{\cal G}_{zy}
\notag\\&
\approx\ \varphi_x \left(\frac{\partial}{\partial\varphi}\,\Pi_{xx}\big(r(\varphi)\big)\right){\cal G}_{xy}\;.
\end{align}
This is a linearization which would correspond to an exact spatial translation
if ${\cal G}(x,y)\propto \partial_\mu \varphi(x)$,
but which is not the case in general. In an adiabatic expansion, the relative size of the first correction
for $j=0$ is
\begin{align}
\label{eq:est:local}
\frac{(\partial_\mu m_{\rm loop})^2}{m_{\rm loop}^4}\ 
\sim\ \frac{2(\partial_\mu\varphi)^2}{\kappa \varphi^4}\ =\ -\:\frac{\lambda r^2}{6 \kappa R^2}\;,
\end{align}
where $m_{\rm loop}^2=(\kappa/2)\varphi^2$ and $\kappa=\lambda,\alpha$, depending on which field we consider
in the loop.
Alternatively, we can make this estimate by realizing that, in momentum space, the coincident limit corresponds
to a zero external momentum approximation. Taking for the external momentum
$(\partial_\mu\varphi)/\varphi$ and comparing the square of this with the squared mass in the loop,
we arrive at the same estimate as in Eq.~\eqref{eq:est:local}.

At the centre of the bubble, for small $r$, the local approximation should therefore be accurate, and, for
$j\not=0$, we should replace $m_{\rm loop}^2\to\omega_{\rm loop}^2=(\kappa/2)\varphi^2+j(j+2)/r^2$ in  Eq.~\eqref{eq:est:local}, such that the approximation further improves for the higher modes. On the other hand, for
large $r$, the relative size of the nonlocal diagram compared to the coincident contribution
$\kappa^2\varphi^2/\kappa$
decreases because $\varphi\to 0$. We can account for this by adding a factor to the error estimate~\eqref{eq:est:local}, which then
becomes
\begin{align}
\label{eq:est:local:improved}
\frac{\lambda r^2}{6 \kappa R^2}\,\frac{\kappa^2\varphi^2}{\kappa^2\varphi^2+\kappa}\;.
\end{align}
We find this to be smaller than, e.g., $25\%$ for all values of $r$ and the
parameters chosen in Sec.~\ref{sec:results}.

While the local approximation is a considerable simplification in the derivation of the first numerical results
presented in this work, we will address an improvement on this procedure (at least for loops from small $j$) in the future.


\subsection{Mode $j=1$}

The case $j=1$ requires special treatment because of the apparent singularity in the tree-level Green's function~\eqref{eq:Gj} and because of the presence of the translational zero modes. Besides the divergence in the denominator, we also note that $P_2^{-j-1}(u)$ is proportional to  $P_2^{j+1}(u)$ in the limit $j\to 1$, such that we cannot use it along with $P_2^2$ in order
to construct the Green's function as in Eq.~\eqref{eq:fgr}. One may instead choose to work with $Q_2^{j+1}(u)$ as an additional basis solution. Explicitly, these functions read
\begin{subequations}
\begin{align}
P_2^{2}(u)\ &=\ 3(1-u^2)\;,\quad P_2^{-2}(u)\ =\ \frac18(1-u^2)\;,\\
Q_2^{2}(u)\ &=\ \frac{5u-3u^3}{1-u^2}\:+\:\frac{3}{2}\, (1-u^2)\ln\frac{1+u}{1-u}\;,
\end{align}
\end{subequations}
but we note that $Q_2^{2}$ does not satisfy the boundary conditions, i.e.~it diverges at both ends of the interval $(-1,1)$. Nevertheless, these functions are useful in order to construct an
approximation to the full solution that only relies on a single numerical coefficient, and
we will return to this point at the end of this section.

The subtraction of the translational zero modes turns out to be possible only when accounting
for the deviation of the bounce from the scale-invariant Fubini-Lipatov form.
Note that in the scale-invariant limit $\varphi\equiv\varphi^{(0)}$, $\tilde\phi^{\rm tr}(r)=P_2^2(u)/(2\sqrt{6} R)$, such that we recover the translational zero mode given in Eq.~\eqref{eq:transmode}.

Next, we need to solve Eq.~\eqref{eq:GF:loopcorrected} for $j=1$.
The solution takes the general form
\begin{align}
\label{eq:Gperp1}
 &\tilde {\cal G}_{1}(r,r^\prime)\ =\ -\:\vartheta(u-u^\prime)\Big[f_-(u)\tilde\phi^{\rm tr}(r^\prime)\:+\:f_+(u^\prime)\tilde\phi^{\rm tr}(r)\Big]\notag\\
 &\qquad -\:\vartheta(u^\prime-u)\Big[f_-(u^\prime)\tilde\phi^{\rm tr}(r)\:+\:f_+(u)\tilde\phi^{\rm tr}(r^\prime)\Big]\notag\\
 &\qquad+\:a \,\tilde\phi^{\rm tr}(r)\tilde\phi^{\rm tr}(r^\prime)\;,
\end{align}
where $\tilde{\cal G}$ is defined in analogy with Eq.~\eqref{eq:Gtilde},
$f_\pm$ are solutions to the inhomogeneous equation
\begin{align}
\label{eq:inhomogeneous}
\tilde O_{j\,=\,1}f_\pm(u)\ =\ \tilde\phi^{\rm tr}(r)\;,
\end{align}
with $f_+(u)$ being regular for $u\to 1$ and $f_-(u)$ for $u\to -1$.  The solutions $f_\pm$ can be obtained numerically and are, in general, not orthogonal to $\tilde\phi^{\rm tr}$. The coefficient $a$ can be determined by imposing orthogonality to the zero mode
\begin{align}
\label{eq:ortho:transl}
\int\limits_0^\infty {\rm d}r\;r \tilde\phi^{\rm tr}(r) \tilde{\cal G}_1(r,r^\prime)\ =\ 0\;.
\end{align}
This condition can be solved for $a$ as
\begin{align}
a\ &=\ \frac{1}{\tilde\phi^{\rm tr}(r)}
\Big\{
\Big[C_+(r)\:+\:C_-(r)\Big]\tilde\phi^{\rm tr}(r)\nonumber\\&\qquad 
+\:f_-(u)D_+(r)\:+\:f_+(u)D_-(r)\Big\}\;,
\end{align}
which is independent of $r$ and where
\begin{subequations}
\begin{align}
C_{+,-}(r)\ &=\ \int\limits_{0,r}^{r,\infty}{\rm d}r^\prime\;r^\prime f_{+,-}(r^\prime)\tilde\phi^{\rm tr}(r^\prime)\;,\\
D_{+,-}(r)\ &=\ \int\limits_{0,r}^{r,\infty}{\rm d}r^\prime\;r^\prime  \big[\tilde\phi^{\rm tr}(r^\prime)\big]^2\;.
\end{align}
\end{subequations}

We can now comment on why this procedure is not applicable to the solutions in the tree-level
Fubini-Lipatov background, where the scaled translational mode is
\begin{align}
\label{eq:trans:tree}
\tilde\phi^{\rm tr}(r)\ =\ \sqrt{\frac{3}{8}}\,\frac{1-u^2}{R}
\end{align}
and where the inhomogeneous equation can be cast to
\begin{align}
&\bigg[
\frac{\rm d}{{\rm d}u}\,(1-u^2)\,\frac{{\rm d}}{{\rm d} u}\:-\:\frac{4}{1-u^2}\:+\:6
\bigg]
f_{\pm}(u)\ =\ \sqrt{\frac{3}{8}}\,\frac{1-u}{1+u}\;.
\end{align}
The particular solutions observing the boundary conditions are
\begin{subequations}
\label{eq:fpm:analytical}
\begin{align}
f_+(u)\ &=\ -\:\sqrt{\frac{3}{8}}\,\frac{1-u}{1+u}\,\frac{5u^2+10u+3}{12}\;,\\
f_-(u) \ &=\ f_+(u)\:-\:\frac{1}{\sqrt{24}}\,Q_2^2(u)\;.
\end{align}
\end{subequations}
Note that $Q_2^2(u)$ is a nonnormalizable solution to the homogeneous equation. We can substitute these results into Eq.~\eqref{eq:Gperp1}. However, it is now not possible to project out the zero mode because the integral
\begin{align}
\int\limits_{-1}^{u^\prime}\frac{{\rm d}u}{(1+u)^2}\;f_-(u)\tilde\phi^{\rm tr}(r)
\end{align}
is logarithmically divergent. This problem does not occur when we account for the deviation
from the Fubini-Lipatov form. In particular, the mass term leads to an exponential decay of
the modes for $r\to\infty$, such that the integral Eq.~\eqref{eq:ortho:transl} is convergent.

Nevertheless, we can compare the result~\eqref{eq:Gperp1} for $\tilde{{\cal G}}_ 1$, based on the numerical solutions $f_\pm(u)$, with one where we use the solutions~\eqref{eq:trans:tree} and~\eqref{eq:fpm:analytical} and determine the parameter $a$ by matching with the numerical ${\cal G}_{1}$, cf.~Fig.~\ref{fig:jeq1} below.


\subsection{Mode $j=0$}

If the case $j=0$ were to be treated in complete analogy with $j=1$, we would need to subtract the contribution from the dilatational zero mode from the $j=0$ Green's function and integrate over the dilatations of the critical bubble as a collective coordinate, implying that bubbles of all radii are nucleated at the same rate. Such a procedure can, however, not be valid because the scale invariance is broken by radiative effects. As a consequence, a unique extremum of the action corresponding to a soliton configuration may emerge or, in the opposite case, no such solution may exist. In the former case, a Gau{\ss}ian evaluation of the functional integral over the dilatational mode becomes possible.

Since, in the perturbatively well-defined case (where, at the point when $\varphi=0$, an infrared divergence in the quartic coupling arising at one-loop order should be avoided), radiative corrections necessarily imply $m^2>0$ in the renormalized potential, there is no zero mode for $j=0$ once these effects are included. Therefore, there is no ambiguity in the solution to Eq.~\eqref{eq:GF:loopcorrected}.

The numerical calculation of the Green's function for $j=0$, including the effects breaking scale invariance, is straightforward, up to matters concerning
the renormalization that require special care and that are explained in Sec.~\ref{sec:resum:renorm:local}. Similar to the case $j=1$, it is nonetheless interesting to construct approximate solutions that rely on a single numerical parameter only. We can find these when considering
\begin{align}
\label{eq:j0}
\tilde O_{j\,=\,0}\,\tilde {\cal G}_{0}(r,r^\prime)
\ &=\ \frac{\delta(r-r^\prime)}{r'}
\end{align}
and the dilatational mode at tree-level
\begin{align}
\tilde\varphi^{\rm dil}_0(r)\ =\ \partial_R\varphi(r)\ =\ -\:\sqrt{-\frac{12}{\lambda}}\,\frac{u(1+u)}{R^2}\;.
\end{align}
Now, this mode is not normalizable, i.e.
\begin{align}
\int\limits_{-1}^1\frac{{\rm d}u}{(1+u)^2}\;\big[r \partial_R\varphi(r)\big]^2
\end{align}
diverges logarithmically for $u\to -1$ ($r\to\infty$). Even so, when neglecting radiative corrections, Eq.~\eqref{eq:j0} is solved by
\begin{align}
\label{eq:G_0:analytical}
&\tilde G_0(u,u^\prime)\ =\ \vartheta(u-u^\prime)\frac{1}{6}\, P_2^1(u)Q_2^1(u^\prime)
\nonumber\\&\qquad+\:\vartheta(u^\prime-u)\frac{1}{6}\, P_2^1(u^\prime)Q_2^1(u)
\:+\: b\,P_2^1(u) P_2^1(u^\prime)
\;.
\end{align}
Note that $rr^\prime \tilde G_0(r,r^\prime)$ is regular everywhere, but the parameter $b$ remains arbitrary. The zero mode cannot be projected out because it is not normalizable. However, by comparing with the numerical result accounting for effects breaking scale invariance, $b$ can be determined such that we obtain an approximate solution that only relies on a single numerical parameter, cf.~Figs.~\ref{fig:jeq0} and~\ref{fig:jeq0-2} below.


\subsection{The negative eigenmode in the loop expansion}

The functional integral over the negative eigenmode can be defined only by analytic continuation via the method of steepest descent. One might be concerned that this will lead to subtleties in the diagrammatic expansion with respect to the contributions from the negative eigenmode in the $j=0$ mode of the (subtracted) Green's function. As we will now describe, however, there are no modifications necessary.

We can isolate the contribution from the negative eigenmode by coupling it to an independent source $J_0$. We expand the field as
\begin{equation}
\Phi\ =\  \varphi^{(0)}\:+\:\hbar^{1/2}\sum_{\mu\,=\,0}^{4}a_{\mu}\phi_{\mu}\:+\:\hbar^{1/2}\Phi''\;,
\end{equation}
where $\mu=0$ corresponds to the negative eigenmode $\phi_0$. We drop the superscript $(0)$ on the tree-level zero modes in this subsection for notational convenience. The generating functional $\mathcal{Z}'[J]\to\mathcal{Z}'[J,J_0]$ (in which the zero modes have been taken care of, see Sec.~\ref{sec:zeroes}) can then be written as
\begin{align}
&\mathcal{Z}'[J,J_0]\ =\ \exp\bigg[-\:\frac{1}{\hbar}\bigg(S[\varphi^{(0)}]\:-\:\int_xJ_x\varphi^{(0)}_x\bigg)\bigg]\nonumber\\&\quad\times\int\mathcal{D}\Phi''\int\frac{\mathrm{d} a_0}{\sqrt{2\pi}}\;\exp\bigg[-\int_{xy}\frac{1}{2}\Phi''_xG^{-1}_{xy}\Phi''_y\:-\:\frac{1}{2}\,\lambda_0\,a_0^2\nonumber\\&\quad\:+\:\hbar^{-1/2}\int_xJ_x\Phi''_x\:+\:\hbar^{-1/2}a_0\int_x(J_{0})_{x}(\phi_{0})_{x}\nonumber\\&\quad-\:\hbar^{1/2}\int_x\frac{\lambda}{3!}\,\varphi^{(0)}_x\big(a_0\phi_0+\Phi'')_x^3\nonumber\\&\quad-\:\hbar\int_x\frac{\lambda}{4!}\big(a_0\phi_0+\Phi'')_x^4\bigg]\;,
\end{align}
where $\lambda_0<0$ is the negative eigenvalue. The contribution from the negative eigenmode to the tree-level subtracted Green's function $G^{\perp}$ [cf.~Eq.~\eqref{eq:Gperpfunc}] can now be obtained straightforwardly by functional differentiation:
\begin{equation}
\hbar\,(G^{\perp}_0)_{xy}\: =\: \frac{\hbar^2}{\mathcal{Z}^{\prime(0)}[0,0]}\,\frac{\delta}{\delta (J_{0})_{x}}\,\frac{\delta}{\delta (J_{0})_{y}}\,\mathcal{Z}^{\prime(0)}[J,J_0]\bigg|_{J,J_0\,=\,0}\,.
\end{equation}
The superscript $(0)$ on $\mathcal{Z}'$ indicates that the interactions between the fluctuations have been set to zero. Proceeding in this way, we obtain
\begin{align}
&\hbar\,(G^{\perp}_0)_{xy}\ =\ \hbar\,(\phi_{0})_{x}(\phi_0)_{y}\nonumber\\&\quad\times\:\frac{1}{\mathcal{Z}^{\prime(0)}[0,0]}\,\exp\bigg[-\:\frac{1}{\hbar}\bigg(S[\varphi^{(0)}]\:-\:\int_xJ_x\varphi^{(0)}_x\bigg)\bigg]\nonumber\\&\quad\times\big(\mathrm{det}''G^{-1}\big)^{-1/2}\int\frac{\mathrm{d} a_0}{\sqrt{2\pi}}\;a_0^2\exp\bigg[-\:\frac{1}{2}\,\lambda_0\,a_0^2\bigg]\;,
\end{align}
where $\mathrm{det}''$ is the determinant over the positive-definite eigenmodes.  Applying the method of steepest descent, we obtain $1/2$ of the integral over $a_0'=-ia_0\in(-\infty,\infty)$:
\begin{align}
&\hbar\,(G^{\perp}_0)_{xy}\ =\ \hbar\,(\phi_{0})_{x}(\phi_{0})_{y}\nonumber\\&\ \times\:\frac{1}{\mathcal{Z}^{\prime(0)}[0,0]}\,\bigg\{\frac{i}{2}\,\exp\bigg[-\:\frac{1}{\hbar}\bigg(S[\varphi^{(0)}]\:-\:\int_xJ_x\varphi^{(0)}_x\bigg)\bigg]\nonumber\\&\ \times\big(\mathrm{det}''G^{-1}\big)^{-1/2}\bigg\}\int\!\frac{\mathrm{d} a_0'}{\sqrt{2\pi}}\;\big(-a_0^{\prime 2}\big)\exp\bigg[-\,\frac{1}{2}\,|\lambda_0|\,a_0^{\prime 2}\bigg]\;.
\end{align}
Performing the remaining Gaussian integral over $a_0'$ yields
\begin{align}
&\hbar\,(G^{\perp}_0)_{xy}\ =\ \hbar\, (-1)\frac{(\phi_{0})_{x}(\phi_{0})_{y}}{|\lambda_0|}\nonumber\\&\quad \times\:\frac{1}{\mathcal{Z}^{\prime(0)}[0,0]}\,\bigg\{\frac{i}{2}\exp\bigg[-\:\frac{1}{\hbar}\bigg(S[\varphi^{(0)}]\:-\:\int_xJ_x\varphi^{(0)}_x\bigg)\bigg]\nonumber\\&\quad\times\big(\mathrm{det}''G^{-1}\big)^{-1/2}|\lambda_0|^{-1/2}\bigg\}\;.
\end{align}
Recognizing the content of the braces as $\mathcal{Z}^{\prime(0)}[0,0]$, we therefore find
\begin{equation}
\label{eq:Gperpminus}
\hbar\,(G^{\perp}_{0})_{xy}\ =\
\hbar\,\frac{(\phi_0)_{x}(\phi_0)_{y}}{\lambda_0}\;,
\end{equation}
which is the expected contribution from the negative eigenmode.

On including the interactions between the fluctuations, we find that each Wick contraction of $a_0^2$ gives a factor of $1/|\lambda_0|$ and an additional factor of $-1$ from the analytic continuation. Thus, the diagrammatic expansion is built straightforwardly out of $G^{\perp}$, without any modifications to the contributions from the negative eigenmode in Eq.~\eqref{eq:Gperpminus}.


\section{Parametric example}
\label{sec:results}

Now, we numerically solve Eq.~\eqref{eq:bounce:loopcorrected} for the bounce and Eqs.~\eqref{eq:Green:spec} and~\eqref{eq:GF:loopcorrected} for the Green's functions self-consistently by running several iterations over these equations. This procedure can be initialized by calculating the bounce in the Coleman-Weinberg effective potential. Since the iterations are repeated until the self-consistent results converge, no memory of this initial step persists. For the parametric example, we use $\lambda=-1$, $m=1$, $\alpha=1$, $m_\chi=1$ and $N=7$. Note that $m=1$ is therefore the basic unit of the quantities that we 
present in this section. When switching to a dimensionful mass, the dimensionless scales can be interpreted as $\varphi/m$, $\Pi^{\rm 
ren}/m^2$ and $r m$ etc.

\begin{figure}
\includegraphics[scale=0.6]{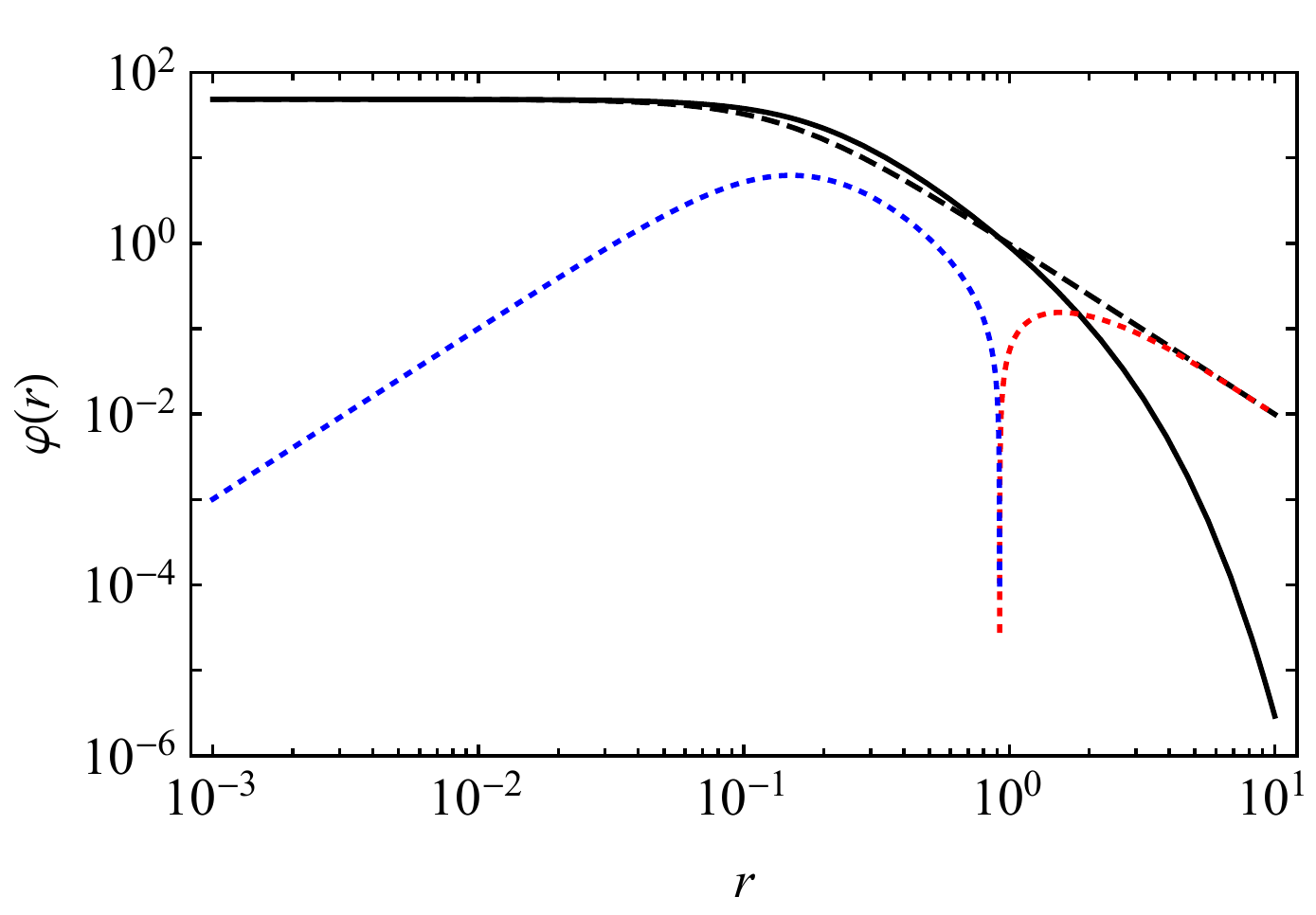}
\caption{\label{fig:bounce_par1}Plots of the numerical bounce (solid black), including the corrections from the breaking of scale invariance, versus the Fubini-Lipatov instanton (dashed black). The dotted lines are the positive (blue) and negative (red) difference between the former and the latter.}
\end{figure}
\begin{figure}
\includegraphics[scale=0.6]{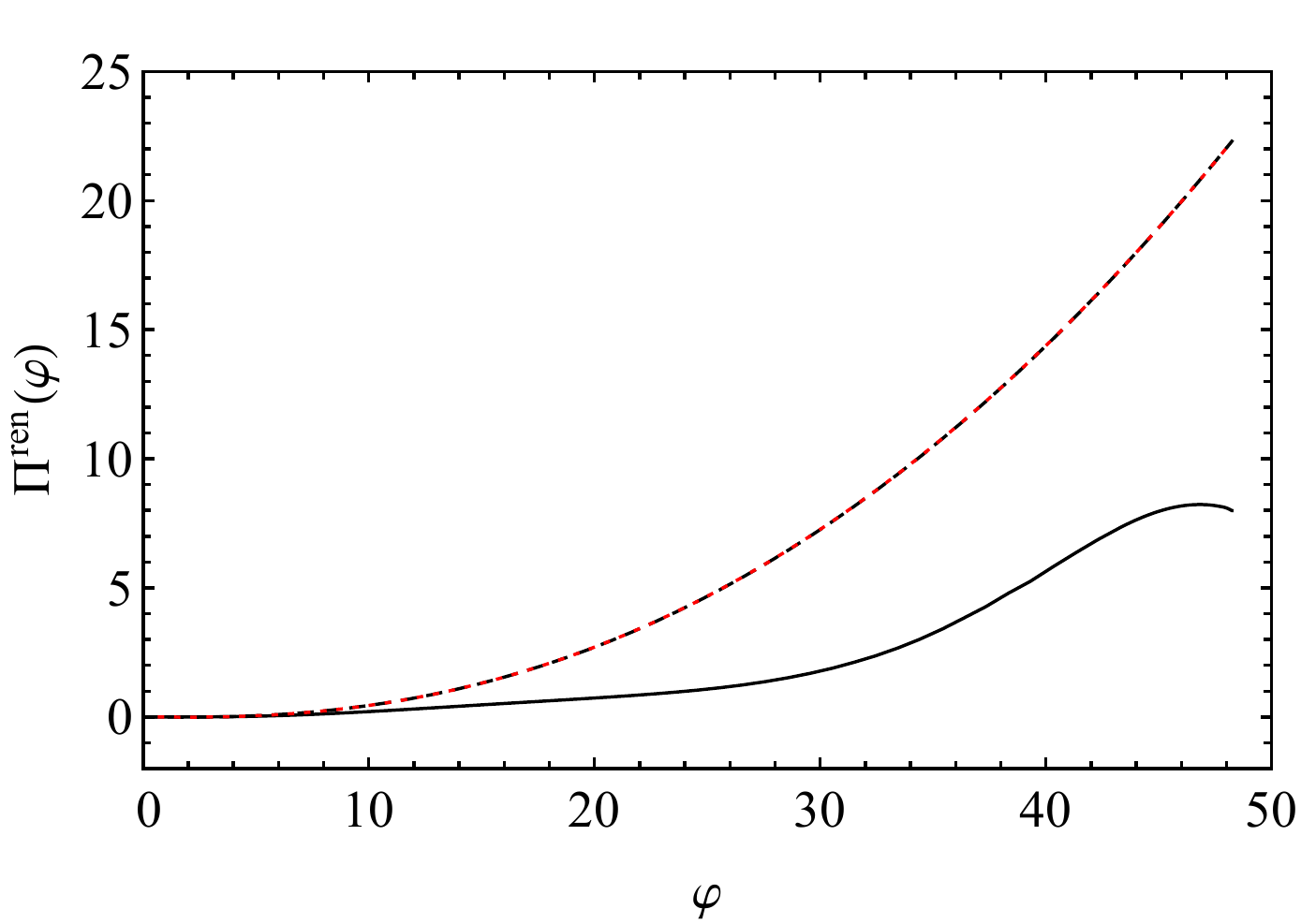}
\caption{\label{fig:Piren_par1}Plots of the renormalized self-energy $\Pi^{\rm ren}(\varphi)$ with (solid black) and without (black dashed) gradients, and from the Cartesian-space computation (red dotted) in Eq.~\eqref{eq:Pi:Cartesian}.}
\end{figure}

In \fref{fig:bounce_par1}, we show the numerical solution for the bounce in comparison with the Fubini-Lipatov instanton. The radius $R$ of the Fubini-Lipatov instanton is fixed by matching to the release value of the numerical bounce. The scale-invariance breaking effects are small in accord with what we expect from the perturbative expansion. Various versions of the renormalized self-energy are shown in \fref{fig:Piren_par1}. The variant without gradients is based on ${\rm Re}[G_j^{\rm hom}]$ given in Eq.~\eqref{eq:Ghom}. We also compare with the self-energy evaluated
as an integral in Cartesian space and without gradients, which is given by
\begin{align}
\label{eq:Pi:Cartesian}
&\Pi^{{\rm ren, hom}}\ =\ \frac{\lambda}{32\pi^2}
\bigg[
-\:\frac{\lambda}{2}\,\varphi^2\nonumber\\&\qquad+\:\bigg(m^2+\frac{\lambda}{2}\varphi^2\bigg)
\ln\bigg(\frac{m^2+\frac{\lambda}{2}\varphi^2}{m^2}\bigg)\bigg]\;.
\end{align}
Apparently, the only residual difference between the two gradient-free versions is due to the different regularization schemes applied in spherical and Cartesian coordinates, and it goes to zero when the respective cutoffs are taken to infinity.

\begin{figure}
\includegraphics[scale=0.6]{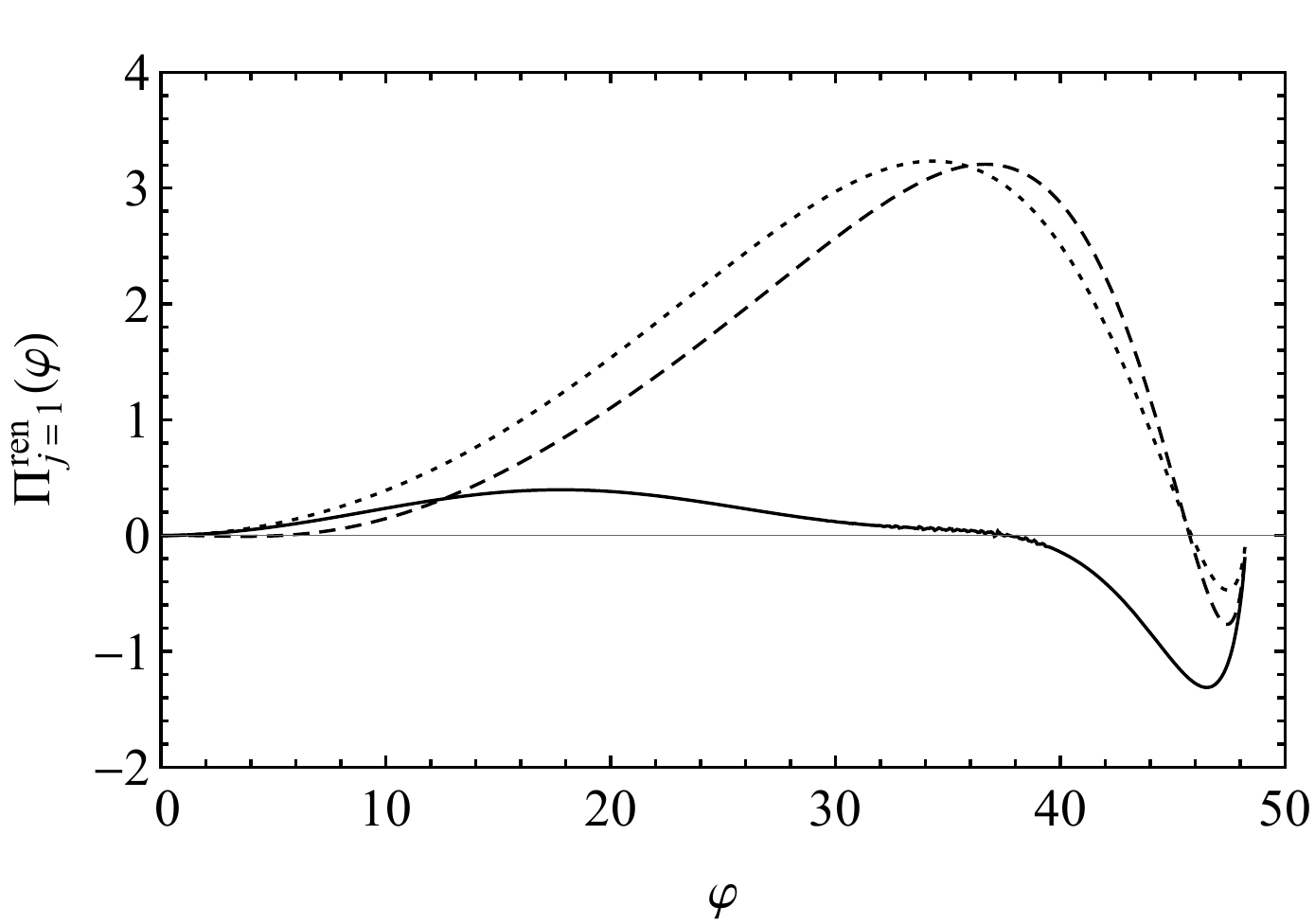}
\caption{\label{fig:jeq1}
Plots of the $j=1$ contribution to the renormalized self-energy $\Pi^{\rm ren}_{j\,=\,1}(\varphi)$ with (solid) and without (dashed) gradients. The dotted line corresponds to the analytic solution based on Eqs.~\eqref{eq:Gperp1} and~\eqref{eq:fpm:analytical} with a fitted value for the parameter $a$.}
\end{figure}
\begin{figure}
\includegraphics[scale=0.6]{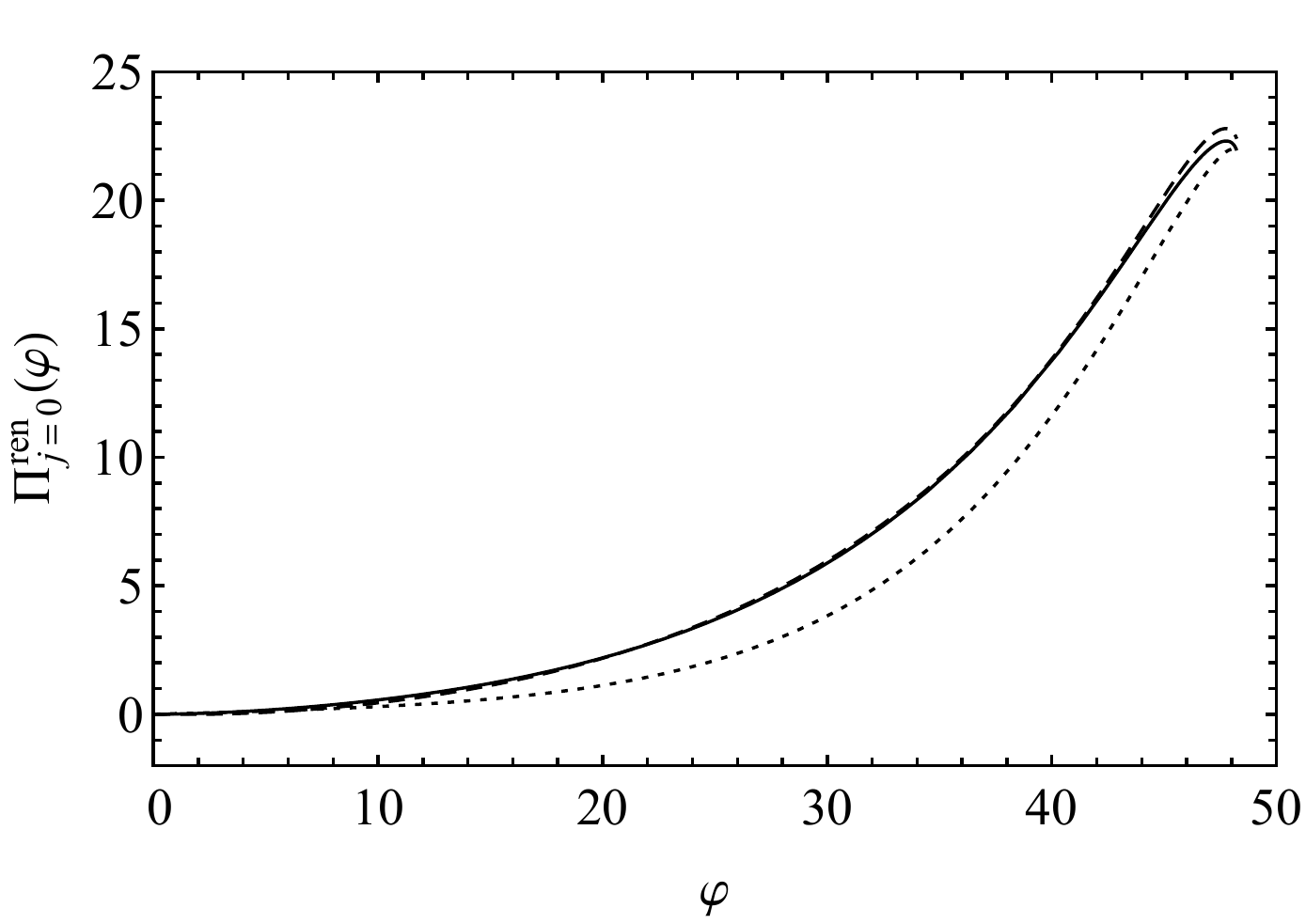}
\caption{\label{fig:jeq0}Plots of the $j=0$ contribution to the renormalized self-energy $\Pi^{\rm ren}_{j\,=\,0}(\varphi)$ with (solid) and without (dashed) gradients. In the case with gradients (solid), the dilatational zero mode has been subtracted by means of the fitted amplitude. The dotted line corresponds to the analytic solution based on Eq.~\eqref{eq:G_0:analytical} with $b=0$.}
\end{figure}
\begin{figure}
\includegraphics[scale=0.6]{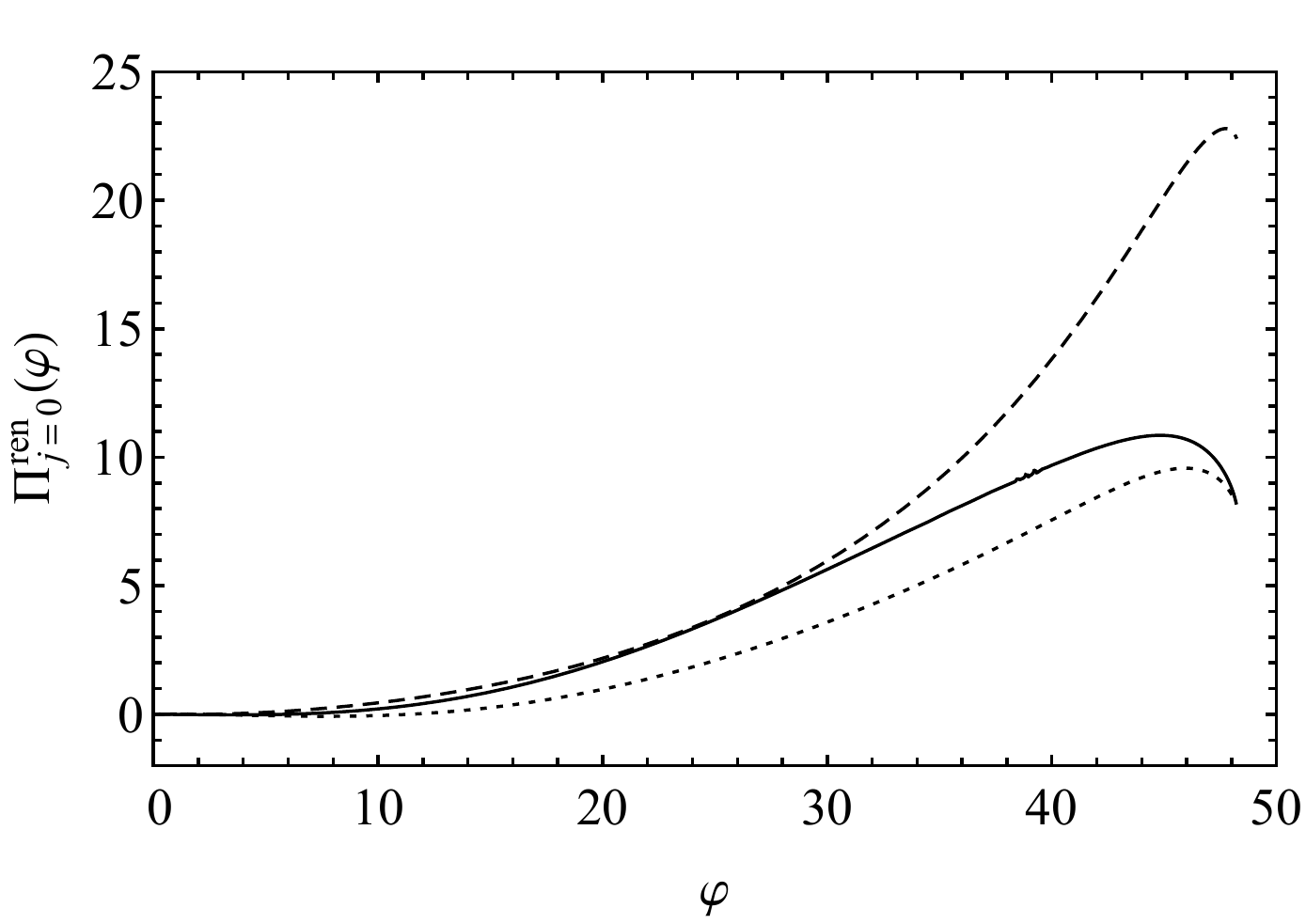}
\caption{\label{fig:jeq0-2}Plots of the $j=0$ contribution to the renormalized self-energy $\Pi^{\rm ren}_{j\,=\,0}(\varphi)$ with (solid) and without (dashed) gradients. The dotted line corresponds to the analytic solution based on Eq.~\eqref{eq:G_0:analytical} with a fitted value for the parameter $b$.}
\end{figure}
\begin{figure}
\includegraphics[scale=0.6]{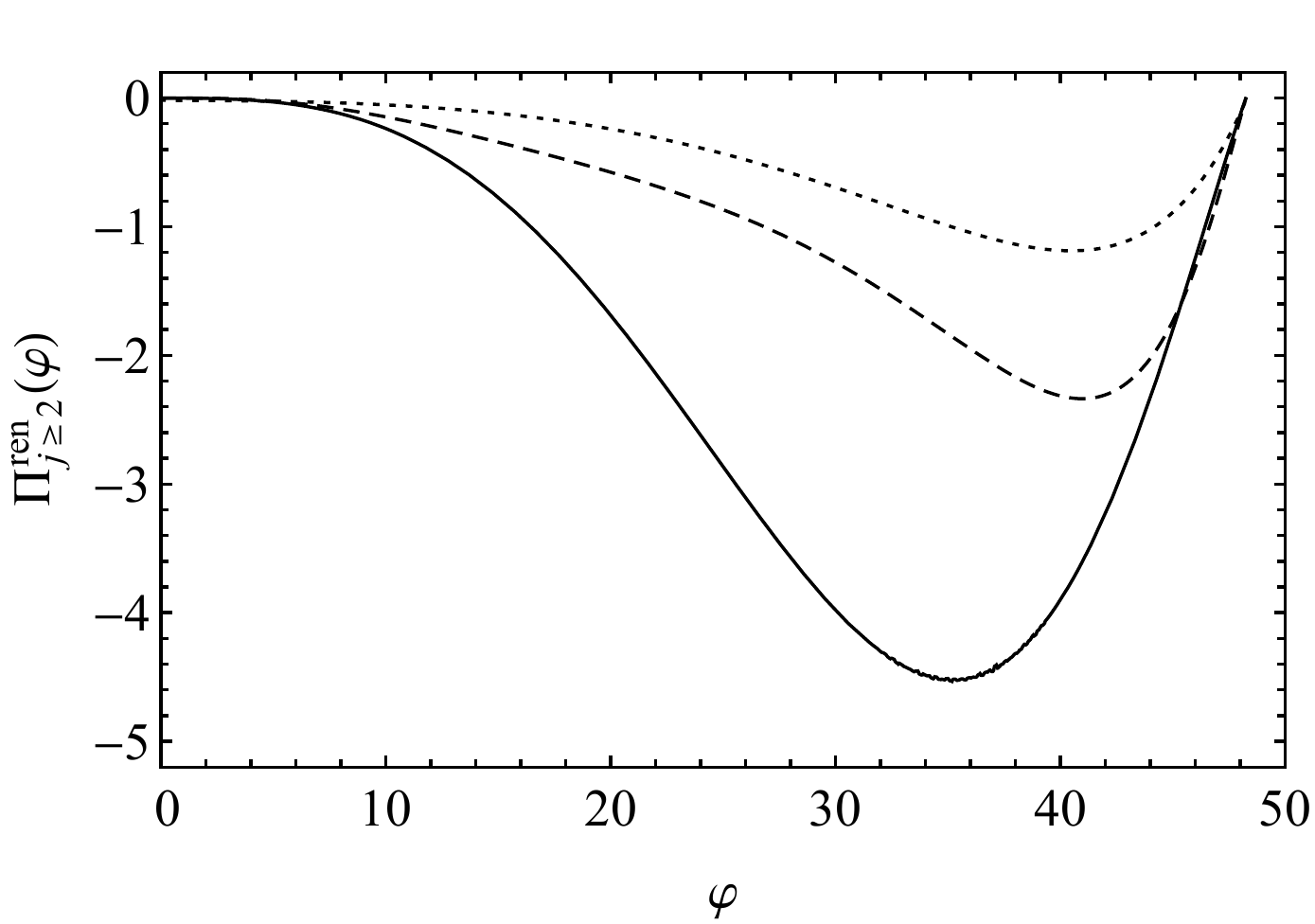}
\caption{\label{fig:jeqgr2}Plots of the contributions to the renormalized self-energy from $j\geq 2$, i.e.~$\Pi^{\rm ren}_{j\,\geq\,2}(\varphi)$ with (solid black) and without (black dashed) gradients, and using the analytic results for the Green's function in the Fubini-Lipatov background (black dotted).}
\end{figure}

In \fref{fig:jeq1}, we present variants of the $j=1$ contribution to the renormalized self-energy $\Pi^{\rm ren}_{j\,=\,1}(\varphi)$. We verify that the gradient effects can be largely isolated in terms of a contribution from the squared translational modes normalized by the parameter $a$, as obtained by a fit to the numerical result. The corresponding results for $j=0$ are presented in \fref{fig:jeq0}, and we compare the self-energies without gradient corrections in \fref{fig:jeq0-2} in order to appreciate the size of the gradient effects and the accuracy of the analytic approximations.

Next, in \fref{fig:jeqgr2}, we make the comparison for the contributions from $j\geq 2$. By comparing with Fig.~\ref{fig:jeq0-2}, we observe that the lion's share of the radiative corrections can be attributed to $j=0$, such that the sector $j\geq 2$ is subdominant. Thus, also the comparably large relative discrepancy apparent in \fref{fig:jeqgr2}
does not invalidate the overall applicability of the analytic approximations.
\begin{figure}
\includegraphics[scale=0.6]{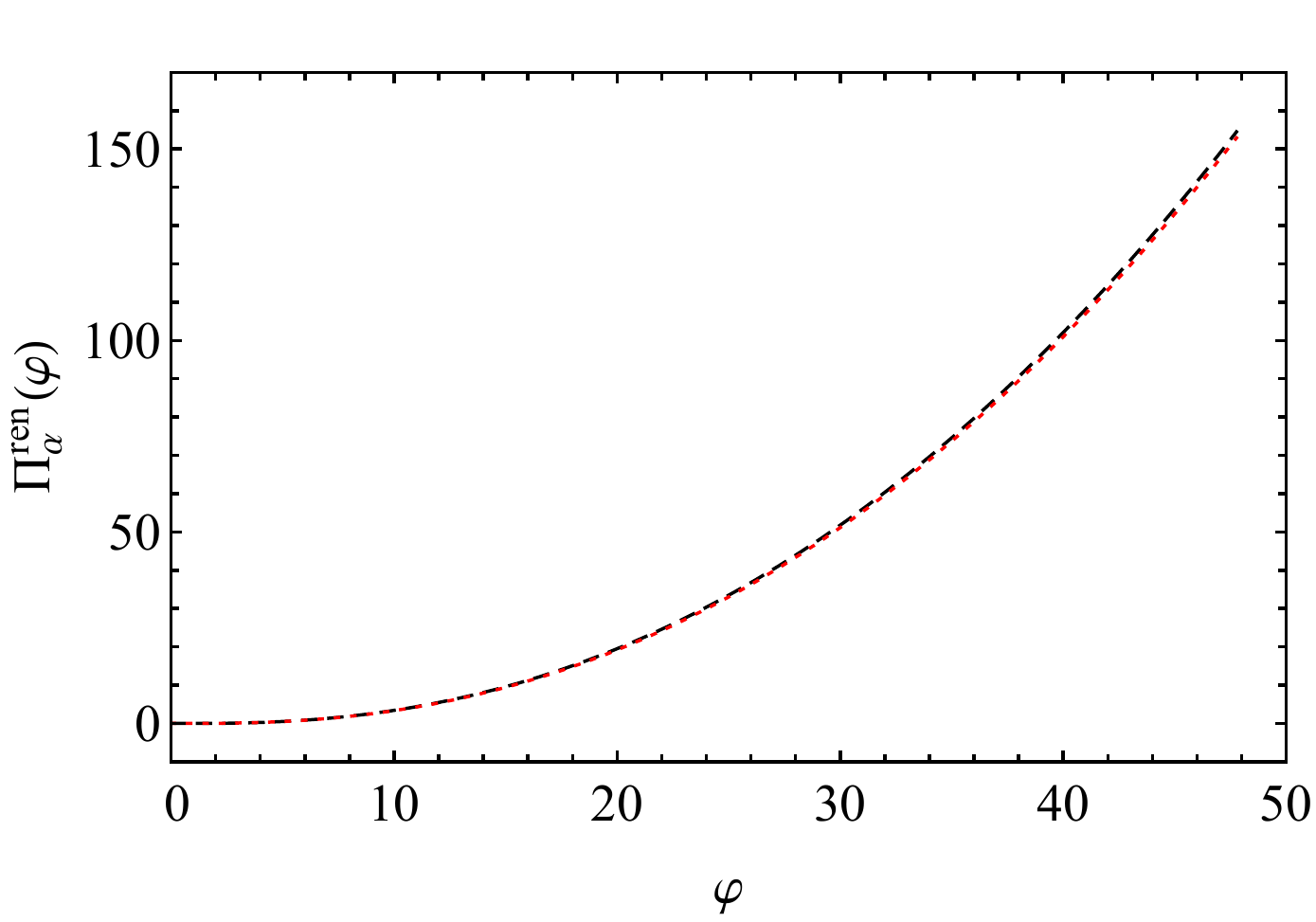}\\[1em]
\includegraphics[scale=0.6]{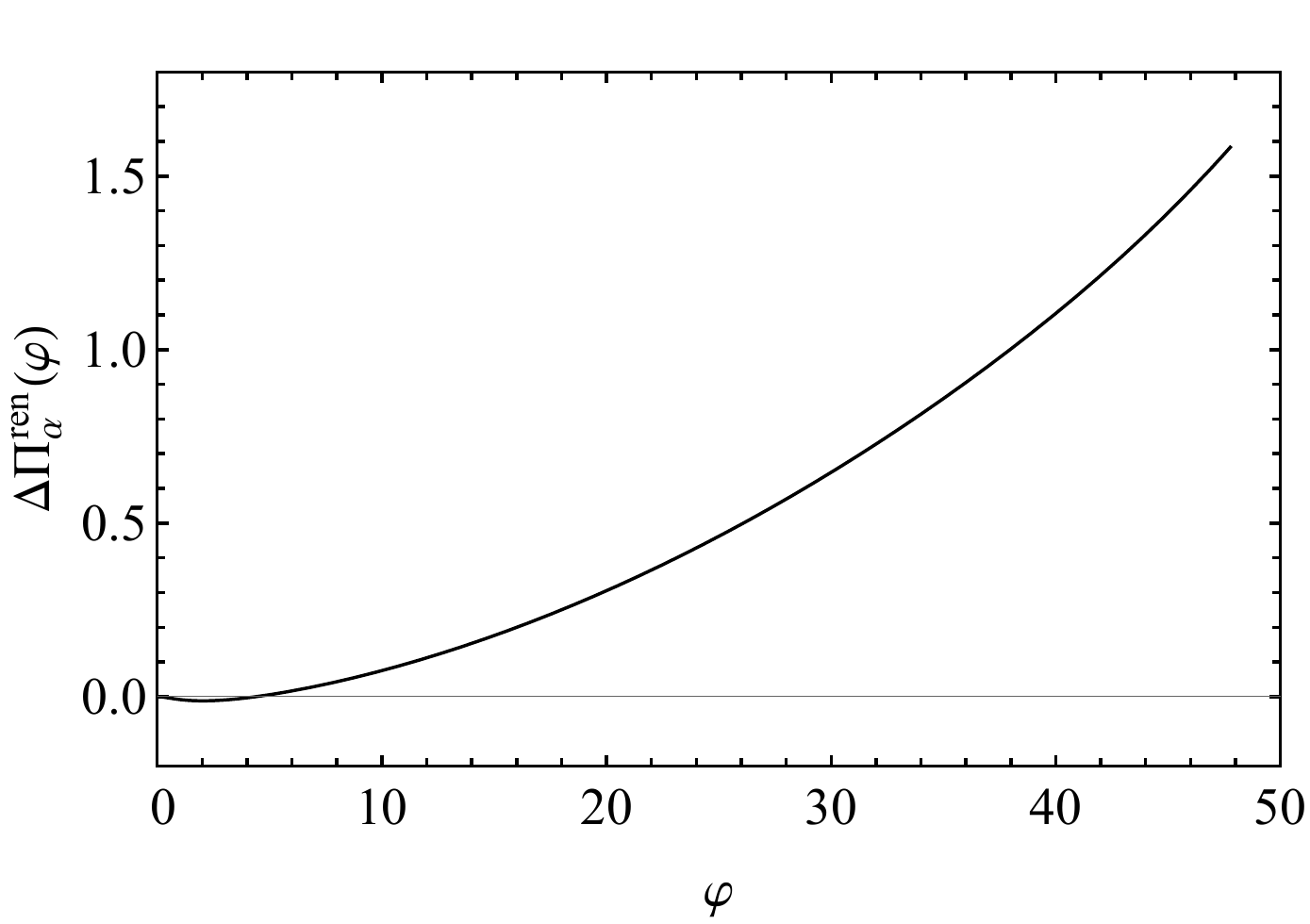}
\caption{\label{fig:Pialpha}Top panel: Plot of the renormalized self-energy $\Pi^{\rm ren}_{\alpha}(\varphi)$ from the self-consistent numerical procedure (black dashed) and based on the analytical results for the Green's function without gradients (red dotted). Bottom panel:  Difference between $\Pi^{\rm ren}_{\alpha}(\varphi)$ with and without gradients.}
\end{figure}

Finally, in \fref{fig:Pialpha}, we show a comparison of  $\Pi^{\rm ren}_{\alpha}(\varphi)$ with and without gradient corrections, where the latter quantity is given by
\begin{align}
&\Pi^{\rm ren, hom}_{\alpha}\ =\ 
\frac{N_\chi\alpha}{32\pi^2}
\bigg[
-\:\frac{\alpha}{2}\,\varphi^2
\nonumber\\&\qquad +\:\bigg(m_\chi^2+\frac{\alpha}{2}\varphi^2\bigg)
\ln\bigg(\frac{m_\chi^2+\frac{\alpha}{2}\varphi^2}{m_{\chi}^2}\bigg)
\bigg]\;.
\end{align}
The gradient effects are considerably smaller than for  $\Pi^{\rm ren}$. This can be attributed to the fact that, unlike the field $\Phi$, the fields $\chi$ do not become tachyonic for large values of $\varphi$, since $\alpha>0$.


\section{Conclusions}
\label{sec:conclusions}

In this paper, we have presented a Green's function method for calculating loop-improved bounce solutions in
classically scale-invariant models.

While the problem of tunneling in classically
scale-invariant scalar theory has been addressed in a number of earlier articles~\cite{Isidori:2001bm,Branchina:2013jra,Chigusa:2017dux,Andreassen:2017rzq,Chigusa:2018uuj}, the present method is
complementary in the following aspects:
\begin{itemize}
\item
\emph{Detemination of the bubble radius.}
Since the bubble radius is not fixed at the classical level, we have determined it through a
self-consistent solution for the bounce and the quantum fluctuations in its background, which
are expressed in terms of Green's functions. This allows us to find solutions that cannot be
constructed perturbatively from classical solutions in the present case because the parameter $R$
for the Fubini-Lipatov instanton is unknown a priori. In contrast, previous methods determine the
radius $R$ by selecting the scale where the running scalar coupling reaches its minimum, which then
also minimizes the tunneling action~\cite{Isidori:2001bm,Chigusa:2017dux,Andreassen:2017rzq}.
Note that such an approach is not applicable to the example discussed in this paper, where
only scalar loops are included, such that the scalar coupling is monotonically increasing. While
this exemplifies the complementarity of the different methods, it would be interesting to compare these when
applied to precisely the same models.

\item
\emph{Role of the approximate dilatational mode in the fluctuation spectrum.}
While the main scope of the present paper is to find self-consistent solutions for the bounce,
a number of articles compute the one-loop determinant around the Fubini-Lipatov instanton using the Gel'fand-Yaglom method. In both approaches, zero modes must be subtracted.
However, we find here that the dilatational mode resides at the end point of a continuum spectrum, such that
common regulation procedures for the functional determinant in the Gel'fand-Yaglom approach that
stipulate this mode to be discrete~\cite{Isidori:2001bm,Chigusa:2017dux,Andreassen:2017rzq} may need to be revisited.

\item
\emph{Resummation of infrared effects.}
Directly related to the previous point, our approach to dealing with the large infrared effects
for approximate dilatations is to perform a one-loop resummation of the Green's function for the angular
momentum $j=0$. In contrast, earlier approaches trade the dilatations for $R$ as a collective coordinate,
such that the integration over $R$ leads to the infrared corrections~\cite{Isidori:2001bm,Chigusa:2017dux,Andreassen:2017rzq}. In the future, it
would be very interesting to compare the two approaches and to chart the respective ranges of applicability.
In addition, we also find large infrared effects in the $j=1$ sector.

\item
\emph{Size of gradient corrections.}
Aside from the infrared effects,
we confirm that the bounce solutions receive contributions from the running coupling that can be computed when neglecting gradient corrections.
The latter are typically consistent with higher-order effects,~cf.~Refs.~\cite{Garbrecht:2015yza,Garbrecht:2017idb} for the thin-wall case. In the classically scale-invariant example of Sec.~\ref{sec:results}, the gradient corrections for the spectator fields, as well as corrections from angular momentum modes $j>1$ in general, appear to be perturbatively suppressed compared to the leading-order running
of the couplings.
However, close to scale invariance, the infrared-enhanced
contributions can become of equal significance compared to the running
of the couplings. It would be interesting to identify the functional form of the infrared
enhancement in future work.

\item
\emph{Higher-order terms from using translational collective coordinates.}
The present method accounts for the next-to-leading-order effects from the Jacobian associated
with the change to translational collective coordinates, which are known from quantum-mechanical problems~\cite{Aleinikov:1987wx,Olejnik:1989id}. However, due to the decomposition~\eqref{eq:Z:decomposition}
of the generating functional, we find a self-consistent solution to the Green's function and the bounce
in the subspace excluding the fluctuations associated with translations. The solutions in the full Hilbert space
can be obtained from these solutions by applying the corrections from the Jacobian. These should be included
in the future when, e.g., aiming to compute the decay rate to leading-loop order.

\item
\emph{Analytic form of the Green's functions and fluctuation spectra.}
We have found an intriguing connection between the archetypal example for tunneling between quasi-degenerate
vacua in field theories~\cite{Coleman:1977py, Callan:1977pt} and the classically scale-invariant models.
More precisely, while the Green's functions for both problems agree up to
the form of $\omega$, cf.~Eqs.~\eqref{eq:m2} and~\eqref{eq:mthickwall},
and an algebraic prefactor,
the eigenspectra and eigenfunctions are very different, which has a profound impact when handling
the translational zero modes and the approximate dilatations.
\end{itemize}

Apart from the comparison between different methods, further desirable developments in the present
framework include the following points:
\begin{itemize}

\item
The local truncation of the convolution integrals appearing in the Schwinger-Dyson equations should
be replaced by an improved approximation.

\item
Fermion and gauge fields should be coupled to the tunneling scalars, requiring the computation
of their Green's functions in the spherically symmetric background of the bounce. This will allow to investigate
the effects of a barrier generated by a negative running of the quartic coupling toward high scales
induced by fermion loops, which is of particular interest because, in the one-loop effective potential
for such a model, no bounces can be found. As for including the gauge fields, based on the present methods,
one may aim for transition rates that are manifestly gauge invariant up to a certain order in
perturbation theory~\cite{Plascencia:2015pga}.

\item
The functional determinant should be computed, possibly using methods similar to those in Ref.~\cite{Garbrecht:2015yza}, applied there to a thin-wall example.

\item
Due to the importance of self-consistent solutions for the bounce, a formulation of the present
problem in the framework of a two-particle-irreducible effective action would be of utility.
Because of the presence of the zero modes in the Green's functions, this may amount to a nontrivial technical step
of general interest.
\end{itemize}

Once these points are partly addressed, the present method will be applicable to a wider class
of models and parameter ranges. Examples are the question of metastability in $\Phi^4$ theory without
couplings to extra spectator fields, but for different renormalization conditions, and also a more direct
comparison with earlier approaches that includes gauge and Yukawa interactions will be possible. Of obvious
interest is the application to the Standard Model and other variants of electroweak symmetry breaking.


\begin{acknowledgments}

The authors would like to thank Mathias Garny, Hiren Patel, Julien Serreau, Edward Shuryak and Carlos Tamarit for helpful discussions. The work of P.M.~was supported by the Science and Technologies Facilities Council (STFC) under Grant No.~ST/L000393/1 and a Leverhulme Trust Research Leadership Award.

\end{acknowledgments}


\begin{appendix}


\section{Coleman-Weinberg effective potential}
\label{app:CW}

The renormalized one-loop Coleman-Weinberg effective potential for the model in Eq.~\eqref{eq:lag} with $m^2=0$ and $g=0$ takes the form
\begin{equation}
\label{eq:UR:eff:formula}
U^{\rm ren}_{\rm eff}\ =\ U\:+\:\delta U\:+\:\frac{1}{2}\int\!\frac{{\rm d}^4 k}{(2\pi)^4}\,\big[\ln\big(k^2\:+\:\lambda\,\varphi^2/2\big)\:-\:\ln\,k^2\big]\;,
\end{equation}
where the one-loop corrections have been normalized with respect to the false vacuum. For the ultraviolet regularization, we apply an ultraviolet cut-off $\Lambda$ to the three-momentum integral.

Now, we first consider the renormalization conditions
\begin{equation}
\label{eq:rc:CW}
\frac{\partial^2 U_{\rm eff}^{\rm ren}(\varphi)}{\partial \varphi^2}\Bigg|_{\varphi\,=\,0}\ =\ 0\;,\qquad \frac{\partial^4 U^{{\rm ren}}_{\rm eff}(\varphi)}{\partial \varphi^4}\Bigg|_{\varphi\,=\,\mu}\ =\ \lambda\;,
\end{equation}
following Coleman and Weinberg~\cite{Coleman:1973jx}, which yield the counterterms
\begin{subequations}
\begin{align}
\delta m^2\ &=\ -\:\frac{\lambda}{16\pi^2}\,\Lambda^2\;,\\
\delta \lambda\ &=\ -\:\frac{3\lambda^2}{32\pi^2}\,\bigg(\ln\bigg|\frac{\lambda \mu^2}{8\Lambda^2}\bigg|+\frac{14}{3}+i\pi\bigg)\;.
\end{align}
\end{subequations}
Note that the coupling-constant counterterm $\delta \lambda$ is complex, since $\lambda<0$. The renormalized one-loop effective potential is
\begin{equation}
\label{eq:CW}
U^{\rm ren}_{\rm eff}(\varphi)\ =\ \frac{1}{4!}\,\lambda\,\varphi^4\:+\:\frac{1}{256\pi^2}\,\lambda^2\,\varphi^4\bigg(\ln\,\frac{\varphi^2}{\mu^2}\:-\:\frac{25}{6}\bigg)\;.
\end{equation}
The potential in Eq.~\eqref{eq:CW} receives loop corrections $\sim\varphi^4\lambda^{n+1}\ln^{n}(\varphi/\mu)$ at $n$-loop order. Therefore, the straightforward perturbation expansion cannot be applied close to the false vacuum around $\varphi=0$, and the tunneling problem is not well defined
for these renormalization conditions.

Due to this problem, instead of Eq.~\eqref{eq:rc:CW}, we take $m^2\neq 0$ and make the choice
\begin{equation}
\label{eq:counters}
\frac{\partial^2 U_{\rm eff}^{\rm ren}(\varphi)}{\partial \varphi^2}\Bigg|_{\varphi\,=\,0}\ =\ m^2\;,\qquad \frac{\partial^4 U^{\rm ren}_{\rm eff}(\varphi)}{\partial \varphi^4}\Bigg|_{\varphi\,=\,0}\ =\ \lambda\;,
\end{equation}
yielding the counterterms
\begin{subequations}
\begin{align}
\delta m^2\ =&\ -\:\frac{\lambda}{32\pi^2}\left(2\Lambda^2+m^2\ln \frac{m^2}{4\Lambda^2}-m^2\right)\notag\\
&-\:\frac{N_\chi\alpha}{32\pi^2}\left(2\Lambda^2+m_\chi^2\ln \frac{m_\chi^2}{4\Lambda^2}-m_\chi^2\right)\;,
\\
\delta \lambda\ =&\ -\:\frac{3\lambda^2}{32\pi^2}\,\left(\ln\frac{m^2}{4\Lambda^2}+2\right)\notag\\
&-\:\frac{3 N_\chi\alpha^2}{32\pi^2}\,\left(\ln\frac{m_\chi^2}{4\Lambda^2}+2\right)
\;,
\end{align}
\end{subequations}
where we have included the contributions from the additional fields as per Eq.~\eqref{eq:speclag}. Substituting these into Eq.~\eqref{eq:UR:eff:formula} yields the result in Eq.~\eqref{eq:U:eff:ren}.

We emphasize that the radiative corrections necessarily require the introduction of one dimensionful scale, which is $\mu$ in Eq.~\eqref{eq:rc:CW} and $m$ in Eq.~\eqref{eq:U:eff:ren}. As long as there are no additional scales introduced, we can therefore refer to a breaking of scale invariance due to radiative effects in both cases. For the potential in Eq.~\eqref{eq:U:eff:ren}, this is especially true of field configurations for which $\varphi^2 \gg -\,m^2/\lambda$.


\section{Fubini-Lipatov Green's function}
\label{app:Greens}

In this appendix, we outline the calculation of the Green's function in the Fubini-Lipatov background. Beginning from the transformed problem in Eq.~\eqref{eq:thickwalltransdiff} with $\omega=j+1$, we recognize the homogeneous equation
\begin{equation}
\Bigg[\frac{\rm d}{{\rm d}u}\,(1-u^2)\,\frac{\rm d}{{\rm d}u}\:-\:\frac{\omega^2}{1-u^2}\:+\:6\Bigg]F_j(u,u')\ = \ 0
\end{equation}
as the associated Legendre differential equation. Splitting around the discontinuity, the general solutions for $u\gtrless u'$ are therefore of the form
\begin{equation}
\label{eq:Fjgen}
F_j^{\gtrless}(u,u')\ =\ A^{\gtrless}(u')P_2^{j+1}(u)\:+\:B^{\gtrless}(u')Q_2^{j+1}(u)\;,
\end{equation}
where $P_{\nu}^{\mu}$ and $Q_{\nu}^{\mu}$ are the associated Legendre polynomials. Note that Eq.~\eqref{eq:Fjgen} is strictly the general solution only for $j=0$ and $j=1$, since  $P_\nu^\mu$ and $Q_\nu^\mu$ are defined only for $\mu\leq \nu$. Nevertheless, we will later be able to extend the solution to $j>1$ by means of the Jacobi polynomials. For the time being, however, it is technically simpler to deal with the associated Legendre polynomials.

Matching around the discontinuity, we require
\begin{subequations}
\begin{gather}
F_j^>(u',u')\ =\ F_j^<(u',u')\;,\\
\lim_{u\,\to\,u'}\Bigg[\frac{{\rm d}}{{\rm d} u}\,F_j^>(u,u')-\frac{{\rm d}}{{\rm d} u}\,F_j^<(u,u')\Bigg]\ =\ -\,\frac{1}{1-u'^2}\;,
\end{gather}
\end{subequations}
from which it follows that
\begin{subequations}
\begin{gather}
A^>\:-\:A^<\ =\ \frac{Q_2^{1+j}(u')}{(2-j)_{2(1+j)}}\;,\\
B^{<}\:-\:B^{>}\ =\ \frac{P_2^{1+j}(u')}{(2-j)_{2(1+j)}}\;.
\end{gather}
\end{subequations}
Here, we have made use of the Wronskian
\begin{equation}
W[P_{\nu}^{\mu}(u),Q_{\nu}^{\mu}(u)]\ = \ \frac{(\nu-\mu+1)_{2\mu}}{1-u^2}\;,
\end{equation}
where $(z)_{\nu}$ is the Pochhammer symbol, defined as
\begin{equation}
(z)_{\nu}\ =\ \frac{\Gamma(z+\nu)}{\Gamma(z)}\;.
\end{equation}
We also require that $F_j(u,u')$ vanish as $u\to\pm\,1$, which implies
\begin{equation}
\frac{A^>}{B^>}\ =\ -\:\frac{\pi}{2}\,\cot (1+j)\pi\;,\qquad B^<\ =\ 0\;.
\end{equation}
Using the identity
\begin{equation}
\frac{\pi(\nu-\mu+1)_{2\mu}}{2\sin (\pi\mu)}\,P^{-\mu}_{\nu}(u)\ =\ \frac{\pi}{2}\,\cot (\pi\mu)\,P_{\nu}^{\mu}(u)\:-\:Q_{\nu}^{\mu}(u)\;,
\end{equation}
we therefore find
\begin{equation}
\label{eq:fgr}
F^>_j(u,u')\ =\ -\,\frac{\pi}{2}\,\mathrm{csc}(j\pi)\,P_2^{-j-1}(u)P_2^{j+1}(u')\;.
\end{equation}
We can extend this result to $j>1$ by reexpressing Eq.~\eqref{eq:fgr} in terms of the Jacobi polynomials \smash{$P_{\nu}^{(\alpha,\beta)}$} via the identity
\begin{equation}
\label{eq:assocLegendrecontinuation}
P_{\nu}^{\mu}(u)\ =\ \bigg(\frac{u+1}{u-1}\bigg)^{\frac{\mu}{2}}(\nu-\mu+1)_{\mu}\,P_{\nu}^{(-\mu,+\mu)}(u)\;,
\end{equation}
which again strictly holds only for $j\leq 1$. For $\nu=2$, the polynomial expansion terminates, and we have
\begin{align}
&P_2^{(\pm \mu,\mp \mu)}(u)\ =\ \frac{1}{2}\Big[(1\pm \mu)(2\pm \mu)\nonumber\\&\qquad -\:3(2\pm \mu)(1-u)\:+\:3(1-u)^2\Big]
\end{align}
for all $\mu$. Substituting this expansion into Eq.~\eqref{eq:fgr} with $\mu=\omega=j+1$ and after some algebra, we arrive at the expression for the hyperradial Green's function in Eq.~\eqref{eq:Gj}.


\section{Orthonormality of the Jacobi polynomials}
\label{app:Jacfuncim}

The associated Legendre polynomials satisfy the familiar orthonormality condition
\begin{equation}
\int_{-1}^{+1}\frac{{\rm d}u}{1-u^2}\;P_{\nu}^{\mu}(u)P_{\nu}^{\mu'}(u)\ =\ \frac{(\nu+\mu)!}{\mu(\nu-\mu)!}\,\delta_{\mu\mu'}\;.
\end{equation}
Using the identity
\begin{equation}
P_{\nu}^{\mu}(u)\ =\ (-1)^{\mu}\frac{(\nu+\mu)!}{(\nu-\mu)!}\,P^{-\mu}_{\nu}(u)\;,
\end{equation}
it follows that
\begin{equation}
\int_{-1}^{+1}\frac{{\rm d}u}{1-u^2}\;P_{\nu}^{+\mu}(u)P_{\nu}^{-\mu'}(u')\ =\ \frac{(-1)^{\mu}}{\mu}\,\delta_{\mu\mu'}\;.
\end{equation}
This can be reexpressed in terms of the Jacobi polynomials via Eq.~\eqref{eq:assocLegendrecontinuation}, giving
\begin{align}
&\int_{-1}^{+1}\frac{{\rm d}u}{1-u^2}\;\bigg(\frac{u+1}{u-1}\bigg)^{+\frac{\mu}{2}}\bigg(\frac{u+1}{u-1}\bigg)^{-\frac{\mu'}{2}}\nonumber\\&\quad\times P_{\nu}^{(-\mu,+\mu)}(u)P_{\nu}^{(+\mu',-\mu')}(u)\nonumber\\&\quad =\ \frac{(-1)^{\mu}}{\mu}\,\frac{\delta_{\mu\mu'}}{(\nu-\mu+1)_{+\mu}(\nu+\mu+1)_{-\mu}}\;.
\end{align}
Making use of the fact that when $\nu\in\mathbb{N}$
\begin{align}
&(\nu-\mu+1)_{+\mu}(\nu+\mu+1)_{-\mu}\nonumber\\&\qquad =\ \frac{(-1)^{\nu}(\nu!)^2\sin(\pi \mu)}{\pi \mu[\nu^2-\mu^2][(\nu-1)^2-\mu^2]\cdots[1-\mu^2]}\;,
\end{align}
we have
\begin{align}
&\int_{-1}^{+1}\frac{{\rm d}u}{1-u^2}\;\bigg(\frac{u+1}{u-1}\bigg)^{+\frac{\mu}{2}}\bigg(\frac{u+1}{u-1}\bigg)^{-\frac{\mu'}{2}}\nonumber\\&\qquad \times P_{\nu}^{(-\mu,+\mu)}(u)P_{\nu}^{(+\mu',-\mu')}(u)\ =\ \frac{(-1)^{\mu+\nu}\pi}{(\nu!)^2\sin(\pi \mu)}\nonumber\\&\qquad \times \: [\nu^2-\mu^2][(\nu-1)^2-\mu^2]\cdots[1-\mu^2]\,\delta_{\mu\mu'}\;,
\end{align}
giving Eq.~\eqref{eq:Jacorthogdisc} for $\nu=2$ and $\mu=n$.

The associated Legendre functions of imaginary order satisfy the following orthonormality condition for integer degree $\nu$~\cite{Bielski} (see also Refs.~\cite{Hutasoit:2009xy,Kalmykov})
\begin{equation}
\int_{-1}^{+1}\frac{{\rm d}u}{1-u^2}\;P_{\nu}^{+i\xi}(u)P_{\nu}^{-i\xi'}(u)\ =\ \frac{2\sinh(\pi \xi)}{\xi}\,\delta(\xi-\xi')\;.
\end{equation}
This can be reexpressed in terms of the Jacobi polynomials with imaginary parameters via Eq.~\eqref{eq:assocLegendrecontinuation}. Namely,
\begin{align}
\label{eq:Jacorthinter}
&\int_{-1}^{+1}\frac{{\rm d}u}{1-u^2}\;\bigg(\frac{u+1}{u-1}\bigg)^{+\frac{i\xi}{2}}\bigg(\frac{u+1}{u-1}\bigg)^{-\frac{i\xi'}{2}}\nonumber\\&\quad\times P_{\nu}^{(-i\xi,+i\xi)}(u)P_{\nu}^{(+i\xi',-i\xi')}(u)\nonumber\\&\quad =\ \frac{2\sinh(\pi \xi)}{\xi}\,\frac{\delta(\xi-\xi')}{(\nu-i\xi+1)_{+i\xi}(\nu+i\xi+1)_{-i\xi}}\;.
\end{align}
For $\nu\in\mathbb{N}$, we have
\begin{align}
&(\nu-i\xi+1)_{+i\xi}(\nu+i\xi+1)_{-i\xi}\nonumber\\&\qquad =\ \frac{(\nu !)^2\sinh(\pi \xi)}{\pi\xi[\nu^2+\xi^2][(\nu-1)^2+\xi^2]\cdots[1+\xi^2]}\;,
\end{align}
and therefore
\begin{align}
&\int_{-1}^{+1}\frac{{\rm d}u}{1-u^2}\;\bigg(\frac{u+1}{u-1}\bigg)^{+\frac{i\xi}{2}}\bigg(\frac{u+1}{u-1}\bigg)^{-\frac{i\xi'}{2}}\nonumber\\&\quad\times P_{\nu}^{(-i\xi,+i\xi)}(u)P_{\nu}^{(+i\xi',-i\xi')}(u)\nonumber\\&\quad =\ \frac{2\pi}{(\nu!)^2}\,[\nu^2+\xi^2][(\nu-1)^2+\xi^2]\cdots[1+\xi^2]\,\delta(\xi-\xi')\;.
\end{align}
We then immediately recover Eq.~\eqref{eq:Jacorthog} for $\nu=2$.


\section{Spectral sum representation}
\label{app:spectral}

In this appendix, we include further details of the derivation of the spectral sum representation of the Green's functions.
We begin by writing the Green's function for the Fubini-Lipatov case in the form
\begin{align}
&G^{\rm FL}(x,x')\ =\ \frac{1}{2\pi^2}\sum_{j\,=\,0}^{\infty}(j+1)U_j(\cos\theta)\nonumber\\&\qquad \times\:\left[\sum_{\lambda^{\rm FL}\,\in\, L^{\rm FL}_{{\rm d}j}}\:+\!\!\int\limits_{\lambda^{\rm FL}\, \in\, L^{\rm FL}_{{\rm c}j}}\!\!\!\!\!\!\!\!\frac{{\rm d}\lambda^{\rm FL}}{2\pi}\right]\frac{\phi^{-}_{\lambda^{\rm FL} j}(r')\phi^+_{\lambda^{\rm FL} j}(r)}{\lambda^{\rm FL}}\;,
\end{align}
where the eigenfunctions \smash{$\phi^{\pm}_{\lambda^{\rm FL} j}(r)$} compose the true eigenspectrum. As described in Sec.~\ref{sec:specsum}, we now define \smash{$\tilde{\phi}^{\pm}_{\lambda^{\rm FL} j}(u)\equiv r\phi^{\pm}_{\lambda^{\rm FL} j}(r)$} and change variables via Eq.~\eqref{eq:cvars}. We then have
\begin{align}
&G^{\rm FL}(x,x')\ =\ \frac{1}{2\pi^2R^2}\bigg(\frac{1+u}{1-u}\bigg)^{\!\frac{1}{2}}\bigg(\frac{1+u'}{1-u'}\bigg)^{\!\frac{1}{2}}\sum_{j\,=\,0}^{\infty}(j+1)\nonumber\\&\times U_j(\cos\theta)\!\left[\sum_{\bar{\lambda}^{\rm FL}\,\in\, \bar{L}^{\rm FL}_{{\rm d}j}}\:+\!\!\int\limits_{\bar{\lambda}^{\rm FL}\, \in\, \bar{L}^{\rm FL}_{{\rm c}j}}\!\!\!\!\!\!\!\!\frac{{\rm d}\bar{\lambda}^{\rm FL}}{2\pi}\right]\!\frac{\tilde{\phi}^{-}_{\lambda^{\rm FL} j}(u')\tilde{\phi}^+_{\lambda^{\rm FL} j}(u)}{\bar{\lambda}^{\rm FL}}\;,
\end{align}
where $\bar{\lambda}^{\rm FL}=R^2\lambda^{\rm FL}$ are the dimensionless eigenvalues. In terms of the thin-wall eigenbasis, viz.~the $f_{\lambda^{\rm TW}j}^{\pm}$ of Sec.~\ref{sec:specsum}, we can write
\begin{align}
\tilde{\phi}_{\lambda^{\rm FL} j}^{\pm}(u)\ &=\ \left[\sum_{\bar{\lambda}^{\rm TW}\,\in\, \bar{L}^{\rm TW}_{{\rm d}j}}\right.\nonumber\\&\qquad\left.+\!\!\int\limits_{\bar{\lambda}^{\rm TW}\, \in\, \bar{L}^{\rm TW}_{{\rm c}j}}\!\!\!\!\!\!\!\!\frac{{\rm d}\bar{\lambda}^{\rm TW}}{2\pi}\right]a_{\lambda^{\rm FL} \lambda^{\rm TW}}^{\pm}\,\tilde{\phi}^{\pm}_{\lambda^{\rm TW}j}(u)\;,
\end{align}
where 
\begin{equation}
\tilde{\phi}^{\pm}_{\lambda^{\rm TW}j}(u)\ \equiv\ \Bigg(\frac{1+u}{1-u}\Bigg)^{\!\pm\frac{\varpi}{2}}f^{\pm}_{\lambda^{\rm TW} j}(u)\;,
\end{equation}
$\varpi$ is defined in Eq.~\eqref{eq:varpidef}, and the amplitudes
\begin{align}
a^{\pm}_{\lambda^{\rm FL} \lambda^{\rm TW}}\ &\equiv\ R^2\int_{-1}^{+1}\frac{{\rm d}u}{(1+u)^2}\;\bigg(\frac{1+u}{1-u}\bigg)\nonumber\\&\qquad\times\:\tilde{\phi}^{\pm}_{\lambda^{\rm TW} j}(u)\tilde{\phi}^{\pm}_{\lambda^{\rm FL} j}(u)\;.
\end{align}
Here, we isolate explicitly in the second parenthesis the weight function $(1+u)/(1-u)$ in the orthogonality of the thin-wall basis [cf.~Eq.~\eqref{eq:innerprod}]. Now, by virtue of the completeness of the thin-wall basis, we can also write
\begin{align}
&G^{\rm FL}(x,x')\ =\ \frac{1}{2\pi^2R^2}\sum_{j\,=\,0}^{\infty}(j+1)U_j(\cos\theta)\nonumber\\&\times\left[\sum_{\bar{\lambda}^{\rm TW}\,\in\, \bar{L}_{{\rm d}j}^{\rm TW}}\:+\!\!\int\limits_{\bar{\lambda}^{\rm TW} \,\in\, \bar{L}_{{\rm c}j}^{\rm TW}}\!\!\!\!\!\!\!\!\frac{{\rm d}\bar{\lambda}^{\rm TW}}{2\pi}\right]\Bigg(\frac{1+u}{1-u}\Bigg)^{\!+\frac{\varpi+1}{2}}\nonumber\\&\times\:\Bigg(\frac{1+u'}{1-u'}\Bigg)^{\!-\frac{\varpi-1}{2}}\frac{f^{-}_{\lambda^{\rm TW} j}(u')f^+_{\lambda^{\rm TW} j}(u)}{\bar{\lambda}^{\rm TW}}\;.
\end{align}
It immediately follows that
\begin{align}
\label{eq:basistrans}
&\left[\sum_{\bar{\lambda}^{\rm FL}\,\in\, L_{{\rm d}j}}\:+\!\!\int\limits_{\bar{\lambda}^{\rm FL}\, \in\, \bar{L}^{\rm FL}_{{\rm c}j}}\!\!\!\!\frac{{\rm d}\bar{\lambda}^{\rm FL}}{2\pi}\right]\frac{a^-_{\lambda^{\rm FL} \lambda^{\rm TW\prime}}a^+_{\lambda^{\rm FL}\lambda^{\rm TW}}}{\bar{\lambda}^{\rm FL}}\nonumber\\&\ \overset{!}{=}\ \frac{1}{\bar{\lambda}^{\rm TW}}\begin{cases} \delta_{\bar{\lambda}^{\rm TW}\bar{\lambda}^{\rm TW\prime}}\;,&\ \bar{\lambda}^{\rm TW},\:\bar{\lambda}^{\rm TW\prime}\ \in\ \bar{L}^{\rm TW}_{{\rm d}j}\;,\\
2\pi\delta(\bar{\lambda}^{\rm TW}-\bar{\lambda}^{\rm TW\prime})\;,&\ \bar{\lambda}^{\rm TW},\:\bar{\lambda}^{\rm TW\prime}\ \in\ \bar{L}^{\rm TW}_{{\rm c}j}\;.
 \end{cases}
\end{align}
In the thin-wall case, the two bases, of course, coincide, and Eq.~\eqref{eq:basistrans} is trivially satisfied. For the Fubini-Lipatov case, only the zero modes of the two bases coincide, as described further in Sec.~\ref{sec:specsum}. On the other hand, the nonzero modes in the Fubini-Lipatov case are a linear combination of the discrete and continuum thin-wall eigenfunctions. This situation is summarized in Table~\ref{tab:eigen}.

We can now write the following representations of both the thin-wall and Fubini-Lipatov Green's functions (in the thin-wall basis) in terms of the dimensionless eigenvalues $\bar{\lambda}^{\rm TW}$:
\begin{align}
&G(x,x')\ =\ \frac{1}{2\pi^2R^2}\begin{Bmatrix} \frac{1}{\gamma R}\\ \Big(\frac{1+u}{1-u}\Big)\Big(\frac{1+u'}{1-u'}\Big)\end{Bmatrix}\nonumber\\& \times\sum_{j\,=\,0}^{\infty}(j+1)U_j(\cos\theta)\left[\sum_{\bar{\lambda}^{\rm TW}\,\in\, \bar{L}^{\rm TW}_{{\rm d}j}}\:+\!\!\int\limits_{\bar{\lambda}^{\rm TW}\,\in\, \bar{L}^{\rm TW}_{{\rm c}j}}\!\!\!\!\!\!\!\!\frac{{\rm d}\bar{\lambda}^{\rm TW}}{2\pi}\right]\nonumber\\&\times\:\Bigg(\frac{1+u}{1-u}\Bigg)^{\!+\frac{\varpi}{2}}\Bigg(\frac{1+u'}{1-u'}\Bigg)^{\!-\frac{\varpi}{2}}\frac{f^{-}_{\lambda^{\rm TW} j}(u')f^+_{\lambda^{\rm TW}}(u)}{\bar{\lambda}^{\rm TW}}\;.
\end{align}
We emphasize that the $\bar{\lambda}^{\rm TW}$ differ between the thin-wall and Fubini-Lipatov cases due to the difference in the values of $\omega$ [cf.~Eqs.~\eqref{eq:m2} and \eqref{eq:mthickwall}]. The \smash{$f^{\pm}_{\lambda^{\rm TW}j}(u)$} are the solutions of the Jacobi differential equation in Eq.~\eqref{eq:Jacobidiff}; namely, the Jacobi polynomials of degree 2:
\begin{equation}
\label{eq:Jacobiexplicit}
P_2^{(\mp\varpi,\pm\varpi)}(u)\ =\ \frac{1}{2}\big(3u^2-1\mp3\varpi u+\varpi^2\big)\;.
\end{equation}
The discrete modes correspond to $\varpi=n\in\{1,2\}$ and the continuum modes to $\varpi=i\xi$  ($\xi\in \mathbb{R}$).

The normalization of the eigenfunctions follows from the orthogonality of the Jacobi polynomials, described in App.~\ref{app:Jacfuncim}. We find that the contribution from the discrete modes is
\begin{align}
\label{eq:discreteG}
G_{\rm d}(x,x')\ &=\ \frac{1}{2\pi^2R^2}\begin{Bmatrix} \frac{1}{\gamma R}\\ \Big(\frac{1+u}{1-u}\Big)\Big(\frac{1+u'}{1-u'}\Big)\end{Bmatrix}\nonumber\\&\quad \times\:\sum_{j\,=\,0}^{\infty}(j+1)U_j(\cos\theta)\nonumber\\&\quad \times\bigg[-\:\frac{3}{2}\,\frac{uu'}{1-\omega^2}\,\sqrt{1-u^2}\sqrt{1-u^{\prime2}}\nonumber\\&\quad \quad-\:\frac{3}{4}\,\frac{1}{4-\omega^2}\,(1-u^2)(1-u^{\prime2})\bigg]\;,
\end{align}
and that from the continuum modes is
\begin{align}
G_{\rm c}(x,x')\ &=\ \frac{1}{2\pi^2R^2}\begin{Bmatrix} \frac{1}{\gamma R}\\ \Big(\frac{1+u}{1-u}\Big)\Big(\frac{1+u'}{1-u'}\Big)\end{Bmatrix}\nonumber\\&\qquad\times\:\sum_{j\,=\,0}^{\infty}(j+1)U_j(\cos\theta)I(u,u')\;,
\end{align}
where
\begin{align}
\label{eq:continuumint}
I(u,u')\ &\equiv\ \int_{-\infty}^{+\infty}\frac{{\rm d}\xi}{2\pi}\;\frac{4}{(4+\xi^2)(1+\xi^2)}\nonumber\\&\qquad\times\:\Bigg(\frac{u+1}{u-1}\Bigg)^{\!+i\xi/2}\Bigg(\frac{u'+1}{u'-1}\Bigg)^{\!-i\xi/2}\nonumber\\&\qquad\times\:\frac{P_2^{(-i\xi,+i\xi)}(u)P_2^{(+i\xi,-i\xi)}(u')}{\omega^2+\xi^2}\;.
\end{align}
Defining
\begin{equation}
L\ \equiv\ \ln\frac{1+u}{1-u}-\ln\frac{1+u'}{1-u'}\ \begin{cases} >0\;,\quad u\ > \ u'\\ <0\;,\quad u\ < \ u'\end{cases}\;,
\end{equation}
and making use of Eq.~\eqref{eq:Jacobiexplicit},
the integral in Eq.~\eqref{eq:continuumint} can be written
\begin{align}
&I(u,u')\ =\ \int_{-\infty}^{+\infty}\frac{{\rm d}\xi}{2\pi}\;\frac{e^{i\xi L/2}}{(1+\xi^2)(4+\xi^2)(\omega^2+\xi^2)}\nonumber\\&\ \times\:\big(1\:-\:3u^2\:+\:3iu\xi\:+\:\xi^2\big)\big(1\:-\:3u^{\prime 2}\:-\:3iu'\xi\:+\:\xi^2\big)\;.
\end{align}
After partial fractioning, we can decompose this as $I=I_1+I_2+I_{\omega}$, where
\begin{subequations}
\begin{align}
I_1\ &=\ \int_{-\infty}^{+\infty}\frac{{\rm d}\xi}{2\pi}\;\frac{3uu'}{1-\omega^2}\big[1\:-\:u u'\:-\:i(u-u')\xi\big]\frac{e^{i \xi L/2}}{1+\xi^2}\;,\\
I_2\ &=\ \int_{-\infty}^{+\infty}\frac{{\rm d}\xi}{2\pi}\;\frac{3}{4-\omega^2}\big[1\:+\:u^2\:-\;4uu'\:+\:u^{\prime 2}\:+\:u^2u^{\prime 2}\nonumber\\&\quad-\:i(u-u')(1-uu')\xi\big]\frac{e^{i \xi L/2}}{4+\xi^2}\;,\\
I_{\omega}\ &=\ \int_{-\infty}^{+\infty}\frac{{\rm d}\xi}{2\pi}\;\frac{1}{(1-\omega^2)(4-\omega^2)}\nonumber\\&\quad \times\:\Big[(1-\omega^2-3u^2)(1-\omega^2-3u^{\prime2})-\:9 u u' \omega^2\nonumber\\&\quad+\:3 i (u-u')(1-\omega^2+3 u u')\xi\Big]\frac{e^{i\xi L/2}}{\omega^2+\xi^2}\;.
\end{align}
\end{subequations}
Making use of the integrals
\begin{subequations}
\begin{gather}
\int_{-\infty}^{+\infty}\frac{\rm{d} \xi}{2\pi}\;\frac{1}{\xi^2+a^2}\,e^{i \xi L/2}\ =\ \frac{1}{2a}\,e^{-a|L|/2}\;,\\
\int_{-\infty}^{+\infty}\frac{\rm{d} \xi}{2\pi i}\;\frac{\xi}{\xi^2+a^2}\,e^{i \xi L/2}\ =\ \frac{1}{2}\,\mathrm{sgn}(L)\,e^{-a|L|/2}\;,
\end{gather}
\end{subequations}
where $\mathrm{sgn}$ is the signum function, we find
\begin{subequations}
\begin{align}
I_1\ &=\ \frac{3}{2}\,\frac{uu'}{1-\omega^2}\,\sqrt{1-u^2}\sqrt{1-u^{\prime 2}}\;,\\
I_2\ &=\ \frac{3}{4}\,\frac{1}{4-\omega^2}\,(1-u^2)(1-u^{\prime 2})\;,\\
I_{\omega}\ &=\frac{1}{2\omega}\,\vartheta(u-u') \bigg(\frac{1-u}{1+u}\bigg)^{\!\omega/2}\bigg(\frac{1+u'}{1-u'}\bigg)^{\!\omega/2}\nonumber\\&\qquad\times\:\frac{3u^2+3 u \omega+\omega^2-1}{(1+\omega)(2+\omega)}\,\frac{3u^{\prime 2}-3 u' \omega+\omega^2-1}{(1-\omega)(2-\omega)}\nonumber\\&\qquad+\:(u\leftrightarrow u')\;.
\end{align}
\end{subequations}
The terms arising from $I_1$ and $I_2$ exactly cancel those terms arising from the discrete modes in Eq.~\eqref{eq:discreteG}. Notice, in particular, that there are no unit step functions in $I_1$ and $I_2$. Putting everything together, and taking the coincident limit, we quickly arrive at the results presented in Sec.~\ref{sec:specsum}.

\end{appendix}


\end{document}